\documentclass[pdflatex,twocolumn,epjc3]{svjour3}          
\pdfoutput=1
\usepackage[T1]{fontenc}

\smartqed  
\usepackage{graphicx}

\usepackage{mathptmx}      
\usepackage{flushend}

\usepackage[numbers,sort&compress]{natbib}
\usepackage[citecolor=blue,urlcolor=blue,linkcolor=blue]{hyperref}
\usepackage{amssymb}
\usepackage{dcolumn}
\usepackage{latexsym}
\usepackage{rotating}
\usepackage{booktabs}
\usepackage{tabularx}
\usepackage{color}
\usepackage{tabularx}

\def\be{\begin{equation}}
\def\ee{\end{equation}}
\def\bea{\begin{eqnarray}}
\def\eea{\end{eqnarray}}

\newcommand{\il}{~}

\journalname{Eur. Phys. J. C}

\begin{document}
\title{Observers in Kerr spacetimes:  the ergoregion on  the equatorial plane}

\author{
D. Pugliese\thanksref{e1,addr1}, H. Quevedo\thanksref{addr2,addr3,addr4}
}

\thankstext{e1}{e-mail:d.pugliese.physics@gmail.com}

\institute{Institute of Physics, Faculty of Philosophy \& Science,
  Silesian University in Opava,
 Bezru\v{c}ovo n\'{a}m\v{e}st\'{i} 13, CZ-74601 Opava, Czech Republic\label{addr1}
 \and
          Dipartimento di Fisica, Universit\`a di Roma ``La Sapienza", I-00185 Roma, Italy\label{addr2}
\and Instituto de Ciencias Nucleares, Universidad Nacional Aut\'onoma de M\'exico,  AP 70543, M\'exico, DF 04510, Mexico\label{addr3}
\and Department of Theoretical and Nuclear Physics, Kazakh National University, Almaty 050040, Kazakhstan
\label{addr4}
}

\date{Received: date / Accepted: date}

\maketitle

\begin{abstract}
We perform a detailed analysis of the properties of stationary observers located on the equatorial plane of the ergosphere in a Kerr spacetime, including light-surfaces.
 This study highlights crucial differences between black hole and the super-spinner sources.
In the case of Kerr naked singularities, the results allow us to distinguish  between ``weak''  and ``strong '' singularities,
corresponding to spin values close to or distant from the limiting case of extreme black holes, respectively.
We derive important limiting  angular frequencies for naked singularities. We especially study very weak singularities
as resulting from the spin variation of black holes.
We also explore the main properties of zero angular momentum  observers for different classes of black hole and naked singularity spacetimes.
\end{abstract}

\maketitle

\section{Introduction}
%

The physics of black holes (\textbf{BHs}) is probably one of the most complex and still controversial aspects  of  Einstein's  geometric theory of gravitation.
Many processes of   High Energy Astrophysics   are supposed to involve singularities and  their formation from a stellar progenitor collapse  or from the merging of a binary \textbf{BH} system.
The interaction of these sources with  the matter environment, which can lead to accretion and jets emission, is the basis for many observed phenomena.
As a consequence of this interaction, the singularity properties,  determined generally  by the  values of their  intrinsic spin, mass or electric charge parameters,
might be  modified, leading to considerable changes of  the singularity itself.
In this work, we  concentrate our analysis  on the ergoregion in the naked singularity (\textbf{NS}) and \textbf{BH} regimes of
the axisymmetric and stationary Kerr  solution.
We are concerned also about the implications of any spin-mass ratio oscillation between the \textbf{BH} and the \textbf{NS} regimes from the viewpoint of stationary observers and  their frequencies,  assuming the invariance of the system symmetries
(axial symmetry   and time independence). 
One of the  goals of this work is to explore the existence of spin transitions in very weak naked singularities \cite{ergon}, which are characterized by a spin parameter $a/M \approx 1$.
If the collapse of a  stellar object  or the   merging of several stellar or   \textbf{BH} attractors  lead  to the formation of a naked singularity,
then a  total or partial destruction of the horizon may occur which should be accompanied by oscillations  of the spin-to-mass ratio.
Naked singularities can also appear in non-isolated  \textbf{BH} configurations as the result of their interaction with the surrounding matter, i. e., { in some transient process of the evolution of an interacting black hole}.
Indeed, the interaction can lead to modifications of
characteristic \textbf{BH}  parameters, for instance, through a  spin-up or spin-down process which can also alter the spacetime symmetries. The details of such spin transitions, leading possibly to the destruction of the horizon, and their consequences are still an open problem.

 In this work,  keeping the Kerr spacetime symmetries unchanged,  we focus on the  variation of the dimensionless spin parameter  in the region within the static limit on the equatorial plane of the attractor, this being the plane of symmetry of the Kerr solution. This special plane of the  axisymmetric geometry  has many interesting properties; for instance,  constants of motion emerge due to  the symmetry under reflection with respect to this plane; the geometry has   some peculiarities that make it immediately comparable with  the limiting  static Schwarzschild solution, in particular,
the location of the outer ergoregion boundary is  independent of the spin value, and  coincides with  the location of the Schwarzschild horizon.
There is also a clear astrophysical interest in the exploration of such a plane, as the large majority  of accretion disks are considered to be  located on the equatorial plane of their attractors.

From a methodological viewpoint, our analysis represents a  comparative study of stationary and static observers in Kerr spacetimes for any range of the  spin parameter.
The findings in this work highlight major differences between the  behavior of these observers in \textbf{BH} and \textbf{NS} geometries.
 These issues are clearly related to the most general and widely discussed problem of defining \textbf{BHs}, their event horizon and their  intrinsic thermodynamic properties
\cite{BB,WW,Fiola:1994ir,Nikolic:2009ju,Bradler:2013gqa,LSS}.
Further, it seems compelling to clarify the role of the static limit and of the ergoregion in some of the well-analyzed astrophysical processes such  as the singularity formation,  through the gravitational collapse of a stellar ``progenitor'' or  the  merging of two
\textbf{BHs}. Similarly, it is interesting to analyze the role of the frame-dragging effect in driving the accretion processes.
In fact, the ergosphere plays  an important
role in the energetics of rotating black holes.

The dynamics inside the ergoregion  is relevant in Astrophysics for  possible observational effects, since in this region the Hawking radiation can be analyzed and the  {Penrose energy extraction  process occurs
\cite{Penrose71,[3],[4],[5],[6]}\footnote{The Hawking process is  essentially due to the vacuum fluctuation happening  in the regions close to the \textbf{BH} horizon; it is  not related to the properties of the ergoregion itself.  The Hawking radiation is  the (spontaneous) emission of thermal radiation which is {created in the vacuum regions surrounding } a \textbf{BH}, and leads to a decrease of the mass. Connected in many ways to the   Unruh  effects, it generally leads to the production of pairs of particles, one  escaping to infinity while the other is  trapped by the \textbf{BH} horizon.
On the other hand, the Penrose  energy extraction, or its wave-analogue of super-radiance, is related essentially to  a classical (i.e. non quantum) phenomenon occurring in the ergoregion, $]r_+,r_{\epsilon}^+[$, due to the frame-dragging of the spinning spacetime. In this way, energy can be extracted from the source, lowering  its angular momentum. For a study of the Hawking radiation in Kerr and Kerr-Newman spacetimes see also \cite{Gray:2015pma}}}.
{ For the actual state of the Penrose process, see \cite{Bejger:2012yb}. Another interesting effect connected directly to the ergoregion is discussed in \cite{Stuchlik:2004wk}. }
 The mechanism, by which energy from compact spinning objects is extracted,
is of  great astrophysical interest and  the effects occurring inside the ergoregion of black holes are essential
for understanding the central engine mechanism of these processes \cite{Meier,Fro-Z}.
Accreting matter  can even get out, giving rise, for example, to jets of matter or radiation \cite{Meier,Gariel:2013hwa} originated inside the ergoregion.
Another possibility is the extraction of  energy from a rotating black hole  through the Blandford-Znajek mechanism (see,
for instance, \cite{RZN,PVP,P11,AO02,PCGG,BSZ10,FS10,IHK85,SKC,HHJEPJC}). {An interesting alternative scenario for the role of the Blandford-Znajek process in the
acceleration of jets is presented in\cite{Pei:2016kka}. Further discussions on the Penrose and Blandford-Znajek  processes may be found in	\cite{Komissarov:2008yh,Lasota:2013kia}}.
In general, using orbits entering the ergosphere, energy can be extracted from
a Kerr black hole or a naked singularity.
On the other hand, naked singularity solutions have been studied in different contexts in \cite{Stuchlik:2006in,Stuchlik:2014jua,Kotrlova:2014ana,Boshkayev,Kolos:2013bca,Schee:2013bya,Stuchlik:2013yca,Torok:2005ct,Poit,A-wen,Stuchlik:2012zza,Stuchlik:2010zz}.  Kerr naked singularities as particle accelerators are considered  in \cite{Patil}-- {see also \cite{Stuchlik:2012zza,Stuchlik:2013yca}.
More generally, Kerr naked singularities can be relevant in  connection to superspinars, as
discussed in \cite{Stuchlik:2012zza}. The stability of
Kerr superspinars has been analyzed  quite recently in \cite{Nakao:2017rgv}, assenting the importance of boundary conditions in dealing with perturbations of \textbf{NSs}.}

An interesting perspective exploring duality between elementary particles and black holes,
pursuing
quantum black holes as the link between
microphysics and macrophysics, can  be found in \cite{Lake:2015pma,Carr:2017grh,Carr:2015nqa,Carr:2014mya}-- {see also \cite{Prok:2008ev}}.
A general discussion on the similarities between characteristic  parameter values of \textbf{BHs} and \textbf{NSs}, in comparison with particle like objects, is addressed  also in  \cite{Pu:Charged,Pu:class,Pu:KN,Pu:Kerr}.
Quantum evaporation of \textbf{NSs} was analyzed  in \cite{Goswami:2005fu},  radiation in  \cite{Vaz:1998gd}, and gravitational radiation
in  \cite{Iguchi:1999ud,Iguchi:1998qn,Iguchi:1999jn}.

Creation and  stability of naked singularities are still intensively debated  \cite{Sha-teu91,ApTho,J-S09,Jacobson:2010iu,Esitenza,Giacomazzo:2011cv}. A  discussion on the ergoregion stability can be found in \cite{Cardoso:2007az,CoSch}.
 However, under quite general conditions  on the progenitor, these analysis  do not exclude   the possibility
that { considering instability processes}  a naked singularity can be
produced as the result of a  gravitational collapse.
These studies,  based upon a numerical integration of the corresponding field equations, often consider  the   stability of the progenitor models and investigate  the  gravitational collapse
 of differentially rotating neutron stars in full general relativity \cite{miller}.
Black hole formation is then  associated with the formation of trapped surfaces. As a consequence of this, a singularity without trapped surfaces,  as the result of a numerical integration, is usually considered as a proof of its  naked  singularity nature.
 However, the non existence of trapped surfaces after or  during  the gravitational collapse is not in general a  proof of the existence of a naked singularity.  As  shown in \cite{Wald:1991zz}, in fact, it is possible to choose a very particular slicing of spacetime during the formation of a spherically symmetric black hole  where  no trapped surfaces exist (see also \cite{Joshi-Book}).
 Eventually, the process of gravitational collapse towards the formation of \textbf{BHs} (and therefore, more generally, the issues concerning the formation or not of a horizon and hence of \textbf{NSs}) is still, in spite of several studies, an open problem. There are
transition periods   of   transient dynamics, possibly involving  topological  deformations of the spacetime, in which we  know  the past and future asymptotic regions of the spacetime, but it is still  in fact largely unclear what happens during that process.
The problem is wide and involves many factors as, especially in non-isolated systems,  the role of matter and  symmetries during collapse.
Another  major process that leads to  black hole formation  is the merging of  two (or more) black holes, recently  detected for the first time in the gravitational waves sector \cite{GW}. {See also \cite{TheLIGOScientific:2017qsa,Monitor:2017mdv} for the first observation  of the probable formation of a \textbf{BH} from the coalescence of two neutron stars}.
{An interesting and detailed analysis of  Kerr and  Kerr-Newman naked singularities  in the broader context of  braneworld Kerr-Newman
(\textbf{B-KN}) spacetimes can be found in \cite{Blaschke:2016uyo},
 where a new kind of instability, called mining instability,  of some \textbf{B-KN} naked singularity spacetimes   was found. In there, the exploration of the   ``causality violation region'' 
is also faced. This  is  the region where the angular  coordinate becomes  timelike,    leading eventually to    closed timelike curves. Details on the relation between this region and the Kerr ergoregion can be found in the aforementioned reference.}

{In  \cite{Pu:Charged,Pu:class,Pu:Kerr,Pu:KN,ergon,Pu:Neutral}, we  focused on the  study of axisymmetric  gravitational fields,
exploring different  aspects of spacetimes with \textbf{NSs} and \textbf{BHs}. The results of this analysis show a clear difference between naked singularities and black holes from the point of view
of the stability properties of circular orbits\footnote{Test particle motion can be used to determine the topological properties of
general relativistic spacetimes \cite{Malament,Hawkingt,geroch}.  Moreover, we proved that in certain \textbf{NS} geometries different regions of  stable timelike circular orbits are separated from each other by empty regions; this means that an accretion disk made of test
particles will show a particular ring-like structure with specific topological properties.}.
This fact  would have significant consequences for the extended matter surrounding the central source
and, hence,  in all processes associated with energy extraction.
Indeed, imagine an accretion disk made of test particles which are
moving along circular orbits on the equatorial plane of a Kerr spacetime. It turns out that in the case of a black hole the accretion
disk is continuous whereas in the case of a naked singularity it is discontinuous. This means that we can  determine the values of intrinsic physical parameters of the central attractor by analyzing
the geometric and topological properties of the corresponding Keplerian accretion disk.
 In addition, these disconnected  regions, in the case of a naked singularities, are a consequence of the repulsive gravity properties found  also in many other black hole  solutions and in some extensions or modifications of   Einstein's theory.
 The effects of repulsive gravity  in the case of the Kerr geometry were considered in  \cite{deFelice} and \cite{FDEFELICE1}. Analogies between the  effects of repulsive gravity and the presence of a
cosmological constant was shown also to occur in  regular black hole spacetimes or in strong gravity objects without horizons \cite{zS,Schee:2015nua}}.

Several studies have already shown that it is  necessary to distinguish between
weak $(a/M\approx 1)$ and strong naked singularities $(a/M>>1)$.
It is also possible to introduce a similar classification for black holes; however, we prove here that only in the case of  naked singularities there are obvious fundamental distinctions between these classes which are not present among the different black hole classes.
Our focus is on  strong \textbf{BHs},  and weak and  very weak \textbf{NSs}.
This analysis confirms  the distinction between {strong} and {weak} \textbf{NSs} and \textbf{BHs}, characterized  by  peculiar limiting values for the spin parameters. Nevertheless, the existence  and meaning of such limits is  still  largely unclear, and more investigation is due.
However, there are  indications about  the existence of  such limits in different geometries,  where weak and strong singularities  could appear.
 In \cite{ergon,Pu:Charged,Pu:class,Pu:KN,Pu:Kerr}, it  was established that the motion of test particles on the equatorial plane of black hole spacetimes can be used to derive information about the structure of the central source of gravitation; moreover, typical effects of repulsive gravity  were observed in the  naked singularity ergoregion (see also \cite{Gariel:2014ara,Pelavas:2000za,Herdeiro:2014jaa,Stuchlik:2014jua}).
In addition,   it was pointed out that there exists a  dramatic  difference between black holes and naked singularities  with respect to the zero and  negative energy states in circular orbits
{(stable circular geodesics with negative energy were for the first time discussed in \cite{Stu80}}).
The static limit would act  indeed as a {semi-permeable membrane} separating the spacetime  region, filled with negative energy particles, from the external one, filled with positive energy particles, gathered from  infinity or expelled from the ergoregion with   impoverishment of the source energy.
The membrane is selective because it acts so as to filter the material in transient between the inner  region and outside the static limit.
This membrane wraps and selectively isolates  the horizon in Kerr  black holes and the singularity in  superspinning solutions, partially isolating  it from the outer region by letting selectively rotating infalling or outgoing matter to cross  the static limit.
As mentioned above, the ergoregion is  involved in the \textbf{BH} spin-up and spin-down  processes leading to a radical change of the dynamical structure of the region closest to the source  and,  therefore,  potentially could give rise to detectable effects.
It is  possible that, during the   evolutionary phases of  the rotating object interacting with the orbiting matter, there can be some evolutionary stages of spin adjustment,  for example,  in the proximity of the extreme value ($a\lesssim M$) where  the speculated spin-down of the \textbf{BH} can occur preventing the formation of  a naked singularity with $a \gtrsim M$
(see also \cite{Esitenza,Gao:2012ca,
Poit,
Stuchlik:2011zza,
vanPutten,
Gammie:2003qi,
Abo,
Kesden:2011ma,
Wald74,
J-S09,Pradhan:2012yx}). The study  of extended matter  configurations  in the Kerr ergoregion is faced for example in  \cite{pugtot,ergon}.
 In \cite{pugtot,ringed,open,dsystem,long}, a model of multi-accretion disks, so called ringed accretion disks, both corotating and counterrotating on the equatorial plane of a Kerr \textbf{BH},  has been proposed, and a model for such ringed accretion disks was developed.
 Matter can eventually be captured by the  accretion disk,  increasing or removing part of its energy and angular momentum,  therefore prompting a shift of its spin  \cite{Li:2013sea,Jacobson:2010iu,Stu80,BiStuBa89,BaBiStu89,Kovar:2010ty}.
A further remarkable aspect of this region is that the outer boundary on the equatorial plane of the central singularity is invariant for every spin change, and coincides with the radius of the
{ horizon of the static case. In the limit of zero rotation, the  outer ergosurface  coalesces}
with the event horizon. The {extension} of this  region  increases with the spin-to-mass ratio,  but the outer limit is invariant.
Although on the equatorial plane the ergoregion is invariant  with respect to any transformation involving a change in the source spin  (but not with respect to a change in the mass $M$), the dynamical structure  of the ergoregion  is not invariant with respect to a change in the spin-to-mass ratio. Nevertheless, concerning the invariance of this region with respect to spin shifts  it has been argued, for example in   \cite{Hawking},  that
the  ergoregion  cannot indeed disappear as a consequence of a change in spin, {because  it may  be }  filled by negative energy matter
{provided by the emergence of a  Penrose process\footnote{We note that the  wave analog of the Penrose process is the superradiant scattering.} \cite{Penrose71}.
 The presence of negative   energy particles, a distinctive feature of the  ergoregion of any spinning source in any range of the spin value,   has special properties when it comes to the circular motion in weakly rotating naked singularities. The presence of this special matter  in an  ``antigravity''
sphere,  possibly filled with negative energy formed according to the Penrose process,
  and bounded by orbits with zero angular momentum,   is expected to play an important role in the source evolution.
In this work, we clarify and deepen those results, formulate in detail those considerations, analyze  the static limit, and perform
 a detailed   study  of this region   from the point of view of  stationary observers. In this regards, we mention also the interesting and recent results published in  \cite{Chakraborty:2016mhx} and \cite{Frolov:2014dta}.

In detail, this article is organized as follows: in Sec.\il\ref{Sec:first} we  discuss the main properties  of the Kerr solution and  the features of the ergoregion in the equatorial plane of the  Kerr spacetimes.  Concepts and notation used throughout this work are also introduced. Stationary observers in \textbf{BH} and \textbf{NS}  geometries are introduced in  Sec.\il\ref{Sec:1-st}.
 Then, in Sec.\il\ref{Sec:saz}, we investigate the case of zero angular momentum observers and find
all the spacetime configurations in which they can exist.  Finally, in Sec. \ref{sec:con}, we discuss our results.

\section{Ergoregion properties in the Kerr spacetime}\label{Sec:first}
The Kerr metric is an axisymmetric,   stationary (nonstatic), asymptotically flat exact solution of  Einstein's equations  in  vacuum.
In  spheroidal-like   Boyer--Lindquist  (BL) coordinates, the line element can be written as
\bea\nonumber && ds^2=-dt^2+\frac{\rho^2}{\Delta}dr^2+\rho^2
d\theta^2+(r^2+a^2)\sin^2\theta
d\phi^2\\\label{alai}&&+
\frac{2M}{\rho^2}r(dt-a\sin^2\theta d\phi)^2\ ,
\\
&&
\Delta\equiv r^2-2Mr+a^2,\quad\mbox{and}\quad\rho^2\equiv r^2+a^2\cos^2\theta \ .
\eea
%
%
%
The parameter $M\geq0$  is      interpreted as  the mass  parameter, while  the rotation parameter $a\equiv J/M\geq0$
(\emph{spin}) is  the  \emph{specific} angular momentum, and   $J$ is the
\emph{total} angular momentum of the gravitational source.
The spherically symmetric (static)  Schwarzschild solution  is a   limiting case for  $a=0$.

  A Kerr black hole (\textbf{BH}) geometry is  defined by the range of the spin-mass ratio   $a/M\in ]0,1[ $, the extreme black hole case corresponds to $a=M$, whereas a super-spinner Kerr compact object or  a naked singularity (\textbf{NS}) geometry occurs  when $a/M>1$.

The Kerr solution  has several symmetry properties.
The Kerr metric tensor (\ref{alai}) is invariant under the application of any two different transformations:
\(\mathbf{\mathcal{P}}_{\mathbf{Q}}:\mathbf{Q}\rightarrow-\mathbf{Q},
\)
where $\mathbf{Q}$  is one of the coordinates $(t,\phi)$ or the metric parameter $a$ while  a single transformation
leads to a spacetime with an opposite  rotation with respect to the unchanged metric.
The metric element is  independent of the coordinate  $t$ and the angular coordinate
$\phi$. The solution is stationary due to the presence of the   {Killing} field $\xi_{t}=\partial_{t} $ and the geometry is  axisymmetric as shown by the presence of the rotational Killing field  $\xi_{\phi}=\partial_{\phi} $.

An observer orbiting, with uniform angular velocity, along the curves $r=$constant and $\theta=$constant
will not see the spacetime  changing during its motion.
{As a consequence of this, the covariant
components $p_{\phi}$ and $p_{t}$ of the particle four--momentum are
conserved along the   geodesics\footnote{We adopt the
geometrical  units $c=1=G$ and  the  signature $(-,+,+,+)$,
Greek indices run in $\{0,1,2,3\}$.  The   four-velocity  satisfy $u^\alpha u_\alpha=-1$.
The radius $r$ has units of
mass $[M]$, and the angular momentum  units of $[M]^2$, the velocities  $[u^t]=[u^r]=1$
and $[u^{\phi}]=[u^{\theta}]=[M]^{-1}$ with $[u^{\phi}/u^{t}]=[M]^{-1}$ and
$[u_{\phi }/u_{t}]=[M]$. For the sake of convenience, we always consider a
dimensionless  energy and effective potential $[V_{eff}]=1$ and an angular momentum per
unit of mass $[L]/[M]=[M]$.}
and we can introduce the constants of motion
\be\label{Eq:after}
{\mathcal{E}} \equiv -g_{\alpha \beta}\xi_{t}^{\alpha} p^{\beta},\quad \mathcal{L} \equiv
g_{\alpha \beta}\xi_{\phi}^{\alpha}p^{\beta}.
\ee
 The constant of motion (along  geodesics) $\mathcal{L}$ is interpreted as the angular momentum  of the particle as measured by an observer at infinity,
and
we may interpret $\mathcal{E}$, for
timelike geodesics, as  the total energy of a test particle
coming from radial infinity, as measured  by  a static observer located  at infinity.}

As a consequence of the metric tensor  symmetry under reflection with respect to  the  equatorial hyperplane $\theta=\pi/2$, the  equatorial (circular) trajectories   are confined in the equatorial geodesic plane.
Several remarkable surfaces characterize these geometries:
For black hole and extreme black hole spacetimes the radii
\bea
r_{\pm}\equiv M \pm\sqrt{M^2-a^2}:\;  g^{rr}=0
\eea
{are the  event  outer and inner (Killing) horizons}\footnote{A Killing horizon is a \emph{null} surface, $\mathcal{S}_0$, whose \emph{null} generators coincide with the orbits of an
one-parameter group of isometries (i. e., there is a Killing field $\mathcal{L}$
which is
normal
to $\mathcal{S}_0$). Therefore, it  is  a lightlike hypersurface (generated by the flow of a Killing vector) on
which the norm of a Killing vector goes to zero. In static  \textbf{BH} spacetimes, the
event, apparent, and Killing horizons  with respect to the  Killing field   $\xi_t$ coincide.
In the Schwarzschild spacetime, therefore, $r=2M$ is the  Killing horizon with \emph{respect} to the  Killing vector
$\partial_t$.  The event horizons  of a spinning \textbf{BH}  are   Killing horizons   with respect to  the Killing field $\mathcal{L}_h=\partial_t +\omega_h \partial_{\phi}$, where  $\omega_h$ is  defined as the angular velocity of the horizon. In this article we shall extensively discuss this special vector in the case of \textbf{NS} geometries.
We note here that the surface gravity of a \textbf{BH} may be defined as the  rate at which the norm of the Killing vector vanishes from the outside.
The surface gravity, $\mathcal{SG}_{Kerr}= (r_+-r_-)/2(r_+^2+a^2)$, is a conformal invariant of the metric, but it rescales with the conformal Killing vector.
Therefore, it  is not the same on all generators (but obviously it is constant along one specific generator because of the symmetries).
 }},
 whereas
\bea
r_{\epsilon}^{\pm}\equiv M\pm
\sqrt{M^2-a^2cos^2\theta}: \, g_{tt}=0
\eea
are the outer and inner \emph{ergosurfaces},  respectively\footnote{{
In the  Kerr solution,  the Killing vector $\partial_t$, representing  time translations at infinity, becomes null  at the outer  boundary of the ergoregion, $r_{\epsilon}^{+}$, which is however a timelike  surface; therefore, $r_{\epsilon}^{+}$ is \emph{not} a Killing horizon.
More precisely, on the ergosurfaces the time translational Killing vector becomes null.}},
with $r_{\epsilon}^{-}\leq r_-\leq r_+\leq r_{\epsilon}^{+}$.
In an  extreme \textbf{BH} geometry, the horizons coincide, $r_-=r_+ =M$, and  the relation
$
r_{\epsilon}^{\pm}=r_{\pm}$ is valid on the rotational axis
(i.e., when  $\cos^2\theta=1$).

In this work, we will deal particularly with  the  geometric  properties  of the  \emph{ergoregion}
$\Sigma_{\epsilon}^+:\;]r_+,r_{\epsilon}^+]$; in this region, we have that  $g_{tt}>0$  on the
\emph{equatorial plane} $(\theta=\pi/2)$ and also  { $\left. r_{\epsilon}^+\right|_{\pi/2}=\left.r_+\right|_{a=0}=2M$}
and $r_{\epsilon}^-=0$.
The outer boundary $r_{\epsilon}^+$ is known as the \emph{static} (or also stationary)   limit \cite{GriPod09}; it is  a \emph{timelike}
surface except on the axis of the Kerr source where it matches the outer horizon and becomes  null-like.
On the equatorial plane of symmetry,  $\rho=r$ and the spacetime singularity is located at $r=0$.
In the naked singularity case, where  the singularity at $\rho=0$ is not covered by a horizon, the region
$\Sigma_{\epsilon}^+$ has a toroidal topology   centered on the  axis with the inner
circle located on the  singularity.
On the equatorial plane, as $a\rightarrow 0$  the geometry ``smoothly" resembles the spherical symmetric case,
$r_+\equiv \left.r_{\epsilon}^+\right|_{\pi/2}$, and the frequency of the signals emitted by an infalling particle in motion
towards $r=2M$, as seen by an observer at infinity, goes to zero.

In general, for $a\neq0$ and  $r\in\Sigma_{\epsilon}^+$, the metric component $g_{tt}$ changes its sign and vanishes
for $r=r_{\epsilon}^{+}$ (and $\cos^2\theta\in]0,1]$).
In the  ergoregion,  the Killing vector $\xi_t^{\alpha} = (1, 0, 0, 0)$ becomes spacelike, i.e.,  {$g_{\alpha\beta}\xi_t^\alpha\xi_t^\beta=g_{tt}>0$}.
{As the quantity  $\mathcal{E}$,  introduced in  Eq.\il(\ref{Eq:after}), is  associated to the   Killing field $\xi_{t}=\partial_{t} $, then  the particle energy   can be also negative inside $\Sigma_{\epsilon}^+$.}
For stationary spacetimes ($a\neq0$)     in $\Sigma_{\epsilon}^+$,  the motion with $\phi=const$ is \emph{not} possible and all particles are forced to rotate with the source, i.e.,  $\dot{\phi}a>0$.
This fact implies in particular that  an observer with  four-velocity  proportional
to $\xi_t^\alpha$ so that $\dot{\theta}=\dot{r}=\dot{\phi}=0$, {(the dot denotes  the derivative with respect to the  proper
time $\tau$ along the trajectory)},
cannot exist inside the ergoregion.
Therefore, for any infalling matter (timelike or photonlike) approaching  the horizon $r_+$ in the region $\Sigma_{\epsilon}^+$,
it holds that $t\rightarrow\infty$ and $\phi\rightarrow\infty$, implying   that the world-lines  around the horizon, as long as $a\neq0$, are subjected to  an  infinite twisting.
On  the other hand, trajectories  with $r=const$ and $\dot{r}>0$ (particles  crossing the static limit and escaping outside
in the region $r\geq r_{\epsilon}^+$) are possible.

Concerning the frequency of a signal emitted by a source in motion along the boundary of the ergoregion $r_{\epsilon}^+$,
it is clear that the proper time of the source particle
is not null\footnote{However, since  $g_{tt}(r_{\epsilon}^{\pm})=0$, it is also known as an infinity redshift surface;
 see, for example, \cite{GriPod09}.}.
Then, for an observer at infinity, the particle  will reach and penetrate the  surface $r=r_{\epsilon}^+$, in general, in a finite time $t$. For this reason, the ergoregion boundary is \emph{not} a surface of infinite redshift, except for the axis of rotation where the ergoregion coincides {with the event horizon}
\cite{Landau,ergon}. This means that an observer at infinity
will see a non-zero emission frequency. In the spherical symmetric case $(a=0)$, however, as $g_{t\phi}=0$
the  proper time interval $d\tau=\sqrt{|g_{tt}|}dt$  goes to zero as one approaches $r=r_+=r_{\epsilon}^+$.
For a timelike particle with positive energy (as measured by an observer at infinity), it is  possible to cross the static limit and
to escape towards infinity.
In Sec.\il\ref{Sec:1-st}, we introduce  stationary observers in \textbf{BH} and \textbf{NS}  geometries. We find the explicit expression for the angular velocity of stationary observers,
and perform a detailed analysis of its behavior in terms of the radial distance to the source and of the angular momentum of the gravity source. We find all the conditions that must be satisfied for a  light-surface to exist.
\section{Stationary observers and light surfaces}
\label{Sec:1-st}
We start our analysis by considering  \emph{stationary observers} which  are defined as
 observers whose  tangent vector is a spacetime  Killing vector; their  four-velocity  is therefore a
linear combination of the two Killing vectors $\xi_{\phi}$ and $\xi_{t}$,  i.e.,  the coordinates  $r$ and $\theta$ are constants along  the worldline of a stationary observer    \cite{Poisson}. As a consequence of this property, a stationary observer does not see the spacetime changing along its trajectory. It is convenient to introduce  the (uniform) \emph{angular velocity} $\omega$ as
\be\label{Eq:spectrum}
{d\phi}/{dt}={u^{\phi}}/{u^t}\equiv\omega,\quad\mbox{or}\quad
u^\alpha=\gamma(\xi_t^\alpha+\omega \xi_\phi^\alpha),
\ee
which is a dimensionless quantity. Here, $\gamma$ is a normalization factor
\be\label{Eq:sic-six}
\gamma^{-2}\equiv-\kappa(\omega^2 g_{\phi\phi}+2\omega g_{t\phi}+g_{tt}) ,
\ee
where $g_{\alpha\beta} u^\alpha u^\beta=-\kappa$.
The particular case  $\omega=0$ defines {\em static observers}; these   observers  cannot exist in the ergoregion.

The angular velocity of a timelike stationary  observer ($\kappa=+1$) is defined within the interval
\bea\label{Eq:ex-ce}
\omega\in]\omega_-,\omega_+[ \quad\mbox{where}\quad \omega_{\pm}\equiv \omega_{Z}\pm\sqrt{\omega_{Z}^2-\omega _*^2},
\\\nonumber
\omega _*^2\equiv \frac{g_{tt}}{g_{\phi \phi}}=\frac{g^{tt}}{g^{\phi\phi}},\quad \omega_{Z}\equiv-\frac{g_{\phi t}}{g_{\phi\phi}},
\eea
as illustrated in
{
Figs.\il\ref{Fig:Pilosrs1} and \ref{Fig:Conf3Dna}-\emph{right},  where the frequencies $\omega_{\pm}$ are plotted for fixed values of
$r/M$ and as functions of the spacetime spin  $a/M$ and radius $r/M$, respectively.}
In particular, the combination
\be\label{Eq:comb}
\mathcal{L}_{\pm}\equiv \xi_{t}+\omega_{\pm}\xi_{\phi}
\ee
{defines null curves, $g_{\alpha\beta}\mathcal{L}^\alpha_{\pm}\mathcal{L}^\beta_{\pm}=0$,  and, therefore, as we shall see in detail below, the frequencies
 $\omega_{\pm}$ are limiting angular velocities for physical observers, defining a family of {null} curves,
rotating  with the velocity $\omega_{\pm}$ around the axis of symmetry.}
{The Killing vectors $\mathcal{L}_{\pm} $ are also generators of Killing event  horizons.
The Killing vector $\xi_t+\omega\xi_{\phi}$ becomes null at r = $r_+$.
At the horizon $\omega_+=\omega_-$ and, consequently, stationary observers cannot exist inside this surface.
}
\subsection{The frequencies $\omega_{\pm}$}
\label{Sec:w-t-1}
We are concerned here with the orbits $r=$const and $\omega=$const,  which are eligible for stationary  observers.
This analysis enlightens the differences between \textbf{NS} and \textbf{BH} spacetimes.   Inside the ergoregion,
 { the quantity in parenthesis in the r.h.s. of Eq.\il(\ref{Eq:sic-six}) is well defined} for any source. However, it becomes null  for photon-like particles and the rotational frequencies $\omega_{\pm}$.
On the equatorial plane, the frequencies $\omega_{\pm}$ are given as
\bea\label{Eq:b-y-proc}
&&
\omega_{\pm}\equiv\frac{2 aM^2\pm M\sqrt{r^2 \Delta}}{r^3+a^2 (2M+r)}\\
&&\nonumber \mbox{with}\quad  \omega_{\pm}(r_+)= \omega_{Z}(r_+)=\omega_{h}\equiv\frac{a}{2 r_+}\equiv \frac{M}{2\omega_0 r_+},
 \\
 && \nonumber\mbox{and}\quad \lim_{r\rightarrow\infty}\omega_{\pm}=0,\quad 
\lim_{r\rightarrow0}\omega_{\pm}=
\omega_0\equiv\frac{M}{a}\ .
\eea

Moreover, for the case of very strong naked singularities $a\gg M$, we obtain that $\omega_{\pm}\rightarrow 0$.

The above quantities are closely related to the main black hole characteristics, and determine also the main features that distinguish
\textbf{NS} solutions from \textbf{BH} solutions.
The  constant $\omega_h$ plays a crucial role for the characterization of black holes, including their thermodynamic properties.
It also determines the  uniform (rigid) angular velocity on the horizon, representing the fact that the black hole rotates rigidly.
This quantity enters directly into the definition of the \textbf{BH} surface gravity and, consequently, into the formulation of the rigidity theorem and into the expressions for the Killing vector (\ref{Eq:spectrum}).
More precisely,
 the  Kerr  \textbf{BH} surface gravity  is defined as $\kappa =\kappa_s-\gamma_a$, where $\kappa_s\equiv { {1}/{4M}}$ is the Schwarzschild surface gravity, while  $\gamma_a=M\omega_{h}^{2}$  (the effective spring constant, according to \cite{Kook}) is the contribution due to the additional component of the
\textbf{BH} intrinsic spin; $\omega_{h}$ is therefore the  angular velocity (in units of $1/M$) on the \emph{event horizon}.
The (strong) rigidity theorem { connects then the event horizon with a Killing  horizon  stating  that,  under suitable conditions,   the event horizon of a stationary (asymptotically flat solution with
matter satisfying suitable hyperbolic equations) \textbf{BH} is a Killing horizon}\footnote{
 Assuming the cosmic censorship validity, the
gravitational collapse should lead to  \textbf{BH} configurations.
The surface area of the \textbf{BH} event horizon
is non-decreasing with time (which is the content of the second   law of black hole thermodynamics).
 The \textbf{BH} event horizon of this
stationary  solution
is a { Killing horizon
with constant surface gravity (zeroth  law)}
\cite{Chrusciel:2012jk,Carter:1997im,WW,Wald:1999xu}. }.

\begin{figure}[h!]
\centering
\begin{tabular}{l}
\includegraphics[scale=.3]{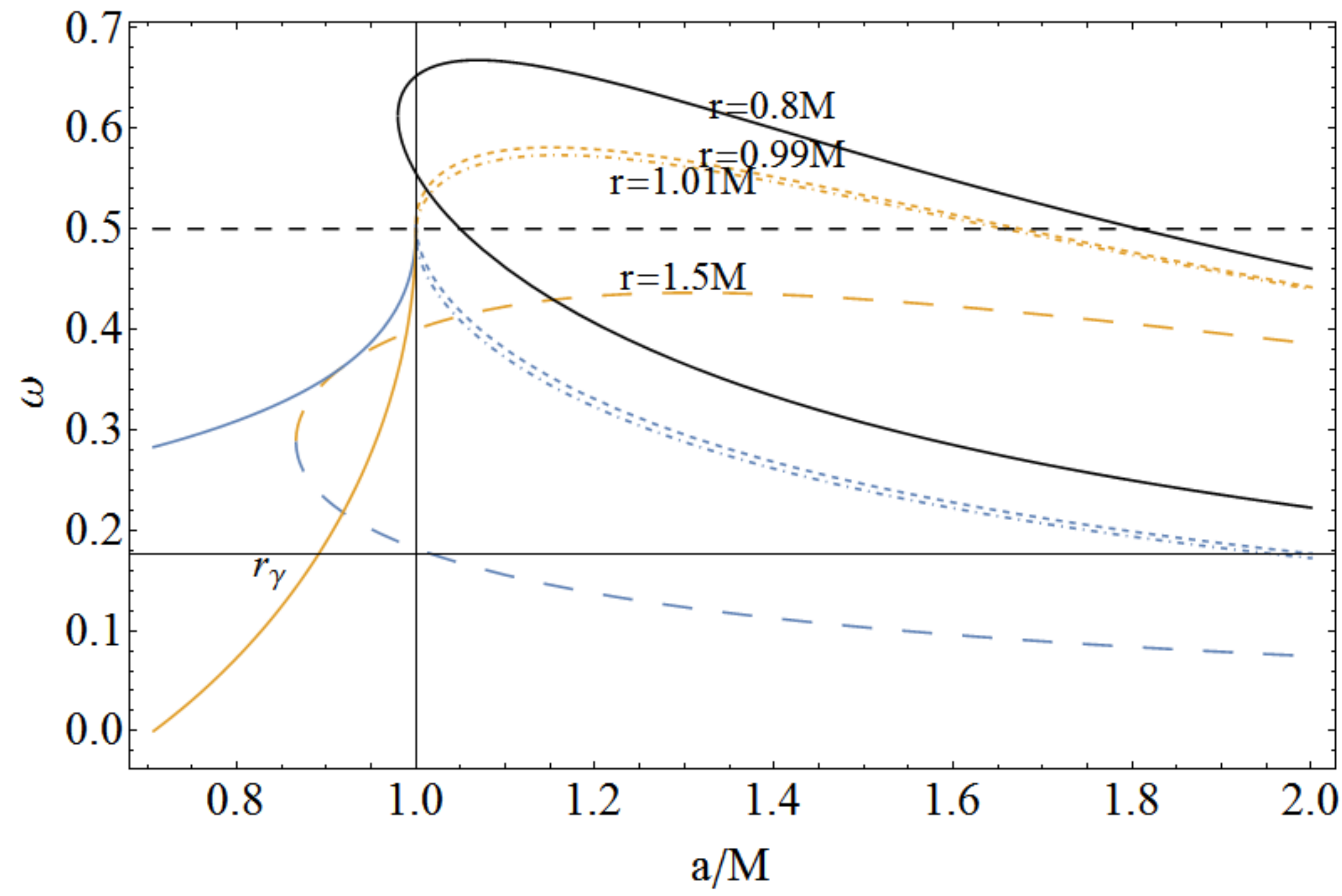}
\end{tabular}
\caption[font={footnotesize,it}]{Plot of the limit frequencies $\omega_{\pm}$ for fixed values of  $r/M$. Frequencies $\omega_{\pm}$, on  $r=r_{\gamma}\in \Sigma_{\epsilon}^+$, photon circular orbit in the \textbf{BHs} ergoregion are also plotted--see Table\il\ref{Table:asterisco1} and \cite{ergon}.}
\label{Fig:Pilosrs1}
\end{figure}

\begin{figure*}[ht!]
\begin{tabular}{lcr}
\includegraphics[width=0.341\hsize,clip]{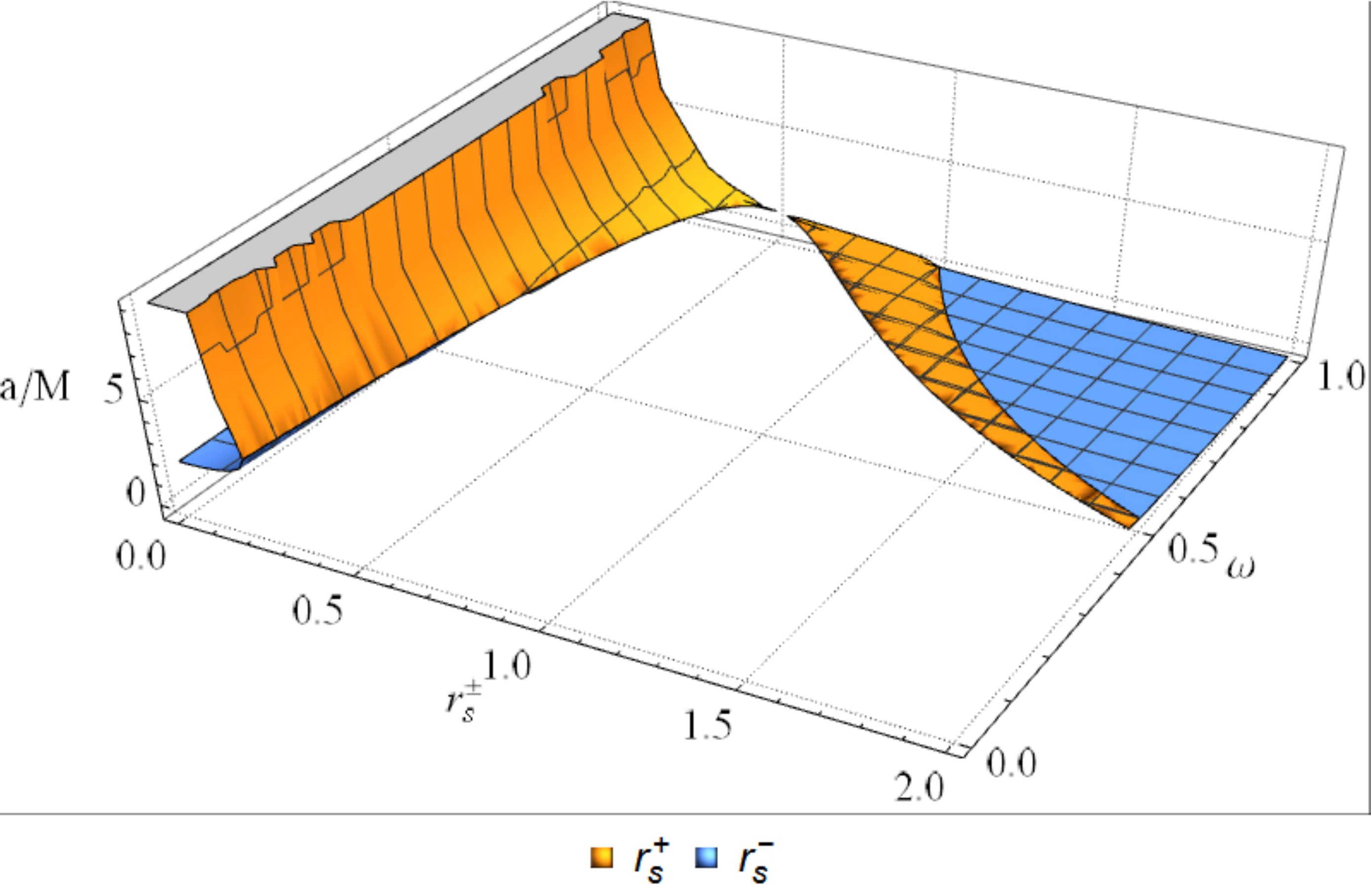}%
\includegraphics[width=0.341\hsize,clip]{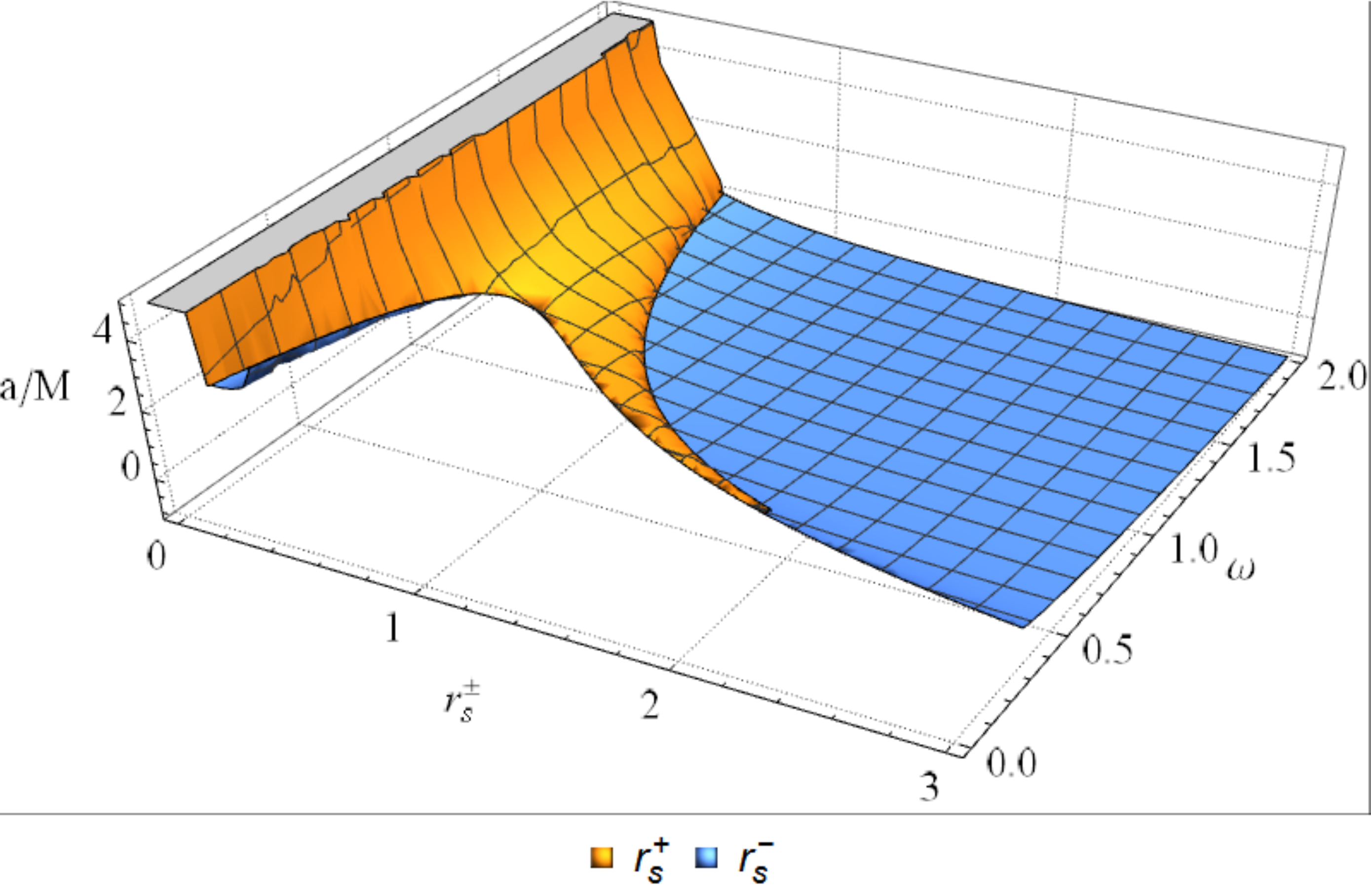}
\includegraphics[width=0.341\hsize,clip]{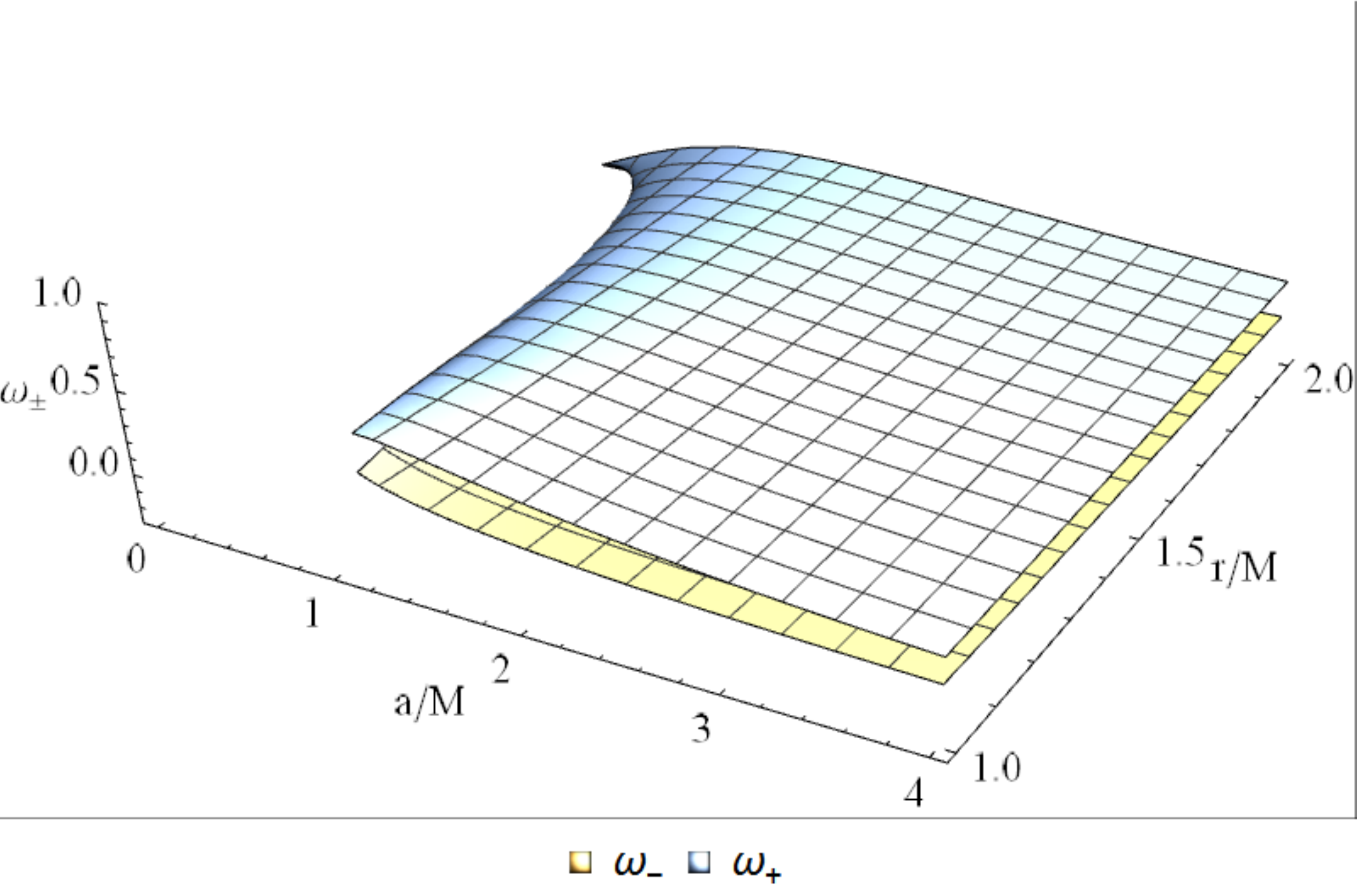}
\end{tabular}
\caption[font={footnotesize,it}]{Left and central panels: Plot of the limit radii $r_s^{\pm}$ as functions of the spacetime spin  $a/M$ and
frequencies $\omega$-see also Figs.\il\ref{Fig:Trav-ling}.
Right panel: Plot of the limit frequencies $\omega_{\pm}$ as functions of the spacetime spin  $a/M$ and radius $r/M$--see also
Fig.\il\ref{Fig:Trav-inr}.}
\label{Fig:Conf3Dna}
\end{figure*}

\begin{figure}[h!]
\centering
\begin{tabular}{cc}
\includegraphics[width=1\columnwidth]{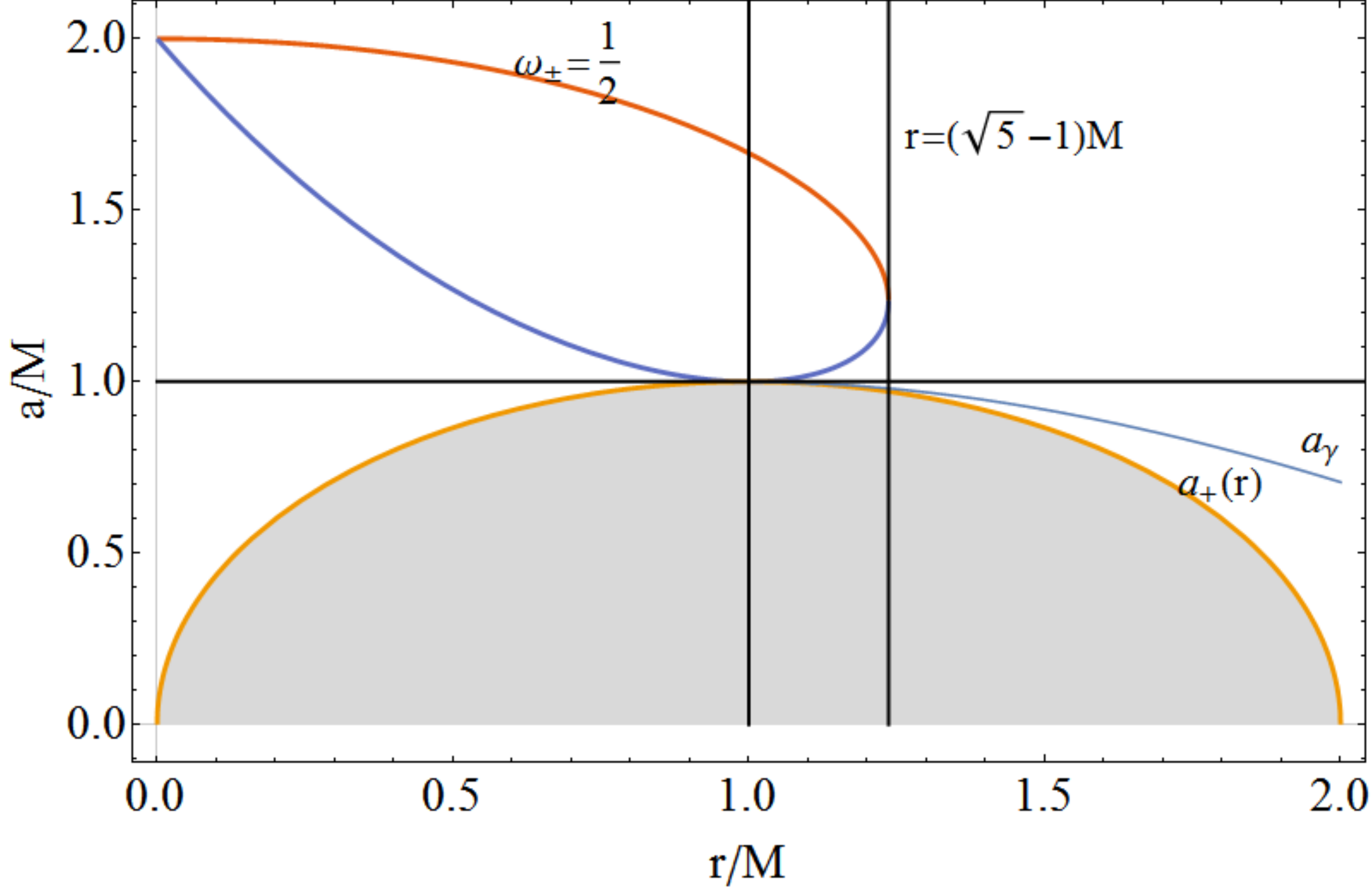}
\end{tabular}
\caption[font={footnotesize,it}]{Plot of the limiting frequency $\omega_{\pm}=1/2$. The spin $a_{+}(r)\equiv\sqrt{r(2M-r)}$, solution of $r=r_+$, and $a_{\gamma}$, solution of $r=r_{\gamma}$ where $r_{\gamma}\in \Sigma_{\epsilon}^+$  is the photon orbit in the ergoregion in a Kerr \textbf{BH}, are also plotted.}
\label{Fig:Pilosrs2}
\end{figure}

\begin{figure}[h!]
\centering
\begin{tabular}{cc}
\includegraphics[scale=.3]{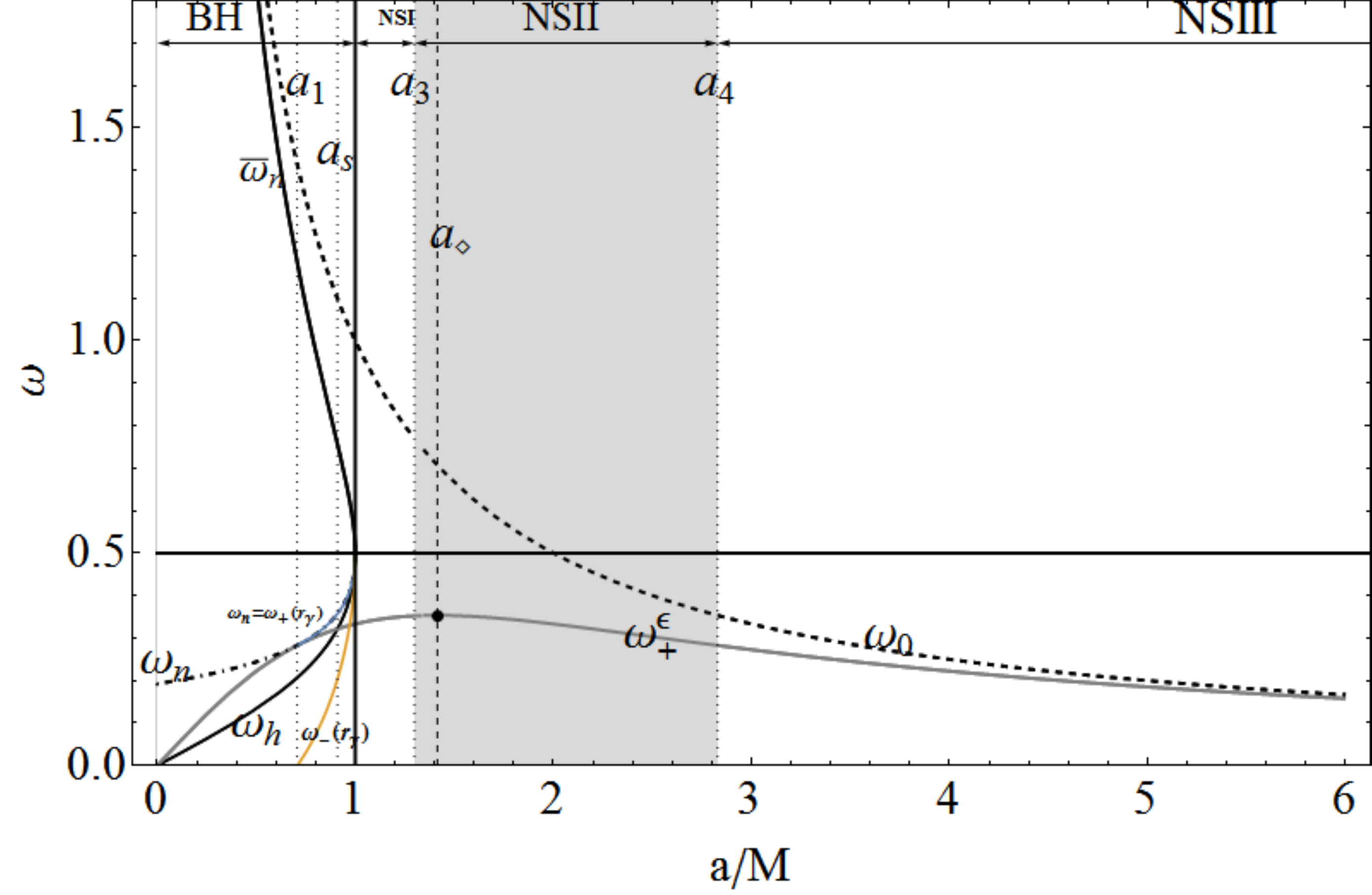}
\end{tabular}
\caption[font={footnotesize,it}]{{Stationary observers: The angular velocities $\omega_{+}^{\epsilon}$ (gray curve), $\omega_{h}$ (black curve), $\omega_n$ (dot-dashed curve), $\omega_0$ (dashed curve), $\bar{\omega}_n>\omega_n>\omega_{h}$ (black thick curve).
Here $\bar{\omega}_n=\omega_n=\omega_{h}=1/2$ at $a=M$, $\omega_{+}^{\epsilon}=\omega_{h}=0.321797$ at $a=a_s$, and $\omega_{+}^{\epsilon}=\omega_n=0.282843$ at $a={a}_1$.
The maximum of $\omega_{+}^{\epsilon}$, at $a=a_{\diamond}=\sqrt{2}M$ (dashed line) where $a_{\diamond}:\;r_e=r_{\epsilon}^+$--see Eq.\il(\ref{Eq:cii-Scli}), is marked with a point. See also Fig.\il\ref{Fig:Conf3Dna}.  The angular velocities $\omega_{\pm}$ on the \textbf{BH} photon orbit $r_{\gamma}\in \Sigma_{\epsilon}^+$ are also plotted. Note that $\omega_n$ it is an extension of $\omega_{+}(r_{\gamma})$ for  $a<a_1$--see Table\il\ref{Table:asterisco2}.
} }
\label{Fig:Trav-ling}
\end{figure}

\begin{figure}[h!]
\centering
\begin{tabular}{lr}
\includegraphics[scale=.3]{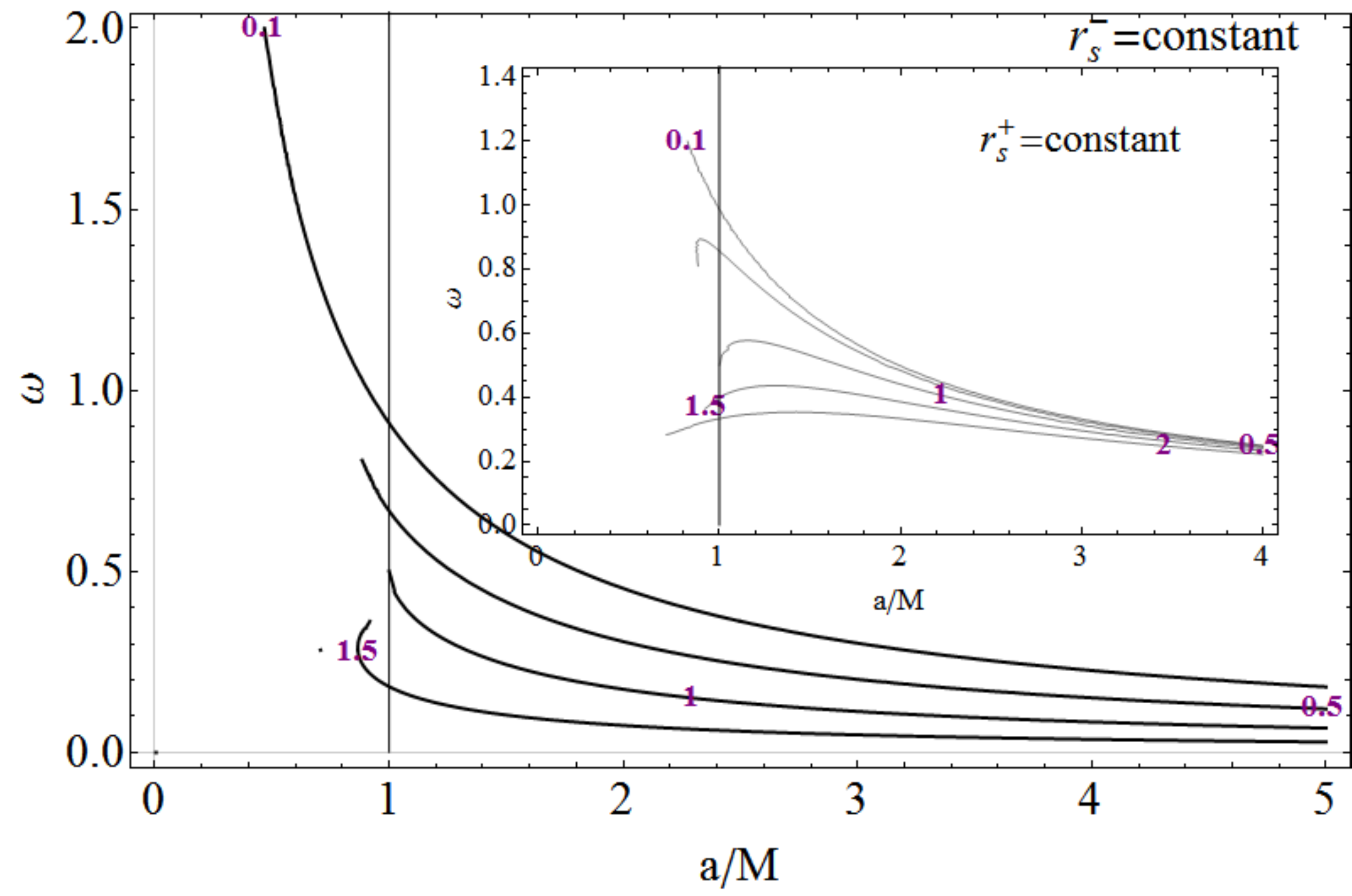}\\
\includegraphics[scale=.3]{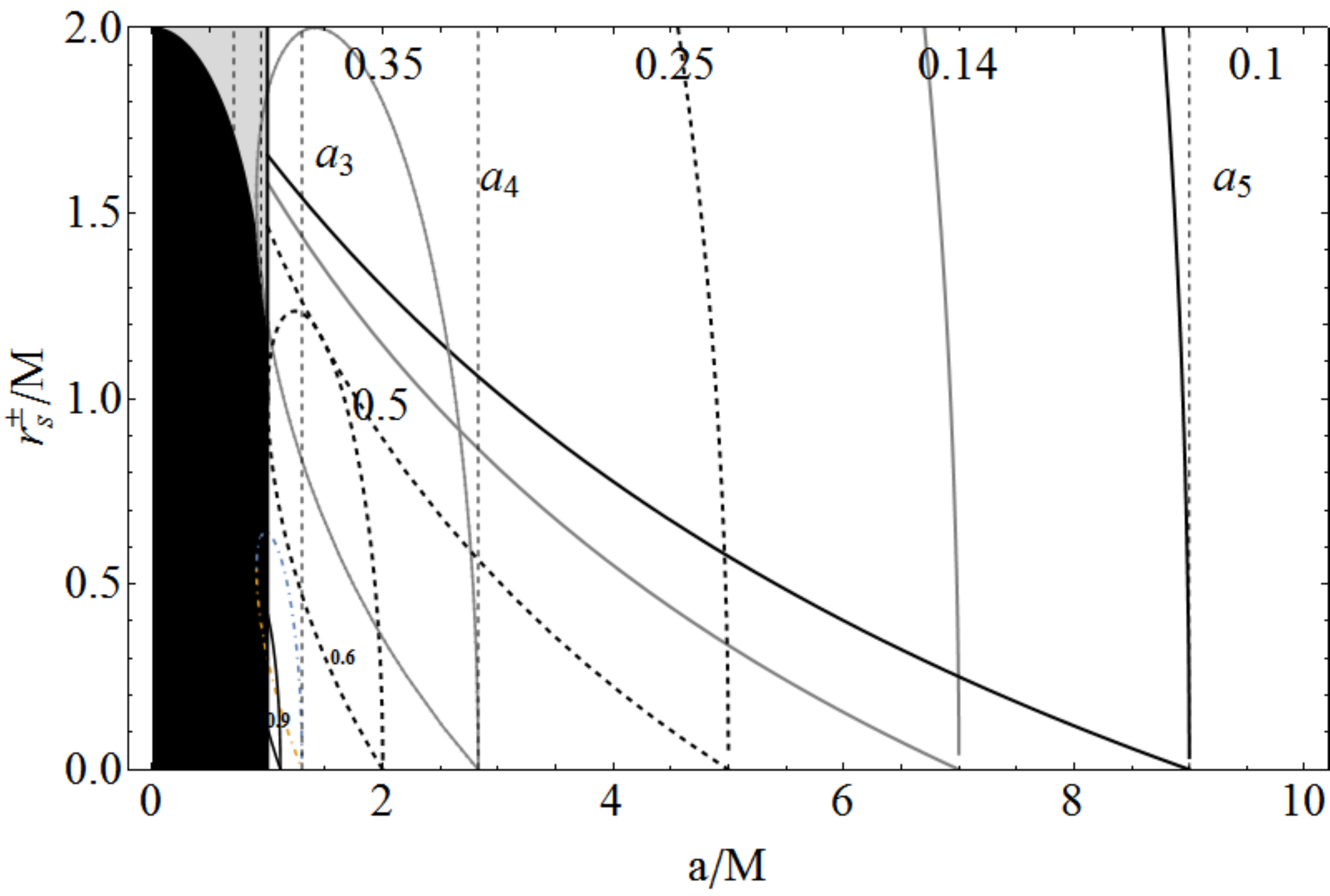}
\end{tabular}
\caption{Upper panel: Plot of the curves $r_s^-=$constant  and $r_s^+=$constant (inside panel) in the plane
$(\omega,a/M)$. The numbers denote the constant radii $r_s^{\pm}/M$  (light cylinders). Bottom panel:
The radii $r_s^{\pm}$ versus the spin $a/M$, for different values of the velocity $\omega$ (numbers close to the curves),
the gray region is $a\in[0,M]$ (\textbf{BH}-spacetime).  The black region corresponds to $r<r_+$.
The dashed lines denote ${a}_1<{a}_2<{a}_3<{a}_4$.
The angular momentum and the velocity  $(a,\omega)$ for $r_s^{\pm}(a,\omega)=0$ are related by $\omega=M/a$.
See also Figs.\il\ref{Fig:Conf3Dna}.}
\label{Fig:Trav-inr}
\end{figure}

\begin{figure}[h!]
\centering
\begin{tabular}{lr}
\includegraphics[scale=.3]{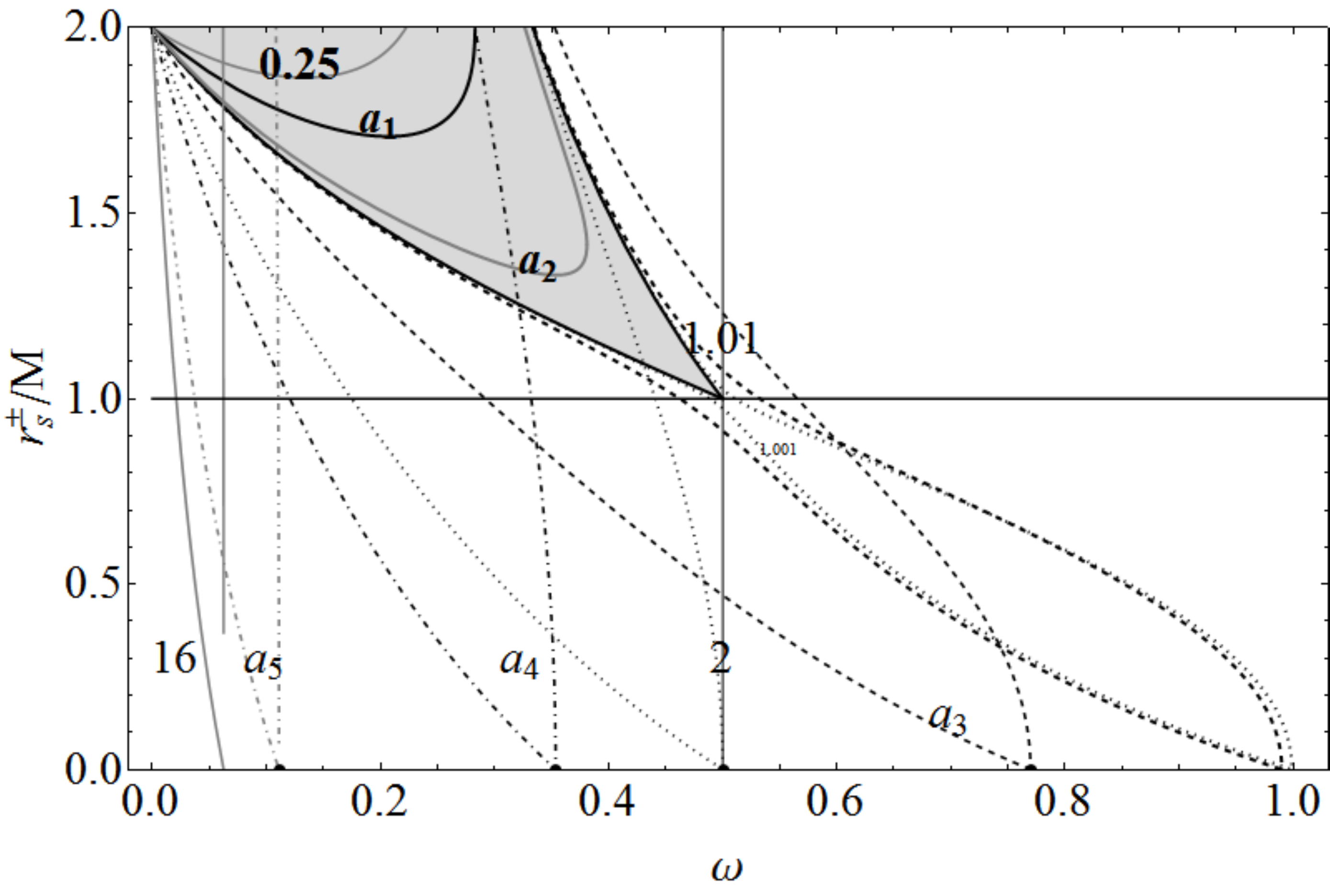}\\
\includegraphics[scale=.3]{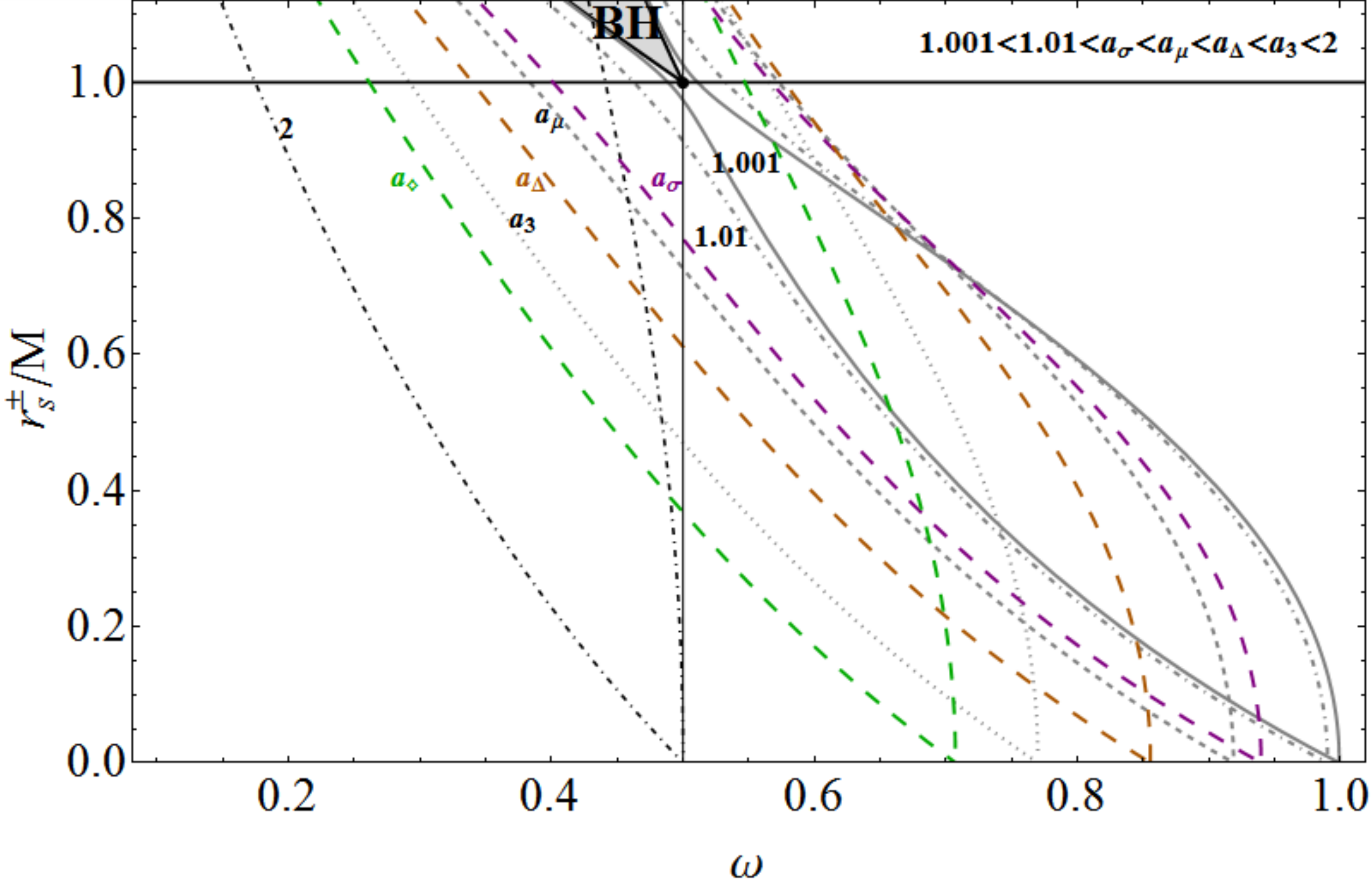}
\end{tabular}
\caption{The radii $r_s^{\pm}$ versus  the frequency $\omega$ for different values of the spin $a/M$
(numbers close to the curves). The gray region is the only region allowed for the case of \textbf{BH} spacetimes.
The surfaces $\hat{r}_{\pm}$ at $a=M$ (extreme-\textbf{BH}-case) are shown in  black-thick. }
\label{Fig:Trav-inr-B}
\end{figure}

The constant limit  $\omega_0\equiv {M}/{a}$ plays an important role because it corresponds to the asymptotic limit for very small values
of $r$  and $R\equiv r/a$. {Note that, on the equatorial plane, $g_{\alpha\beta}\mathcal{L}_0^\alpha \mathcal{L}_0^\beta=R^2$, where $\mathcal{L}_0\equiv \left.\mathcal{L}_{\pm}\right|_{\omega_0}$.}
 {The asymptotic behavior of these frequencies may be deeper investigated by considering the  power series expansion for the  spin  parameter and  the radius  determined by the expression}
\be\label{Eq:okokko}
{\rm for\ } r\rightarrow\infty:\;
\omega_{\pm}=\pm \frac{M}{r}\left(1-\frac{M}{r}\right) + o[r^{-3}] \ ,
\ee
which shows a clear decreasing as the gravitational field diminishes. For large values of the rotational parameter, we obtain
\bea\label{Eq:tend-most}
&& 
\omega_{\pm}= \frac{M}{a} \frac{2M\pm r}{2M+r} +
\\\nonumber
&&\frac{M}{a^3} \frac{r^2}{(2M+r)^2} \left( \mp 2 M^2 -2M r \mp \frac{1}{2}r^2\right) +
o[a^{-5}] \ ,
\eea
so that for extreme large values of the source rotation, the frequencies vanish and no stationary observers exist,
thought differently for the  limiting frequencies $\omega_{\pm}$ (see Figs.\il \ref{Fig:Conf3Dna}).
It is therefore  convenient to introduce the dimensionless radius  $R\equiv r/a$, for which we obtain the limit
\bea\label{Eq:mane-germ}
R\rightarrow0:&&\omega_+=\frac{M}{a}-\frac{M R^2}{2  a}-\frac{M^2 R^3}{4 a^2} +o[R^3];
\\\label{Eq:prod-exape}
&&\omega_-=\frac{M}{a}-R+\frac{\left(M^2+a^2\right) R^2}{2M a}-\\
&&\nonumber\frac{\left(a^4+4M^2 a^2-M^4\right) R^3}{4 a^2M^2} +o[R^3];
\\\nonumber
R\rightarrow\infty:&&\omega_{\pm}=\frac{(\mp M^2+4 M a\mp a^2)M}{2 a^3 R^3}\mp\frac{M^2}{a^2 R^2}\pm\frac{M}{a R} +o[R^{-3}].
\\&&\label{Eq:new-Gov}
\eea
{
Equations \il(\ref{Eq:tend-most}), (\ref{Eq:mane-germ}) and (\ref{Eq:prod-exape}) show  the particularly different behavior
of $\omega_{\pm}$ with respect to the asymptote  $\omega_0$.}
The behavior of the frequencies for fixed values of the radial coordinate $r$ and varying values of the specific rotational parameter $a/M$ is illustrated in Fig.\il\ref{Fig:Pilosrs1}. We see that the region of allowed values for the frequencies is larger for naked singularities than for black holes. In fact, for certain values of the radial coordinate $r$, stationary observers can exist only in the field of naked singularities. This is a clear indication of the observational differences between black holes and naked singularities.
{
The allowed values for the frequencies are bounded by the limiting value $\omega_0=M/a$; for a broader discussion on the role of the dimensionless spin parameter $a/M$ in Kerr geometries, see also
\cite{pugtot}\footnote{For simplicity we use here dimensionless quantities.  We introduce the rotational version of the Killing vectors $\xi_t$ and $\xi_{\phi}$, i.e., the
canonical vector fields
$\tilde{V}\equiv(r^2+a^2)\partial_t +a\partial_{\phi}$
 and $\tilde{W}\equiv\partial_{\phi}+a \sigma^2 \partial_t$. Then, the contraction of
 the geodesic four-velocity with $\tilde{W}$ leads to the (non-conserved) quantity
$\mathcal{L}-\mathcal{E} a \sigma^2$, which is a
 function of the conserved quantities $(\mathcal{E},\mathcal{L})$, the spacetime parameter  $a$ and the polar coordinate $\theta$; on the equatorial plane, it then reduces to
$ \mathcal{L-E} a$.
When we consider the principal null congruence
$
\gamma_{\pm}\equiv\pm\partial_r+\Delta^{-1} \tilde{V}$,
the angular momentum $\mathcal{L}=a \sigma^2$, that is, $\bar{\ell}=1$ (and $\mathcal{E}=+1$, in proper units), every principal null geodesic is then characterized by $\bar{\ell}=1$. On the horizon,
 it is
$\mathcal{L=E}=0$ \cite{Chandra,pugtot}}}
Moreover, for a given value of $\omega_\pm$, the corresponding radius is located at a certain distance from the source, depending on
the value of the rotational parameter $a$. The following configuration of frequencies, radii and spin determines the location structure of stationary observers:
\bea\label{Eq:hu-mas-photon}
&&
\omega_+ \in]0,\omega_0[,\ \mbox{
 for}\ a\in]0,M[\ \mbox{in}\ r\in]0,r_-]\cup[r_+,+\infty[
\\\nonumber
&& \mbox{and for} \ a\geq M  \ \mbox{in}\ r>0
\\
&&
 \omega_- \in]0,\omega_0[
 \ \mbox{
 for}\   a\in]0,M[ \  \mbox{in}\ r\in]0,r_-]\cup[r_+,r_{\epsilon}^+[
 \\\nonumber
 && \mbox{and for}\  a\geq M \ \mbox{in}\   r\in]0,r_{\epsilon}^+[ \ .
\eea
Thus,  we see that in the interval $]0, M/a[$ observers can exist with frequencies  $\omega_{\pm}$; moreover, the frequency
$\omega_-$ is allowed in   $r\in\Sigma_{\epsilon}^+$, while observers with  $\omega_-<0$  can exist in $ r>r_{\epsilon}^+$.
Moreover, it is possible to show that, in \textbf{BH} geometries, the condition  $\omega_{\pm}\ngeq1/2$ must be satisfied outside the outer horizon ($r>r_+$).
The particular value  $\omega_\pm=\omega_h= 1/2$ is {therefore}  the limiting angular velocity in the case of an extreme black hole, i.e., for $a=M$ so that $r=r_+=r_-=M$ in Eq.(\ref{Eq:b-y-proc}).
The behavior of the special frequency $\omega_{\pm}=1/2$ is depicted in Fig.\il \ref{Fig:Pilosrs2} and in Figs.\il\ref{Fig:Trav-ling},
\ref{Fig:Conf3Dna}, \ref{Fig:Trav-inr}, and \ref{Fig:Trav-inr-B}, {where other relevant frequencies are also plotted}.

{Eqs\il(\ref{Eq:hu-mas-photon})   enlighten  some important properties of the light surfaces (frequencies $\omega_{\pm}$) and of stationary observers, associated with frequencies $\omega\in]\omega_{-},\omega_{+}[$ in the regime of strong singularities.
Eqs(\ref{Eq:hu-mas-photon}) also enlighten the dependence of the frequencies on the dimensionless spin $a/M$ and radius $R=r/a$. It is clear that when the frequency interval  $]\omega_{-},\omega_{+}[$ shrinks, depending on the singularity spin $a/M$ or the distance from the source $r/M$, the range of possible  frequencies  for stationary observers reduces.
This occurs in general when $\omega_+\approx\omega_-$.
According to Eqs\il(\ref{Eq:hu-mas-photon}), the    frequencies $\omega_{\pm}$ are bounded from above by the limiting frequencies
$\omega_0=M/a$ and from below by the null value $\omega_{\pm}=0$.
Thus, at fixed radius $r$, for very strong naked singularities $a/M\gg1$, we have that $\omega_0\approx0$ and the range of possible frequencies for stationary observers becomes smaller. This effect will be discussed more deeply in  Sec.\il\ref{Sec:p-lS}, where we shall focus  specifically on the frequency $\omega_0$.}
On the other hand, considering the limits (\ref{Eq:b-y-proc}), together with Eqs\il(\ref{Eq:okokko})-- (\ref{Eq:new-Gov}),
we find that the range of possible frequencies shrinks also in the  following situations: when moving outwardly with respect to the singularity (at fixed $a$),  very close to the source,
approaching the horizon  $r_h$ according to
Eq.\il(\ref{Eq:b-y-proc}), or also for  very large or very small $R=r/a$. The last case points out again the importance of the scaled radius  $r/a$.

Essentially, stationary observers can  be near the singularity only at a particular frequency.
The greater is the \textbf{NS} dimensionless spin, the lower is the limiting frequency $\omega_{\pm}$,
with the extreme limit at $\omega_+=\omega_-$.
In other words, the frequency  range, $]\omega_-,\omega_+[$,   for stationary observers  vanishes as the value $r=0$ is approached.
{The singularity  at $r=0$ in the \textbf{NS} regime is actually related to the  characteristic constant  frequency
$\omega=\omega_0$ in the same way as  in \textbf{BH}-geometries the outer horizon  $r=r_+$ is  related to the constant frequency
$\omega_h$ (cf. Eq.\il(\ref{Eq:b-y-proc})).
Consequently,  a \textbf{NS}  solution must be characterized by the frequency   $\omega_0$ and a  \textbf{BH} solution
by the frequency  $\omega_h$.
Therefore, the frequency  $\omega_0$   may be  seen   actually as  the \textbf{NS} counterpart of the \textbf{BH} horizon angular frequency  $\omega_h$ (see Fig.\il\ref{Fig:Trav-ling}).}
For $r>r_+$, it holds that $\omega_+>\omega_-$.

{Then, in general,  for  \textbf{BHs} and \textbf{NSs} in the static limit $r_{\epsilon}^+=2M$, we obtain that }
\be
\omega_+^{\epsilon}\equiv\omega_{+}(r_{\epsilon}^+)=\frac{aM}{2M^2+a^2}\quad\mbox{with}\quad \omega_-(r_{\epsilon}^+)=0.
\ee
Moreover, $\omega_-<0$ for $r>r_{\epsilon}^+$, and $\omega_->0$ inside the ergoregion $\Sigma_{\epsilon}^+$, while $\omega_+>0$  everywhere.

\begin{figure}[h!]
\centering
\begin{tabular}{c}
\includegraphics[scale=.3]{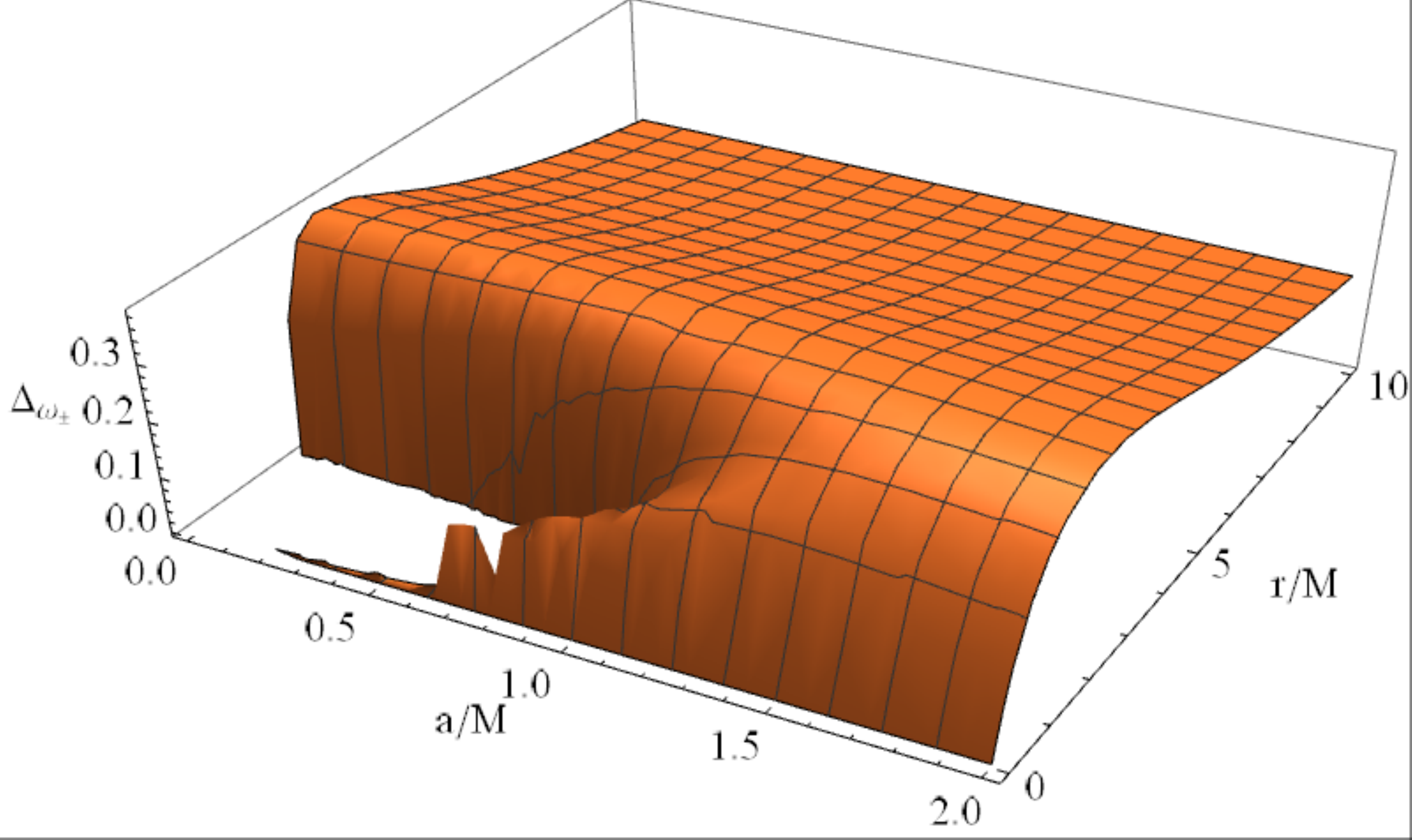}
\\
\includegraphics[scale=.3]{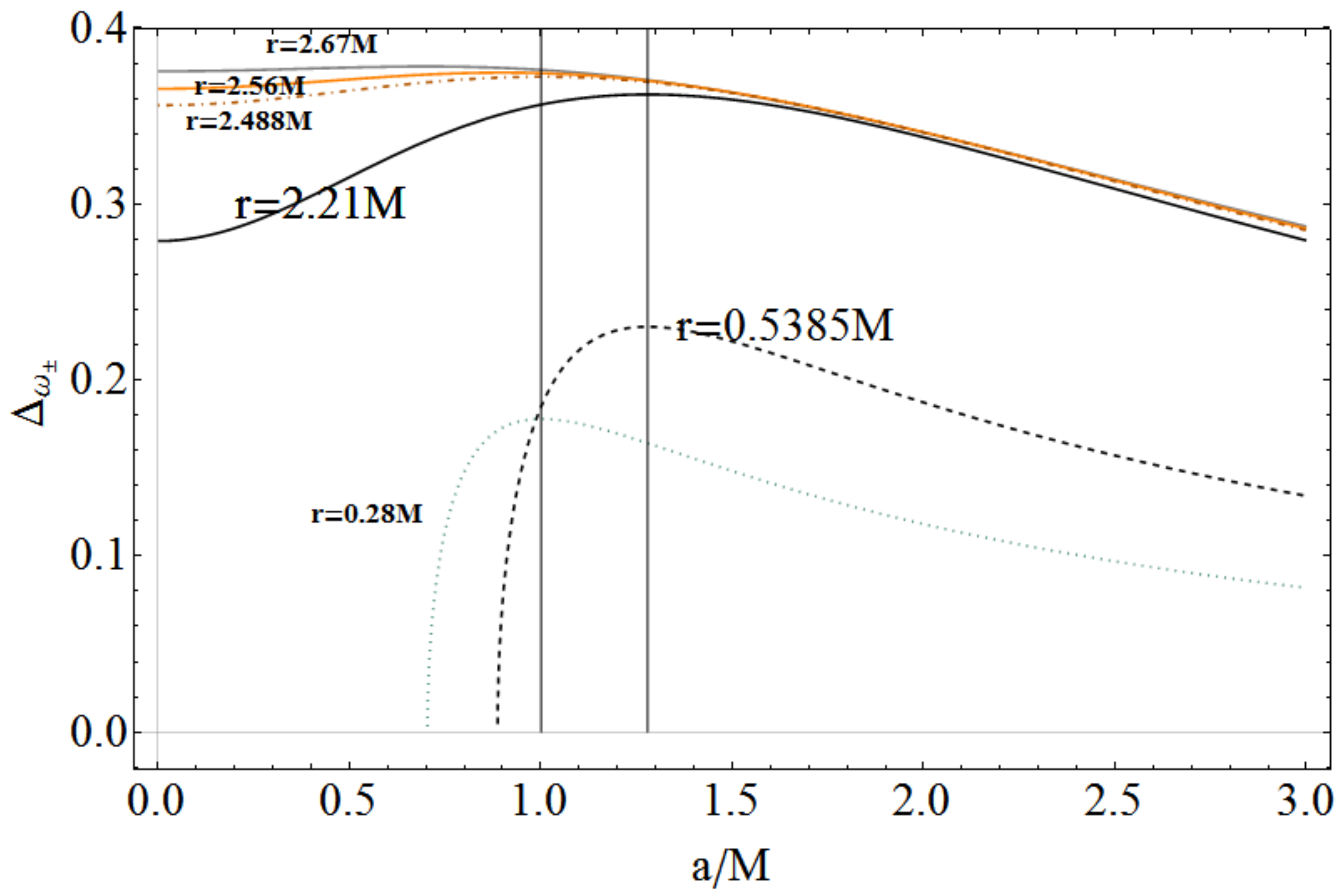}
\end{tabular}
\caption[font={footnotesize,it}]{Upper panel: Plot of the frequency interval $\Delta \omega_{\pm}=\omega_+-\omega_-$ as a
function of the radius $r/M$ and the \textbf{BH}  and \textbf{NS} spin $a/M$.
The extrema $r_{\Delta}^{\pm}$ and $r^{\pm}_{\blacksquare}$ are solutions of  $\partial_r\Delta \omega_{\pm}=0$
and $\partial_{a}\Delta \omega_{\pm}=0$, respectively. Lower panel:
The frequency interval $\Delta \omega_{\pm}=\omega_+-\omega_-$ as a function of $a/M$ for selected values of the orbit radius $r/M$; the
 maximum points are for the radii $r_{\Delta}^{\pm}$ or $r_{\blacksquare}^+$--see Figs.\il\ref{Fig:L0V0Zamos}. }
\label{Fig:desol}
\end{figure}

In general, any frequency value should be contained within the range $\omega_+-\omega_-$; therefore, it is convenient to define
the {\emph{frequency interval}}
\be
\Delta_{\omega_{\pm}}\equiv \omega_+-\omega_-=2\sqrt{\omega_{Z}^2-\omega _*^2},
\ee
which is a  function of the radial distance from the source and of the attractor  spin. { Figs.\il\ref{Fig:desol} show  the frequency interval   $\Delta_{\omega_{\pm}}$ as a function of $r/M$ and $a/M$}.

An analysis of this quantity  makes it possible to derive some key features about the eligible frequencies.
For convenience, we present in Table\il\ref{Table:asterisco1} some special values of the spin-mass ratio, which we will consider in the following analysis. We summarize the obtained results in the following way:
\begin{table*}
\centering
\caption{Classes of  \textbf{BH} and  \textbf{NS} geometries according to their specific spins.
The radii $(r_{\gamma}^-, r^-_{mso})$ corresponds to  the photon circular  orbit (or also last circular orbit) and the  marginally stable circular orbit, respectively, for corrotating orbits in \textbf{BH} geometries.
The \textbf{NS} case is characterized by  the zero angular momentum radii
($\mathcal{L}(\hat{r}_{\pm})=0$) and  the radius of the marginally stable circular orbit  $r_{mso}^{(NS)-}\in \Sigma_{\epsilon}^+$.
The explicit expressions for these radii can be found in  \cite{Pu:Charged,Pu:class,Pu:Kerr,Pu:KN,Pu:Neutral} }.
\label{Table:asterisco1}
\begin{tabular}{lrcl}
\hline
\textbf{Black hole  classes:} $\mathbf{BHI:}\; [0,a_1[;\quad \mathbf{BHII:}\; [a_1,a_2[,\quad \mathbf{BHIII:}\;[a_2,M]$
\\\\
${a}_1/M\equiv1/\sqrt{2}\approx0.707107:\;r_{\gamma}^-(a_1)=r_{\epsilon}^+, \quad a_2/M\equiv {2 \sqrt{2}}/{3}\approx0.942809:\; r_{mso}^-(a_2)=r_{\epsilon}^+$
\\\\
 \hline 
\textbf{Naked singularity classes:}$\mathbf{NSI:}\; ]M,a_{3}],\quad \mathbf{NSII:}\;  ]a_3, a_4],\quad \mathbf{NSIII:}\; ]a_4,+\infty]$
\\ \\
$\quad{a}_3/M\equiv 3\sqrt{3}/4\approx1.29904:\;\hat{r}_{+}(a_3)=\hat{r}_{-}(a_3),\quad
a_4/M\equiv 2 \sqrt{2}\approx2.82843:\;r_{mso}^{(NS)-}(a_4)=r_{\epsilon}^+\;
$
\\\\
\hline
\end{tabular}
\end{table*}

Firstly, for any  \textbf{NS} source  with  $a>a_{\Delta}\equiv 1.16905M$,  the interval
 $\Delta_{\omega_{\pm}}$ increases as the observer (on the equatorial plane) moves inside the ergoregion $\Sigma_{\epsilon}^+$
towards the static limit.

\begin{figure}[h!]
\centering
\begin{tabular}{c}
\includegraphics[scale=.3]{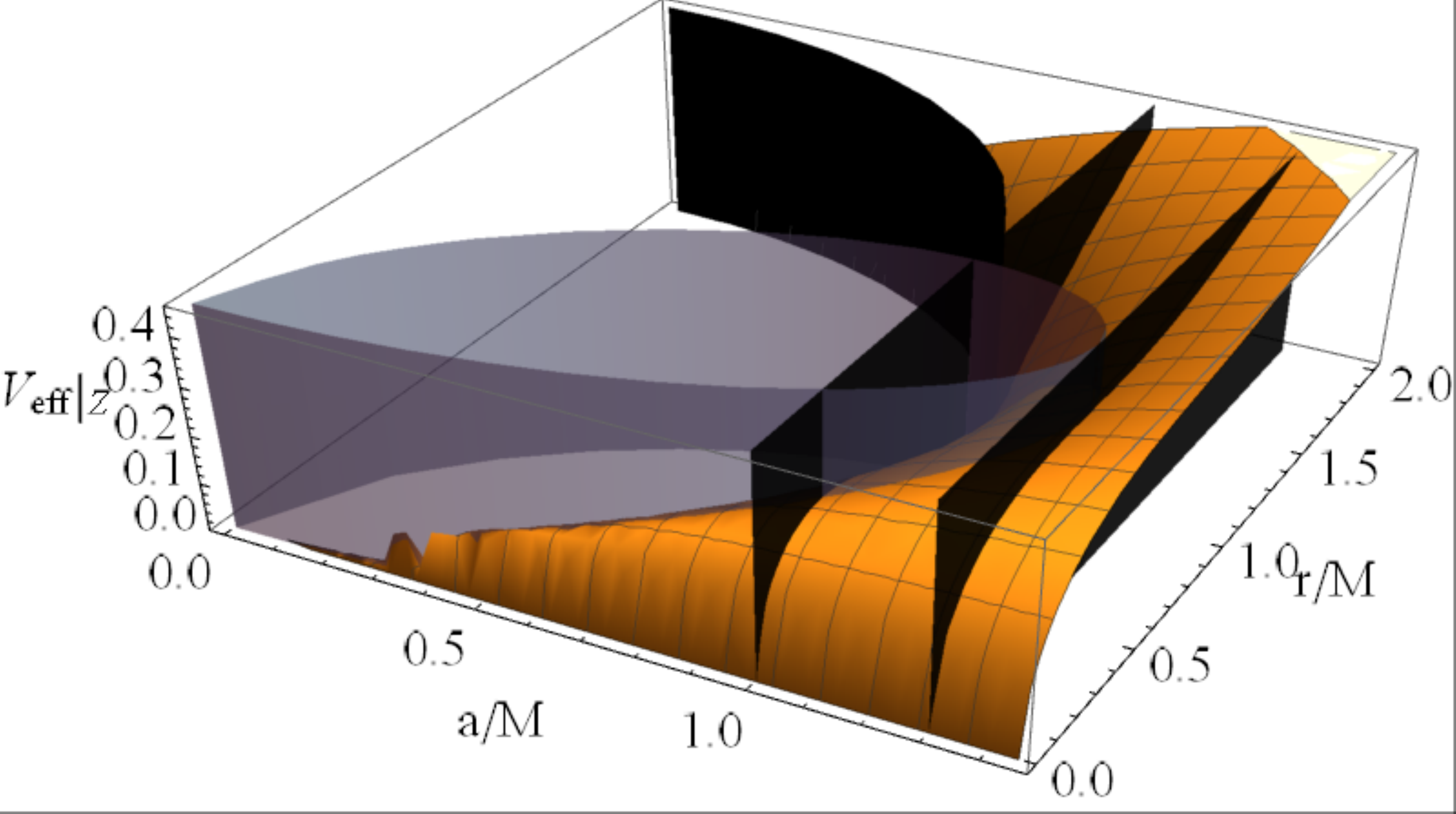}
\\
\includegraphics[scale=.3]{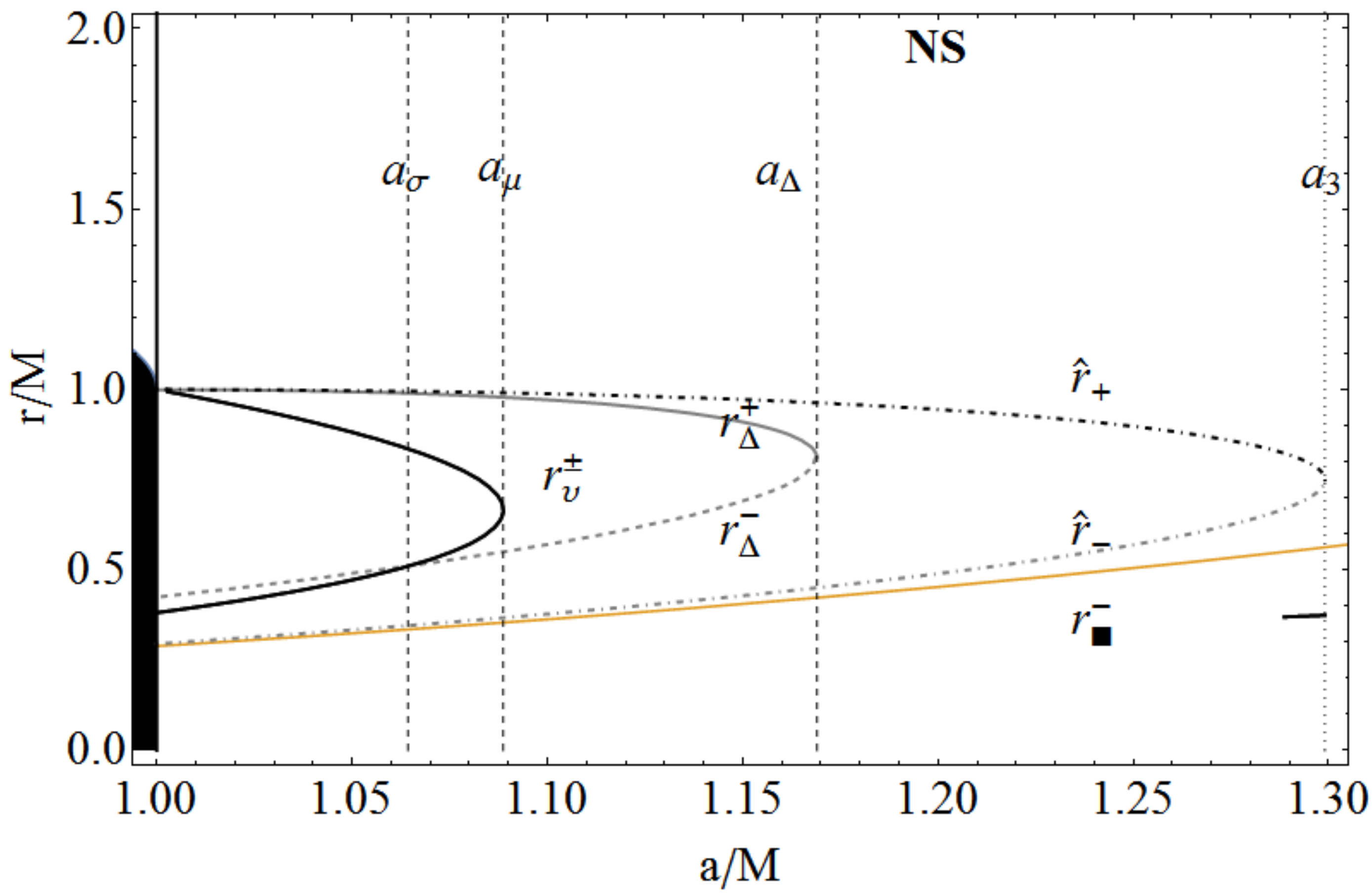}
\\
\includegraphics[scale=.3]{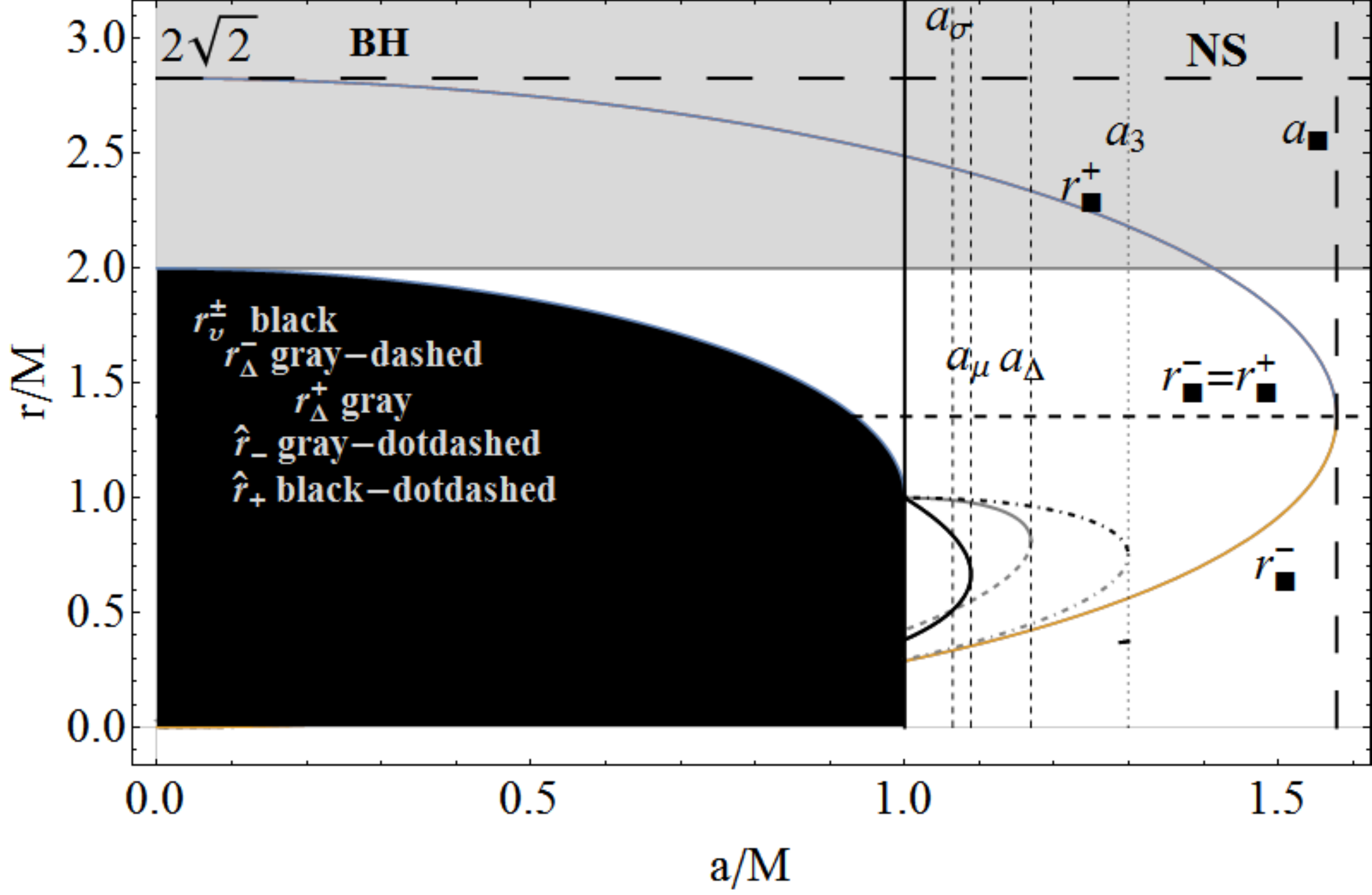}
\end{tabular}
\caption[font={footnotesize,it}]{Upper panel: The effective potential  $\left.V_{eff}\right|_Z$ for the ZAMOS $\mathcal{L}=0$,
for \textbf{BH} and \textbf{NS} sources as a function of the source spin $a/M$ and the radius $r/M$.
The effective potential function
 is the value of $\mathcal{E}/\mu$
at which the (radial) kinetic energy of the particle vanishes.
{Black planes represent  the spin values $a=M$, extreme Kerr \textbf{BH}, and ${a}_3\equiv {3 \sqrt{3}}/{4}M$, a \textbf{NS} geometry, where  $\hat{r}_-=\hat{r}_+$.
The orbits  $\hat{r}_-\leq \hat{r}_+$,  gray surfaces, are for  $a<M$ (\textbf{BH}-case)   inside the horizon  ($r<r_+$). The inner black surface is the  horizon $r_+$}.
Central panel: The radius $r(a)$, solution of $\partial_r\Delta_{\omega_{\pm}}=0$, i.e., it represents the critical points of the separation parameter  $\Delta_{\omega_{\pm}}\equiv\left.(\omega_+-\omega_-)\right|_{\pi/2}$  on the equatorial plane $\theta=\pi/2$. The radius $r_{\upsilon}^{\pm}$, where the orbital energy $\mathcal{E}=0$, and the orbits $\hat{r}_{\pm}$, for which $\mathcal{L}=0$, are also plotted. Dashed lines represent the  spins  $a_{\sigma}=1.064306M$, $a_{\mu}=4\sqrt{2/3}/3 M\approx1.08866M$,   $a_{\Delta}=1.16905M$ and ${a}_3={3 \sqrt{3}}/{4}M$.
The black region corresponds to $r<r_+$. Bottom panel: The radii $r_{\blacksquare}^{\pm}:\partial_{a}\Delta_{\omega_{\pm}}=0$ are plotted as functions of $a/M$--see also Figs.\il\ref{Fig:desol}.}
\label{Fig:L0V0Zamos}
\end{figure}

Secondly, in the case of \textbf{NS}  geometries with   $a\in]M,a_{\Delta}[$, i.e., belonging partially to the class of \textbf{NSI} spacetimes, the situation is very articulated.  There is  a region of  maximum and a minimum frequencies, as the observer moves
from the source towards  the static limit. This phenomenon  involves  an orbital range partially located within the interval
  $]\hat{r}_-, \hat{r}_+[$, which is characterized by the presence of counterrotating circular orbits  with negative  orbital angular momentum $\mathcal{L}=-\mathcal{L}_-$ (cf. Fig.\il\ref{Fig:L0V0Zamos}, {where the radii $\hat{r}_\pm$ are plotted.}).

For the maximum spin, $a=a_{\Delta}$,  {we obtain $\omega^+=\omega^-$ on the radius  $r\equiv r_{\Delta}^{\pm}(a_{\Delta})=0.811587M$} and, therefore, the range of possible frequencies for
 stationary observers vanishes. { The points  $r_{\Delta}^{\pm}(a)$ represent the extrema of the interval  $\Delta_{\omega^{\pm}}$, i.e., the solutions of the equation
$\partial_r \Delta_{\omega^{\pm}}=0$ -- Fig.\il\ref{Fig:L0V0Zamos}}.
This  property is present \emph{only}  in the case of  \textbf{NS}  geometries.
In fact,  there are  the two  critical orbits $r_{\Delta}^+>r_{\Delta}^-$ and $r=r_{\Delta}^{\pm}(a_{\Delta})$, which are the boundaries of a closed region, whose extension reaches a maximum  in the case of the extreme Kerr geometry  $a=M$,
and  is zero for $a=a_{\Delta}$.
 For $r\in]r_{\Delta}^-,r_{\Delta}^+[$, the separation parameter $\Delta_{\omega_{\pm}}$ decreases with the  orbital distance, then on the  inner radius  $r_{\Delta}^-$ it reaches  a maximum value, whereas  on the outer  radius $r_{\Delta}^+$  it reaches a  minimum.
In the outer regions, at $r>r_{\epsilon}^+$, the separation parameter increases with the distance from the source.
This feature   constitutes therefore a major  difference in the  the behavior of stationary observers within and outside the ergoregion of a naked singularity spacetime.
However, a deeper analysis of the  equatorial plane, outside the static limit, shows the existence of a second region for light surfaces  in the \textbf{NS} case.

On the other hand, the angular velocity $\omega_-$  decreases with the orbit in the Kerr  spacetime. The maximum frequency $\omega_+$ also decreases in the  \textbf{NS} spacetimes. In the  \textbf{BH} cases, the angular velocity is always increasing for sources of the class
 \textbf{BHI}, while for the other sources there is a  maximum for the velocity  $\omega_+$ at   $r=r_{\gamma}^-$, which is the circular orbit of a photon or null-like particle corotating with the source. Such a kind of orbit, contained in $\Sigma_{\epsilon}^+$, is a feature of the \textbf{BHII-III} spacetimes \cite{ergon}, this is also know as marginally or last  circular orbit as no circular particle motion is possible in the region $r<r_{\gamma}^-$.
{
We close this section with a  brief discussion on the variation of the frequency interval $\Delta_{\omega_{\pm}}$, following a spin transition  with $a>0$.
In the case of a singularity spin-transition, there are two extreme radii for the frequency interval}
{
\bea&&\label{Eq:def-tru-all-ru}
r_{\blacksquare}^+\equiv\eta  \cos \left[\frac{1}{3} \arccos \left(-\frac{8 a^2}{\eta ^3}\right)\right],
 \\\nonumber
 &&r_{\blacksquare}^-\equiv\eta  \sin \left[\frac{1}{3}\arcsin\left(\frac{8 a^2}{\eta ^3}\right)\right],\quad\eta\equiv\frac{2 \sqrt{8M^2-a^2}}{\sqrt{3}},
 \\\nonumber
 && \mbox{or alternatively }\quad a= \sqrt{-\frac{r \left(r^2-8M^2\right)}{r+2M}}\quad \mbox{for }\quad r\in]0,2\sqrt{2}M[ \ ,
\eea
where  $r_{\blacksquare}^{\pm}:\, \left. \partial_a\Delta_{\omega_{\pm}}\right|_{r_{\blacksquare}^{\pm}}=0$ are maximum points--
see Figs.\il\ref{Fig:L0V0Zamos} and \ref{Fig:desol}}.
\subsection{Light surfaces}
\label{Sec:p-lS}
\begin{figure*}[ht!]
\begin{tabular}{ccc}
\includegraphics[scale=.24]{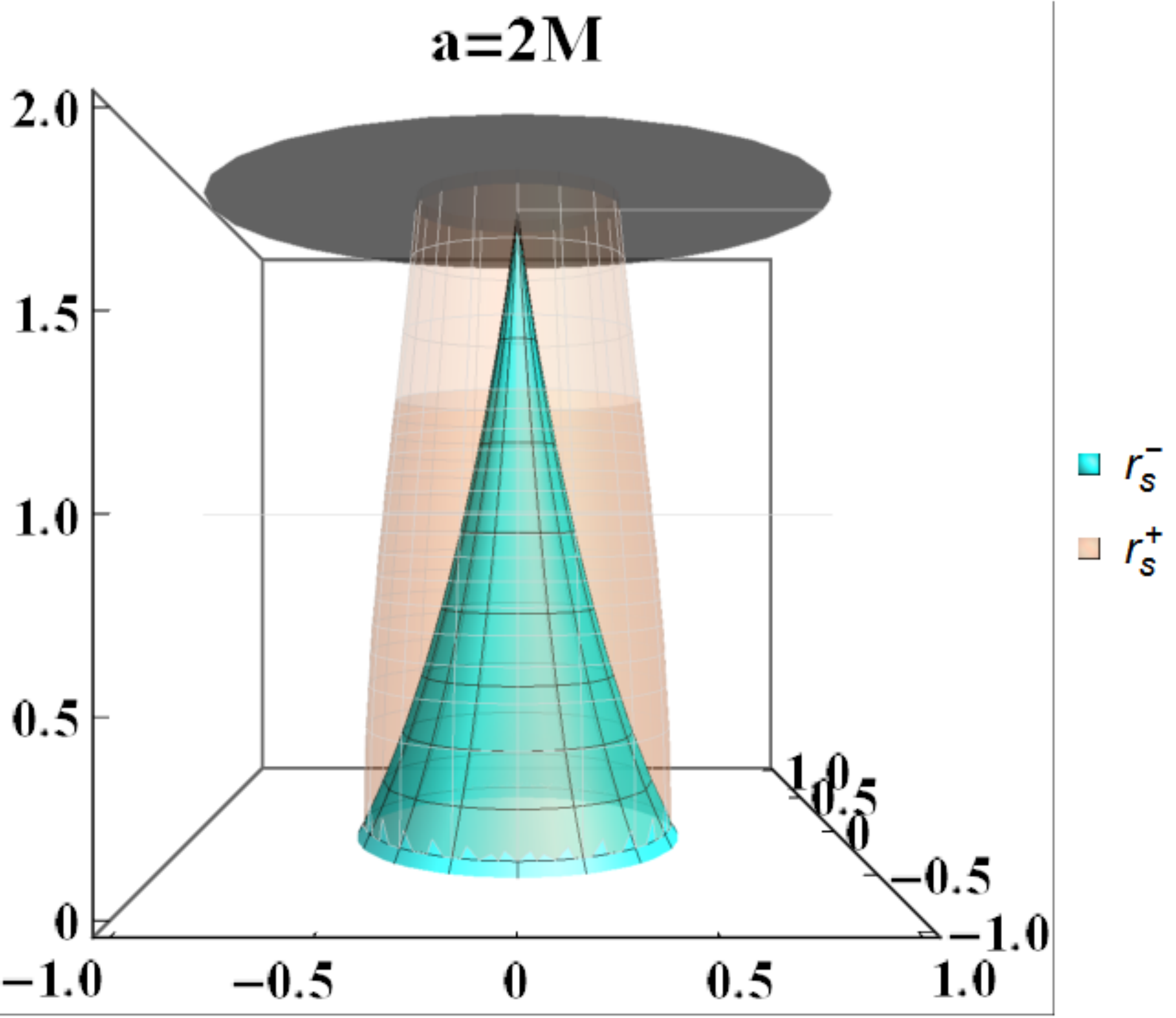}%
\includegraphics[scale=.24]{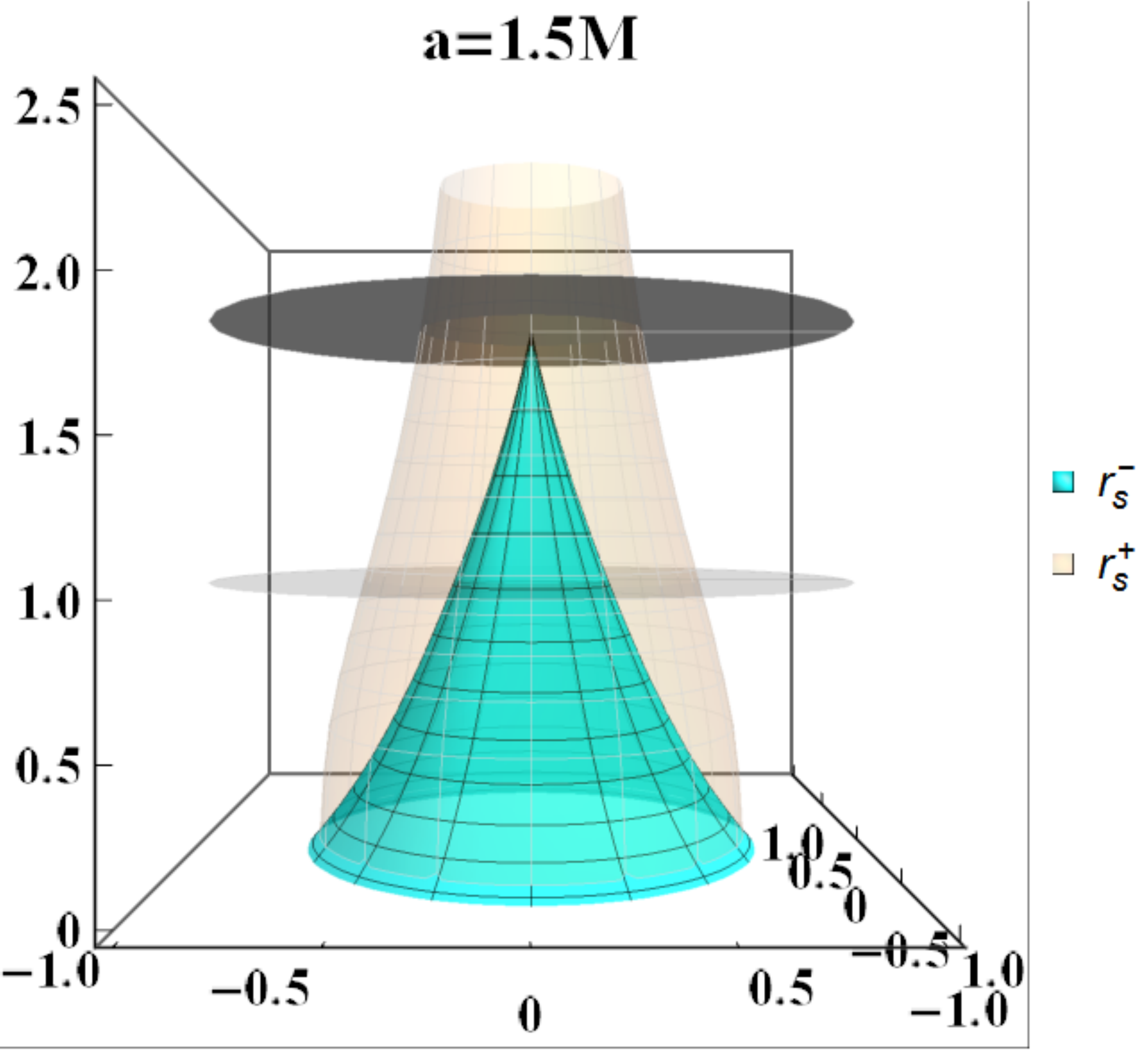}
\includegraphics[scale=.24]{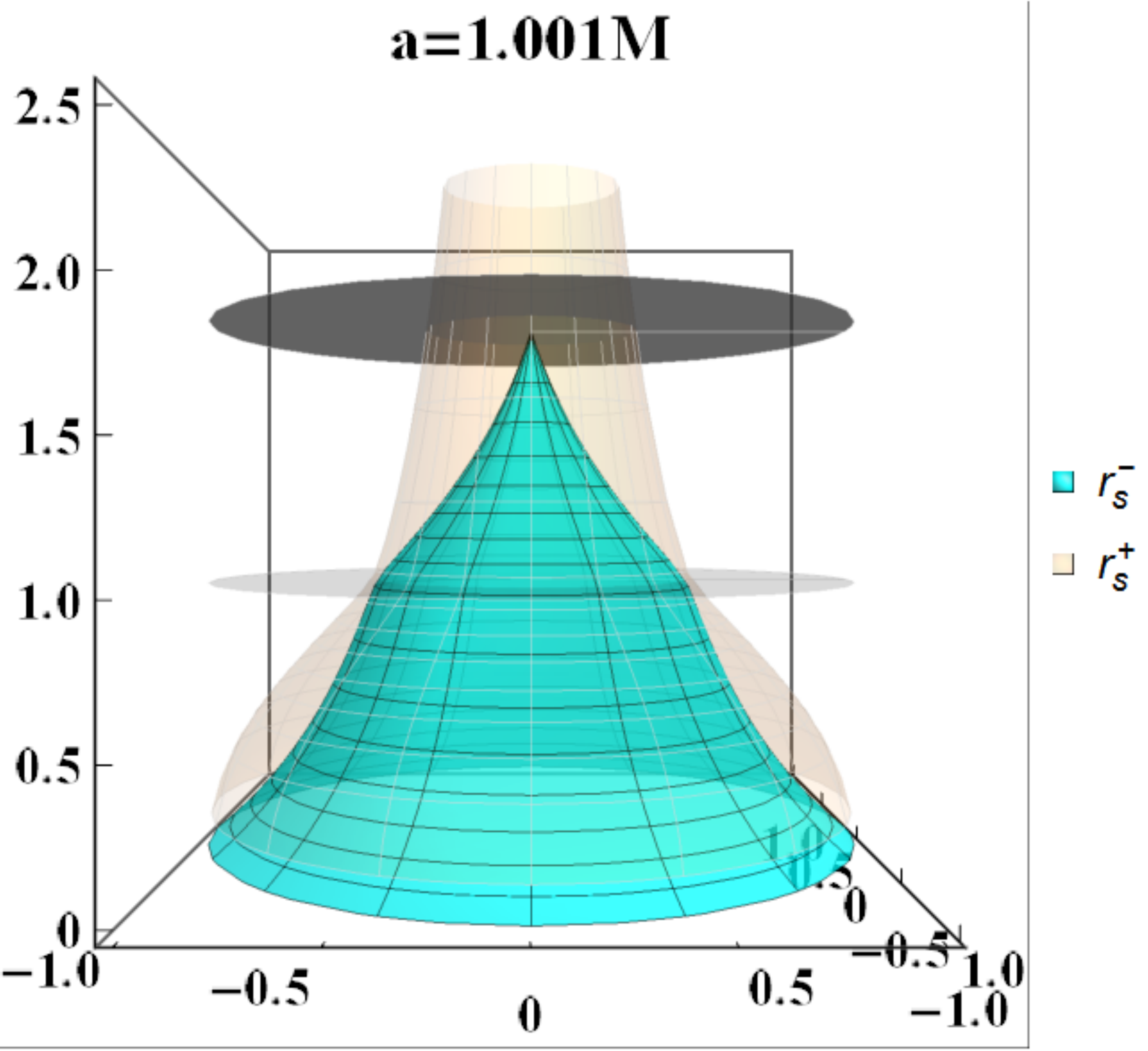}
\\
\includegraphics[scale=.3]{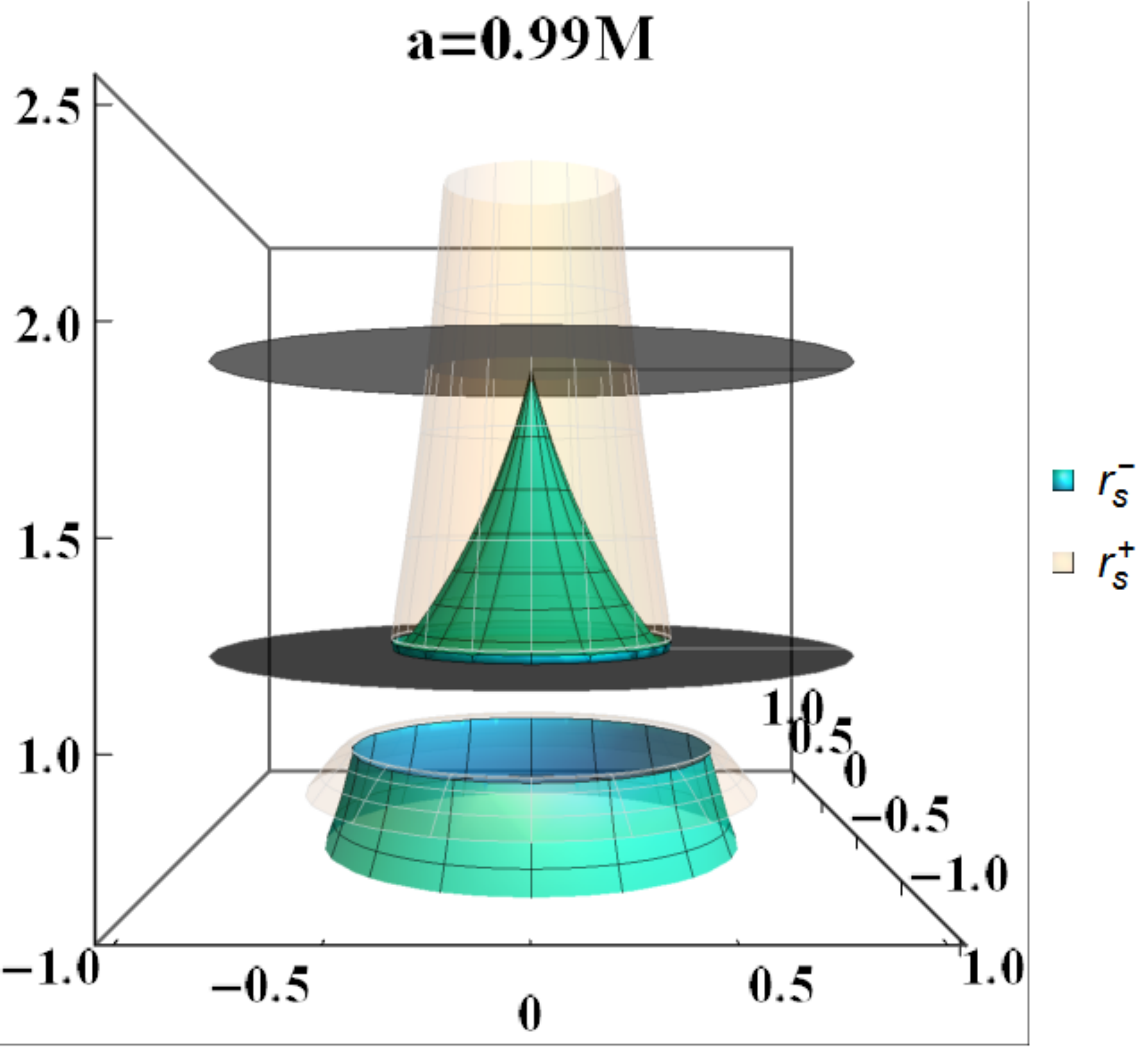}%
\includegraphics[scale=.3]{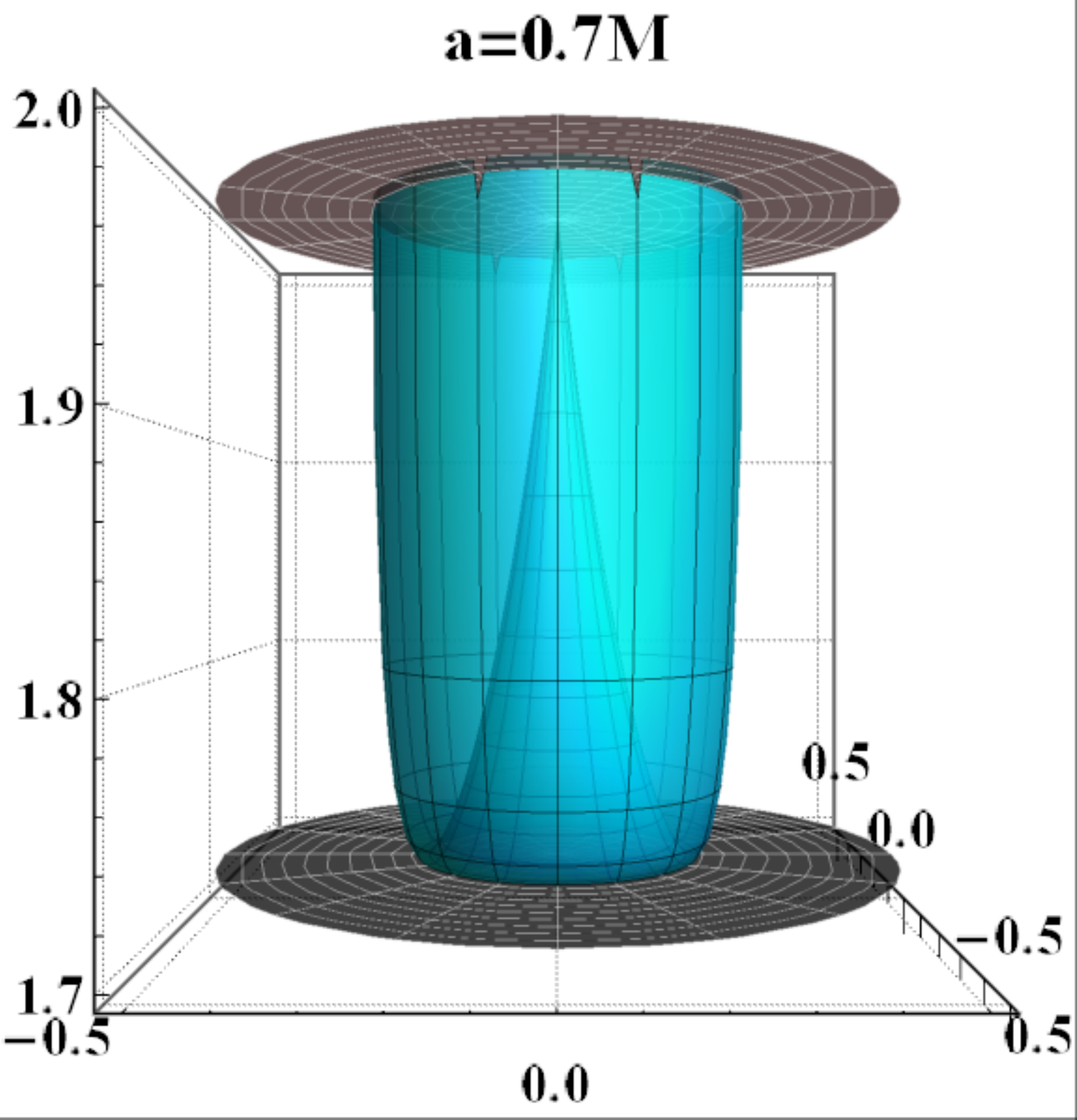}
\end{tabular}
    \caption[font={footnotesize,it}]{{Plots of the surfaces $r_{s}^{\pm}$ (in units of mass) versus the frequency
		$\omega$ for different spin values $a/M$, including \textbf{BH} and \textbf{NS} geometries--see also Figs.\il\ref{Fig:Trav-inr-B}.  The surfaces $r_{s}^{\pm}$ are represented    as  revolution surfaces   with height
		$r_{s}^{\pm}$ (\emph{vertical axes}) and radius $\omega$ (horizontal plane).
		Surfaces are generated by rotating  the two-dimensional curves $r_{s}^{\pm}$ around an axis (revolution of the  function curves $r_{s}^{\pm}$ around the ``z'' axis). Thus,
		$r=$constant with respect to the frequency $\omega$ is represented by a circle under this transformation. The disks in the plots are either $r=M$, $r=r_+$ or $r=r_{\epsilon}^+=2M$. The surfaces $r_{s}^{\pm}$ are green and pink colored, respectively (as mentioned in  the legend). {In the last panel $(a=0.7M)$, both radii $r_s^{\pm}$ are green colored}}}
\label{Fig:QPlot}
\end{figure*}

\begin{figure*}[ht!]
\begin{tabular}{cc}
\includegraphics[scale=.25]{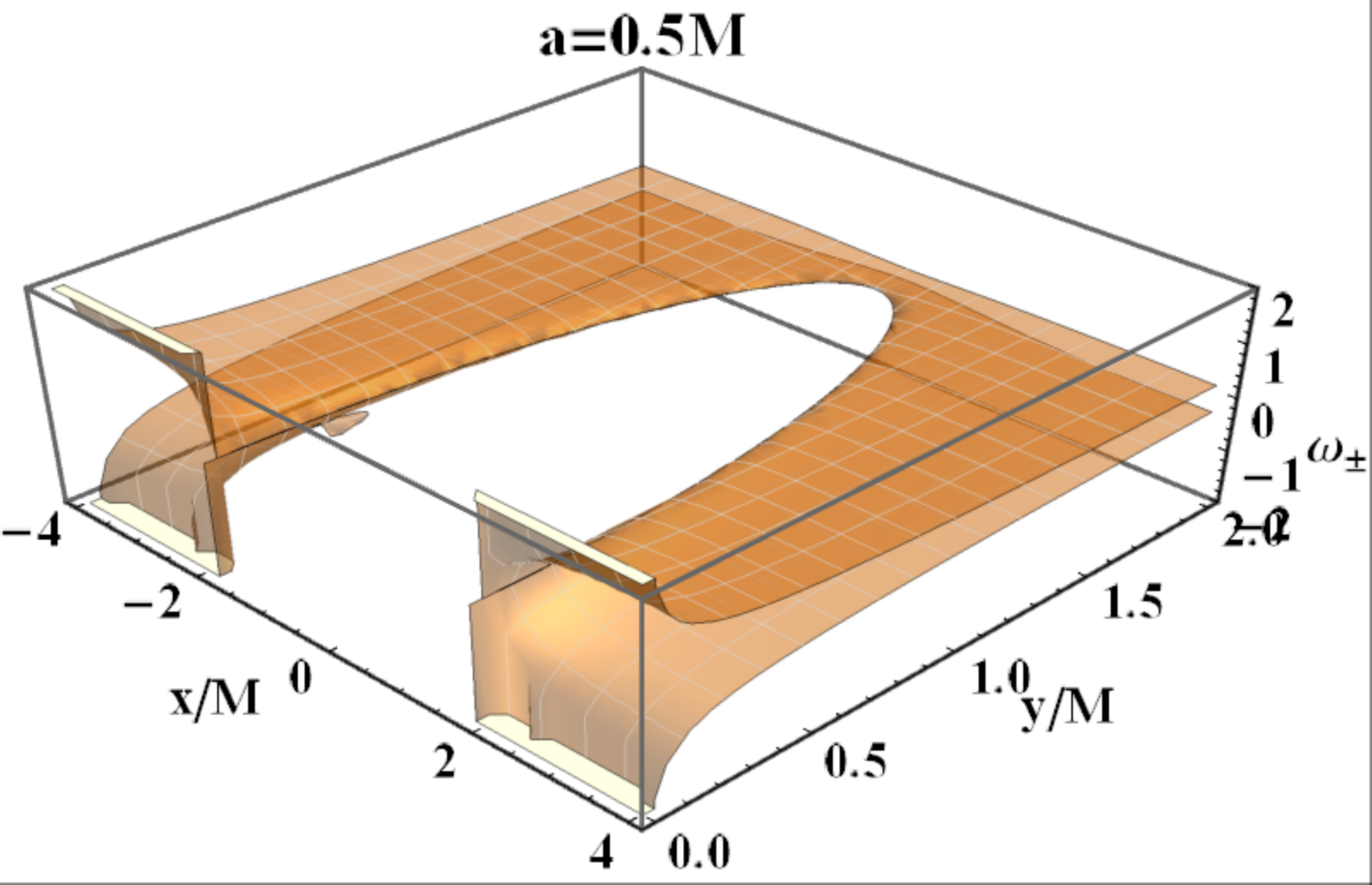}
\includegraphics[scale=.25]{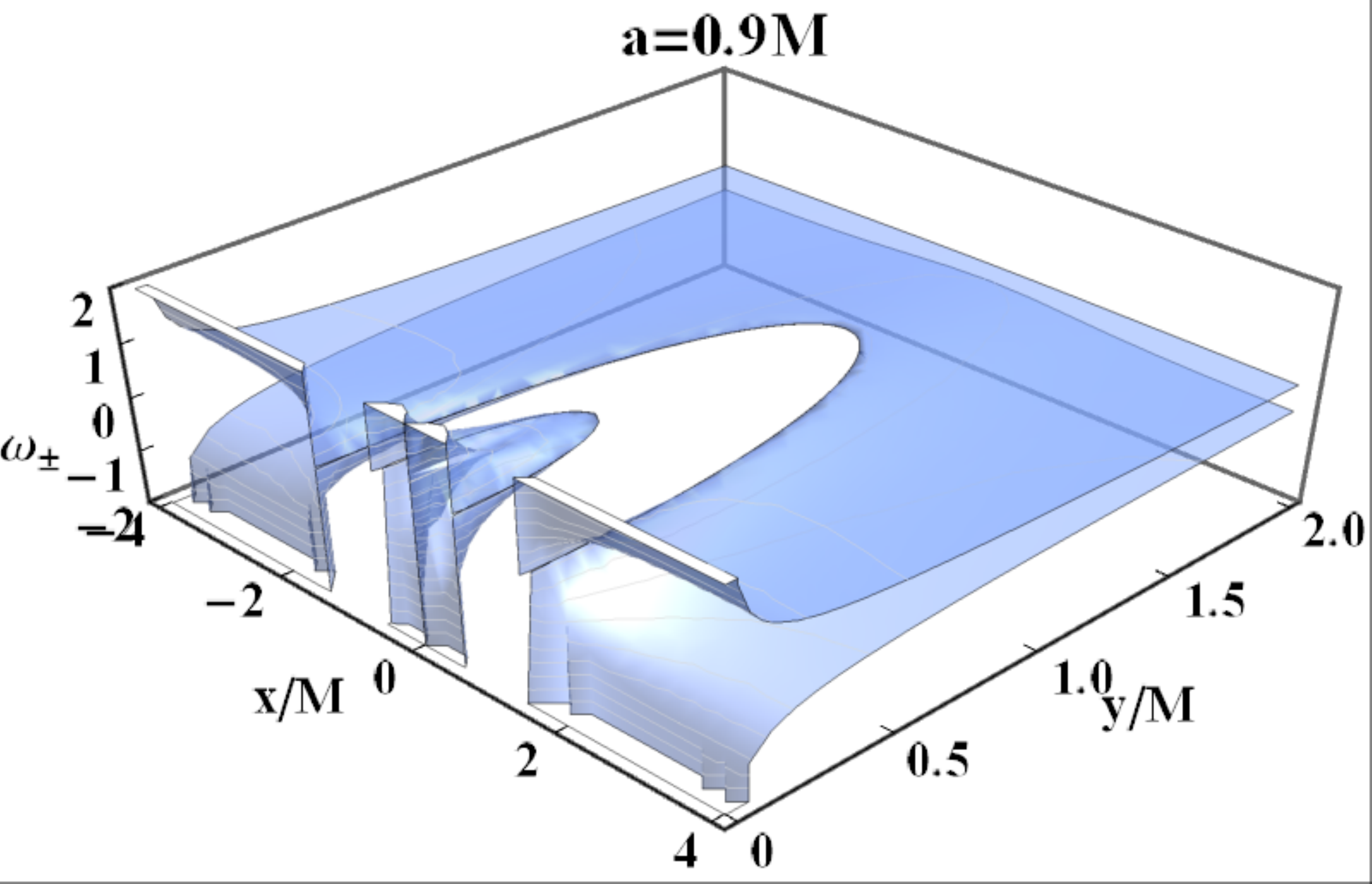}%
\\
\includegraphics[scale=.25]{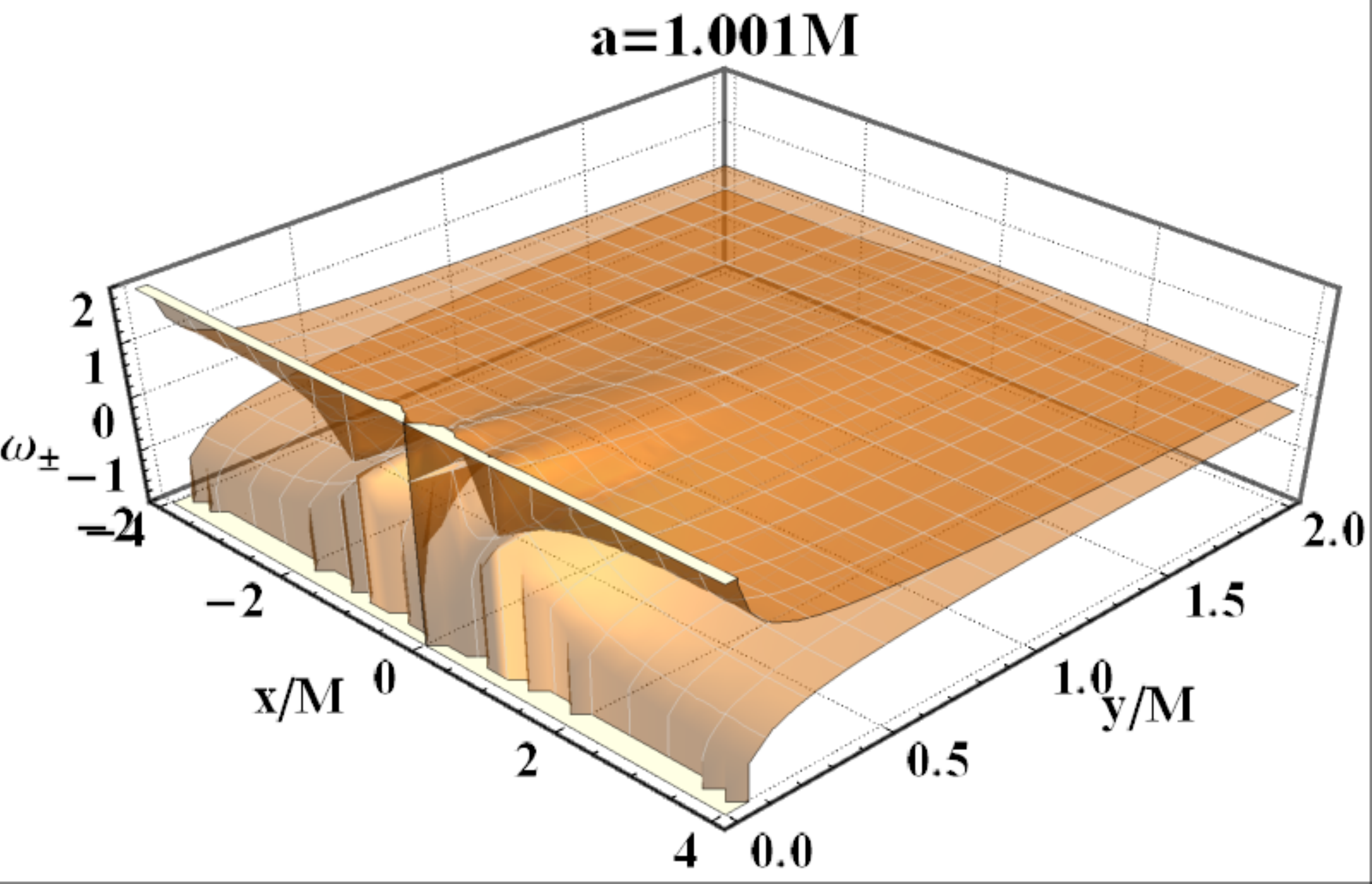}
\includegraphics[scale=.25]{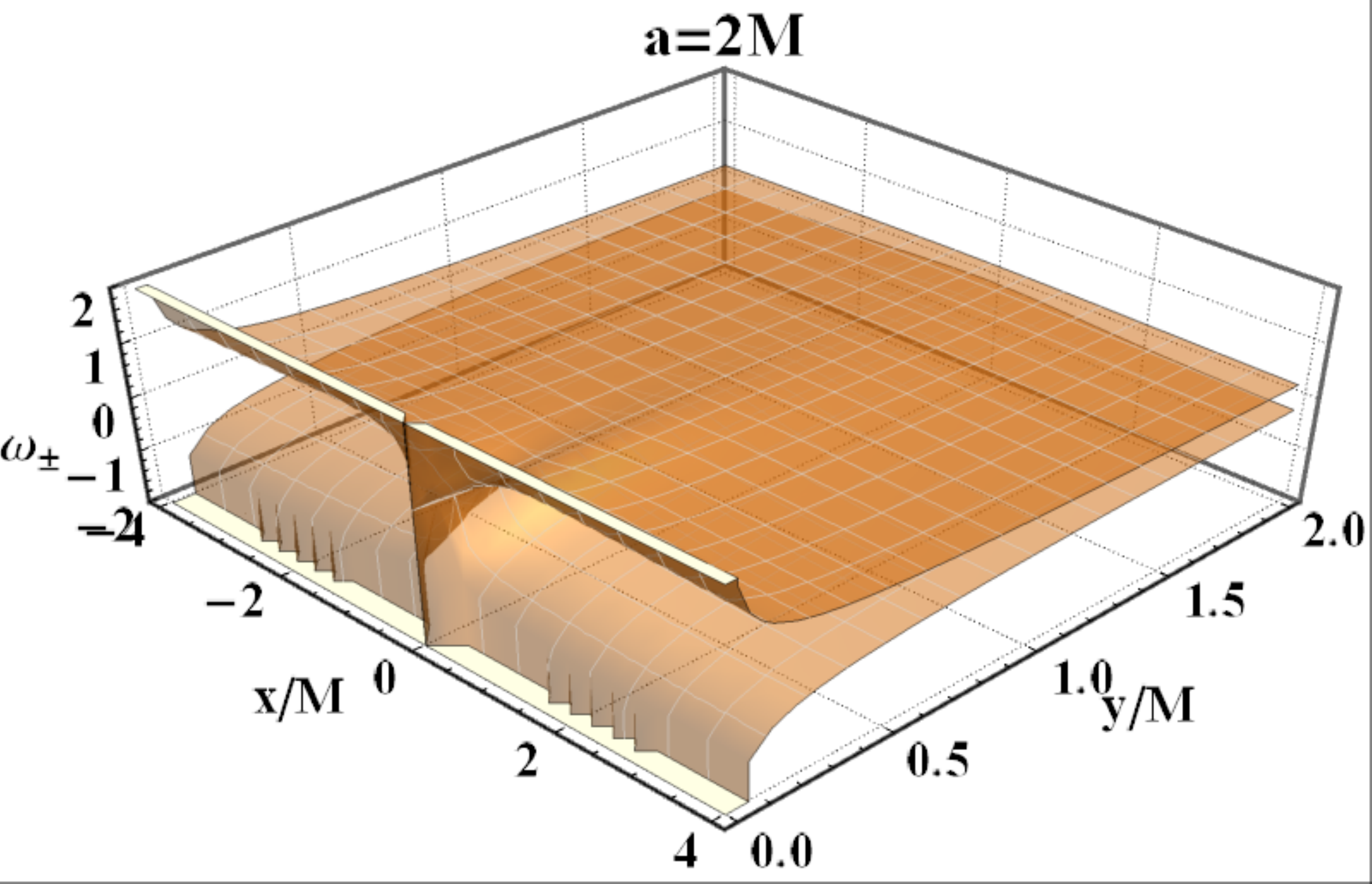}%
\end{tabular}
    \caption[font={footnotesize,it}]{Plots of frequency surfaces $\omega_{\pm}(r,\theta)$ as functions of the radial distance $r$ in Cartesian coordinates $(x,y)$ for different spin values $a$, including \textbf{BH}s and \textbf{NS}s --see also Figs.\il\ref{Fig:Trav-inr-B}.}
\label{Fig:QPlot1}
\end{figure*}

In this section, we briefly study the conditions for the existence   of   \emph{light surfaces} and and their morphology.
The condition (\ref{Eq:ex-ce}),  for the definition of a stationary observer, can be restated in terms of the solutions
$r_{s}^{\pm}$, considering $\omega$ as a fixed parameter.
 Therefore, we now consider the solutions $r_s^{\pm}$ of the   equation for  the light surfaces defined in Eq.\il(\ref{Eq:comb}) in terms of the Killing null generator  $\mathcal{L}_{\pm}$,
{as functions of the frequency $\omega$}.
 We obtain
%
\bea\nonumber
&&\frac{r_s^-}{M} \equiv \frac{2 \beta_1 \sin \left(\frac{1}{3} \arcsin\beta_0\right)}{\sqrt{3}}
,\quad
\frac{r_s^+}{M} \equiv \frac{2 \beta_1 \cos \left(\frac{1}{3}\arccos(-\beta_0)\right)}{\sqrt{3}}\\
&&
\mbox{where}\quad
\beta_1\equiv\sqrt{\frac{1}{\omega ^2}-\frac{1}{\omega_0^2}},\quad \beta_0\equiv\frac{3 \sqrt{3} \beta_1 \omega ^2}{\left(\frac{\omega }{\omega_0}+1\right)^2}\ ,
\eea
where $\omega_0\equiv M/a $ (cf. Eq.\il(\ref{Eq:b-y-proc}) and Fig.\il\ref{Fig:Conf3Dna}).
{For $\omega=1/2$,  in the limiting case of  $a=M$,
we have that $\omega_n=\bar{\omega}_n= \omega_h=1/2$ and  $r_s^{\pm}=M$--see Figs.\il\ref{Fig:Trav-ling}, \ref{Fig:Trav-inr-B}  and
\ref{Fig:Conf3Dna}\footnote{More precisely, it is  $r_s^+=r_s^-=0$ for $a>0$ and $\omega=\omega_0$.
Also,  $r_s^+=r_s^->r_+$ for $a=0$ and
$\omega = \pm\frac{1}{3 \sqrt{3}}$.
In the  extreme Kerr spacetime geometry, we have that
$r_s^+=r_s^->0$ for $a=M$,   $\omega=1/2$ for $r=M$, and  $\omega=-1/7$ for $r=4M$.
 For a Kerr geometry, where  $a/M\in]0,1[$,  it is  $r_s^+=r_s^->r_+$ for  $\omega=\omega_n$ or $\bar{\omega}_{n}$ (one positive and one negative value solution),
while in the naked singularity case  where $a>M$, the condition $r_s^+=r_s^->0$ is valid only for one negative frequency -- see
Figs.\il\ref{Fig:Trav-inr} and \ref{Fig:Trav-inr-B}.}. } Thus, there are  solutions $r_s^+=r_s^{-}=0$
 for $a\in]0,M[$  if $\omega\in (\omega_{n},\bar{\omega}_{n})$ where  (for simplicity we use a dimensionless spin $a\rightarrow a/M$)
{
\bea&&\label{Eq:gent-Mat}
\bar{\omega}_n\equiv\frac{9-a^2+6 \sqrt{9-5 a^2} \cos \left[\frac{1}{3} \arccos\alpha\right]}{a(a^2+27)} 
\eea
\bea&&
\omega_n\equiv\frac{9-a^2-6 \sqrt{9-5 a^2} \sin \left[\frac{1}{3} \arcsin\alpha\right]}{a(a^2+27)}
\\
&&
\alpha\equiv\frac{a^4-36 a^2+27}{\left(9-5 a^2\right)^{3/2}}
 \eea}
The situation is summarized in Table\il\ref{Table:asterisco2}.
\begin{table*}
\centering
\caption{Existence of stationary observers in \textbf{BH}  and  \textbf{NS}  spacetimes, respectively. The spin/mass ratio $a/M$, angular frequencies $\omega$ and orbital ranges $r$ are listed. See also Fig.\il\ref{Fig:Trav-ling}.}
\label{Table:asterisco2}
\textbf{Black holes:}\hspace{2cm} \textbf{Naked singularities:}\\
\begin{tabular}{lcr|lcr}
\hline
 $a\in]0,{a}_1]$& $\omega\in]0,\omega_{+}^{\epsilon}[$ & $r\in  ]r_s^-,r_{\epsilon}^+]\quad $&$\quad a>M$&$\omega\in]0,
\omega_{+}^{\epsilon}[$&$ r\in ]r_s^-,r_{\epsilon}^+]$
 \\
 $a\in]{a}_1,M]$&$\omega\in ]0,\omega_{+}^{\epsilon}[$& $r\in]r_s^-,r_{\epsilon}^+]\quad $& & $\omega=\omega_{+}^{\epsilon}$ &  $ r\in ]r_s^-,r_{\epsilon}^+[$\\
 &
$\omega =\omega_{+}^{\epsilon}$ & $r\in ]r_s^-,r_{\epsilon}^+[ \quad $&&
  $]\omega_{+}^{\epsilon},\omega_0[$& $ r\in ]r_{s}^-,r_s^+[$
 \\
 &$\omega\in ]\omega_{+}^{\epsilon},\omega_n[$ & $r\in]r_s^-,r_s^+[\quad $&&&
\\
\hline
\end{tabular}
\end{table*}
We see that $\omega_n=\omega_{+}^{\epsilon}$ for  $a={a}_1$,
$\omega_n=\bar{\omega}_n=\omega_{h}=1/2$ at\footnote{For a closer look at the role of this special frequency we note that
$\bar{\omega}_n=\omega_n=\omega_{h}=1/2$ at $a=M$ and, clearly, $\omega_0=1/2$ for $a=2M$.
We refer then to Figs.\il\ref{Fig:Pilosrs2}, \ref{Fig:Trav-ling}, \ref{Fig:Trav-inr},
\ref{Fig:Trav-inr-B}, and \ref{Fig:QPlot}.}
$a=M$,
$\omega_{+}^{\epsilon}=\omega_{h}$ at $a=a_s\equiv\sqrt{2 \left(\sqrt{2}-1\right)}M\approx0.91017M$ and $a=0$ (the static solution).
Moreover, we have that  $\omega_0>\bar{\omega}_n>\omega_n>\omega_{+}^{\epsilon}$ and $\omega_n>\omega_{h}$ for \textbf{BH}-sources,
where  $\omega_{h}>\omega_{+}^{\epsilon}$ for $a\in]a_s,M]$.
In the \textbf{NS} case,   there are no crossing points  for the radii $r_{s}^{\pm}$ and   $\omega_0>\omega_{+}^{\epsilon}$
(see Fig.\il\ref{Fig:Trav-ling}).
{The shrinking of the frequency interval  $]\omega_{-},\omega_{+}[$   is also shown in  Figs.\il\ref{Fig:Trav-inr-B}, \ref{Fig:QPlot} and  \ref{Fig:QPlot1}, where the radii  $r_s^{\pm}$ are also plotted as functions of the frequencies}.
Figures\il\ref{Fig:Trav-ling}, \ref{Fig:Trav-inr}, and \ref{Fig:Trav-inr-B}  contain all the information about the differences
between black holes with  $ a <M $, and the case of naked singularities with $a>M$. We summarize the situation in the following statements:
\begin{description}
\item[{Naked singularities spacetimes}:]
 For $a>M$, the  solutions for the equation of the light surfaces  in the limiting case   $\omega=0$ (static observer)  are  located at $r=r_{\epsilon}^+$.
While  for any frequency within the range  $\omega\in]0, \omega_{+}^{\epsilon}[$ there is one solution $r_{s}^-$, for larger frequencies in the range $\omega\in[\omega_{+}^{\epsilon},\omega_0[$ there are two solutions $r_s^{\pm}$.
In the ergoregion $\Sigma_{\epsilon}^+$ of a naked singularity, there exists a  limit $\omega_0\equiv M/a$  for the angular frequency.

\item[Extreme black hole spacetime:]
For $a=M$, we obtain the following set of solutions $(\omega =0, \ r=r_{\epsilon}^+)$,  $(\omega\in]0,1/3[,\;r=r_s^-)$, and
$ ( \omega\in[{1}/{3},{1}/{2}[,\; r=r_s^{\pm})$.

 \item[{Black hole spacetimes}:]
We consider first the class  \textbf{BHI} with   $a\in]0,{a}_1]$. In the limit  $\omega =0$, there exists a solution
for the light surface with $r=r_{\epsilon}^+$.
More generally, the solutions are constrained by the following set of conditions:
\bea
&&\mathfrak{C_1}:\quad\omega\in]0, \omega_{+}^{\epsilon}] \cup \omega\neq\omega_{h}\quad \mbox{with solution}\; r=r_s^-.
\\
&&\mathfrak{C_2}:\quad\omega\in[\omega_{+}^{\epsilon}, \omega_n[\quad \mbox{with solution}\quad r=r_s^{\pm},
\\
&&\qquad\quad \omega=\omega_n,\quad \mbox{with solution}\quad r=r_s^-.
\eea

Then, we consider \textbf{BH} spacetimes   with spin $a\in]{a}_1,a_s[$, where $a_s\equiv\sqrt{2 \left(\sqrt{2}-1\right)}M<{a}_2$.
These spacetimes include a part  of \textbf{BHII}-sources and  the condition $\mathfrak{C_1}$ applies.

%
For spacetimes with rotation $a=a_s$, the conditions $\mathfrak{C_1}$ and $\mathfrak{C_2}$ apply. Then,
in the special case $\omega_{+}^{\epsilon}=\omega_{h}$  or  $\omega =\omega_{+}^{\epsilon}$, there is a solution with $r=r_s^+$.

Finally, for spacetimes with  $a\in]a_s,M[$, which belong to the class of \textbf{BHII} and \textbf{BHIII} sources, the condition
$\mathfrak{C_1}$  holds, whereas   the condition $\mathfrak{C_2}$ applies for frequencies within the interval $\omega_{h}<\omega_n$.
Finally, in the special case  $\omega =\omega_{h}$, there is one solution at $r=r_s^+ $, and for  $\omega =\omega_n$
we have  the solution $r=r_s^-$.
\end{description}

A summary and comparison of these two cases is proposed also in Figs.\il\ref{Fig:Trav-inr} and \ref{Fig:Trav-inr-B},
where the surfaces  $r_{s}^{\pm}$ are studied as functions of  $a/M$ and $\omega$.
It is evident that  the extreme solution  $a/M=1$  is a  limiting case  of both surfaces  $r_{s}^{\pm}$,
varying  both in terms of the spin and the angular velocity $\omega$.
Thus,  the difference between the  regions where stationary observers can exist in the \textbf{BH}  case
(gray regions in  Figs.\il\ref{Fig:Trav-inr-B}) and  in the \textbf{NS} case are clearly delineated.
In \textbf{BH}  spacetimes, the surfaces $r_{s}^{\pm}$  are confined within a restricted radial   and frequency range.
On the other hand, in the naked singularity case, the orbits and  the frequency range is larger than in  the black hole case.
Moreover, the   surfaces $r_{s}^{\pm}$ can be closed  in the case of \textbf{NS} spacetimes, inside the ergoregion,
for sufficiently low values of the   spin parameter,   namely $a\in]M,{a}_4]$.
Furthermore,  in any  Kerr  spacetime, there is a light surface at  $r_{s}^{\pm}=r_{\epsilon}^{+}$ with $\omega=M/{a}_4$.
{
In Sec.\il\ref{Sec:saz}, we complete this analysis by  investigating  the special case of zero angular momentum observers, and we find
all the spacetime configurations in which they can exist. }
\section{Zero Angular Momentum Observers}
\label{Sec:saz}

This section is dedicated to the study of  Zero Angular Momentum Observers (ZAMOs) which are defined by
the condition
\be
\mathcal{L}\equiv u_{\alpha}\xi_{(\phi)}^\alpha=
g_{\alpha\beta}\xi_{\phi}^{\alpha}p^{\beta}= g_{t\phi}\dot{t}+g_{\phi\phi}\dot{\phi}=0.
\ee
In terms of the particle's four--velocity, the condition $\mathcal{L}=0$ is equivalent to
\(
{d\phi}/{dt}=-{g_{\phi t}}/{g_{\phi\phi}}\equiv\omega_{Z}=(\omega_++\omega_-)/2,
\)
where the quantity $\omega_{Z}$ is the   ZAMOs  angular velocity introduced in Eq.\il(\ref{Eq:ex-ce}),
and  the frequency of arbitrary  stationary observers is written in terms of  $\omega_{Z}$ \cite{ergon}.
The  sign of $\omega_{Z}$  is in concordance with  the  source rotation.
The   ZAMOs  angular velocity is a function of the spacetime spin (see Figs.\il \ref{wplotn}  and \ref{rewe}, {where constant ZAMOs frequency profiles are shown}).
In the plane $\theta=\pi/2$, we find  explicitly
\be
\left.\omega_{Z}\right|_{(\theta=\pi/2)} =\frac{2 a M^2}{r^3+a^2 (r+2M)}.
\label{eq:wz}
\ee
As discussed in \cite{Pu:Kerr,Pu:KN,ergon}, ZAMOs along circular orbits with radii $\hat{r}_{\pm}$ are possible only  in the case of ``slowly rotating'' naked singularity spacetimes of class \textbf{NSI}. This is a characteristic of naked singularities which is
interpreted generically as a repulsive effect exerted by the singularity \cite{Pu:Charged,Pu:class,Pu:Kerr,Pu:KN,Pu:Neutral}.
  On the other hand, $\omega_{Z}^2=\omega _*^2$  for $r=r_{\pm}$, while    $\omega_{Z}^2>\omega _*^2$ in the region  $r>r_+$ for \textbf{BH} spacetimes, and in the region  $r>0$ for \textbf{NS} spacetimes (see also Fig.\il\ref{rewe}).

\textbf{ZAMOs angular velocity and orbital regions}

The  ZAMOs angular velocity $\omega_{Z}$   is always positive for  $a>0$, and vanishes only in the limiting case $a=0$.
This  means that the ZAMOs  rotate in the same direction as the source (dragging of inertial frames).

{As can be seen from Eq.(\ref{eq:wz}), the frequency $\omega_Z$ for a fixed mass and $a\neq0$ is strictly decreasing as the radius $r/M$ increases.}

{For the \textbf{NS} regime  it is interesting to investigate the  variation of \textbf{ZAMO}  frequency $\omega_{Z}$
on the orbits   $\hat{r}^{\pm}$. These special radii of the \textbf{NS} geometries do not remain constant under a spin-transition  of the central singularity.
We shall consider this aspect focusing on the curves  $\hat{r}^{\pm}(a)$ of the plane $r-a$  as illustrated in Fig.\il\ref{Fig:L0V0Zamos}.
 This will enable us to evaluate simultaneously  the frequency variation on these special orbits, following a spin variation of the naked singularity in the rage of definition of  $\hat{r}^{\pm}$, and to evaluate the combined effects of a variation in the orbital distance from the  singularity and a change of spin. A similar analysis will be done, from a different point of view, also for stationary observers.}

 In $\Sigma_{\epsilon}^{\pm}$, the velocity $\omega_{Z}=\hat{\omega}^{-}$  (in \textbf{NSs}) always \emph{decreases} with  the  orbital radii $\hat{r}^{-}$, i.e. $\partial_{\hat{r}^{-}}\hat{\omega}^{-}<0$, when the spin increases, i.e. $\partial_{a}\hat{r}^{-}>0$
(see Figs.\il\ref{Fig:L0V0Zamos} and  \ref{rewe}).
As $\hat{r}^+$ monotonically  decreases with the spin  during a  \textbf{NS} spin-up process (see Fig.\il\ref{Fig:L0V0Zamos}), the frequency  $\hat{\omega}^+=\omega_Z(\hat{r}^+)$ decreases in the spin-range   $a\in[M,a_{\omega}[ $, and increases  in the range
$]a_{\omega},a_3]$; therefore,  the special value  $a_{\omega}=1.1987M$ is  a minimum point of the ZAMOs frequency $\hat{\omega}_Z^+$--
see Fig.\il\ref{rewe}.
\emph{Viceversa}, as  $\hat{r}^-$ increases after a \textbf{NS}  spin-up, the corresponding ZAMOs frequency
$\hat{\omega}^-=\omega_Z(\hat{r}^-)$ decreases as the observer moves along the curve   $\hat{r}^-(a)$.
Thus,   we can say that, if the \textbf{NS} spin increases, the frequency  $\hat{\omega}^+$ decreases, approaching, but never reaching,
the singularity, i. e.,  $\partial_a\hat{r}^+<0$  for $a\in[M,a_{\omega}[$.
{
\emph{Viceversa},  increasing the \textbf{NSs}  spin in  spacetimes with   $a\in]a_{\omega},a_3[$, the  frequency  $\hat{\omega}^+$ increases again and the orbit $\hat{r}^+$  moves towards  the  central singularity.
On the other hand, the   frequency  $\hat{\omega}^-$ monotonically decreases with the naked singularity spin, i.e.
$\partial_a\hat{r}^->0$; therefore,  for a fixed \textbf{NS} spin,   the frequency interval decreases, i.e.
 $\hat{\omega}^->\hat{\omega}^+$.
In fact, the velocity $\omega_{Z}$ is strictly \emph{decreasing} with  the  radius ${r}$ in the \textbf{BH} and \textbf{NS} regimes with
$a\neq0$ (i.e. $\partial_r \omega_Z<0$).
Moreover, in general  $\omega_{Z}$  increases   as the observer approaches the black hole at fixed spin, and  it  decreases as  the observer moves far   away from the center of rotation.}

In the static limit, we have that $\omega_{Z}(r_{\epsilon}^+)=\omega_{+}^{\epsilon}/2$.
In fact, the asymptotic behavior of the frequency is determined by the relations
\bea
\lim_{r\rightarrow r_+}\omega_{Z}=\lim_{r\rightarrow r_+}\omega_{\pm}=\omega_{h},
\;\lim_{r\rightarrow+\infty}\omega_{Z}=0, \;\lim_{r\rightarrow0}\omega_{Z}={\omega_0}.
\eea

\textbf{Change in the intrinsic spin}
The angular velocity of the ZAMOs inside $\Sigma_{\epsilon}^{+}$ varies according to the source spin.
This might be especially important in a possible process of spin-up or spin-down as a result of the interaction, for example,
 with the surrounding matter. In \cite{ergon},  this phenomenon and its implications were investigated, considering different regions close to the singularity.
For a fixed orbital radius $r$, the ZAMOs angular velocity strongly depends on the value of the spacetime spin-mass ratio. In particular, depending on the value of the ratio $a/M$, there can exist a radius of maximum frequency $r_e$ given by

\bea\label{Eq:cii-Scli}
{r_e}\equiv
&&
\frac{\sqrt[3]{3} a^2+\Upsilon^2}{3^{2/3}\Upsilon },\quad \Upsilon\equiv\sqrt[3]{9M a^2+\sqrt{3} \sqrt{a^4 \left(27M^2-a^2\right)}}
\eea
{
that are solutions of the equation $\left.\partial_a \omega_Z\right|_{\pi/2}=0$
at which the frequency is denoted by $\omega_e\equiv\omega_{Z}(r_e)$ (see Figs.\il\ref{wplotn} and \ref{rewe}).}
A detailed analysis of the expression for the radius $r_e$ shows that in can exist in spacetimes that belong  to the
class \textbf{BHII}  with spin  $a=a_s$, where  $r_e(a_s)=r_+(a_s)$, and to the classes  \textbf{BHIII}, \textbf{NSI}, and \textbf{NSII} {with the limiting value $a=a_{\diamond}=\sqrt{2}M$}, where $a_{\diamond}:\;r_e=r_{\epsilon}^+$
(see Figs.\il\ref{wplotn} and \ref{Fig:Conf3Dna}).
Spacetimes with spin $a_s$ belong to the class \textbf{BHII}, as defined in   Table\il\ref{Table:asterisco2}, and have been analyzed in the context of stationary observers in  Sec.\il\ref{Sec:p-lS} and  Sec.\il\ref{Sec:w-t-1} (Figs.\il\ref{Fig:Trav-ling}, \ref{wplotn},
\ref{rewe} and \ref{Fig:L0V0Zamos1} {show the behavior of several quantities  related to ZAMOs in relation to other frequencies.}). In this particular case, we have that
\be
\omega_{+}^{\epsilon}=\omega_{h}=0.321797\quad\mbox{and}\quad
r_e(a_s)=r_+(a_s).
\ee
{We   focus our attention on ergoregion   $\Sigma_{\epsilon}^+$, bounded from above by the radius $r_{\epsilon}^+$ and from
below by  $r=0$ and $r=r_+$ for  \textbf{NSs} and \textbf{BHs}, respectively.
We consider the role of the radius $r_{e}$, as the maximum point of the ZAMO frequency,  as a function of the source spin-mass ratio.
 Thus, for black holes with $a\in[0,a_s]$, the frequency $\omega_{Z}$  increases with $a/M$ always  inside the ergoregion; this holds for any orbit inside
$\Sigma_{\epsilon}^+$ (i.e. for a fixed value $\bar{r}\in \Sigma_{\epsilon}^+$, if a \textbf{BH} spin-up shift occurs
 in the range $[0,a_s]$, the function $\omega_Z(\bar{r},a)$ increases with $\bar{r}$). For spins $a\in]a_s,M]$,  instead, the frequency
$\omega_{Z}$ grows with the spin  only for $\bar{r}\in] r_e, r_{\epsilon}^+[$; on the contrary, for radii located close to the horizon,
$\bar{r}\in ]r_+,r_e]$, $\omega_Z(\bar{r},a)$  \emph{decreases} following a spin up in the range  $\in]a_s,M]$ (i.e,
 $\partial_a r_+<0$ and $\partial_a r_e>0$).
In the case of \textbf{NS}-spacetimes,   the frequency $\omega_{Z}(\bar{r}, a)$  is an increasing function of the dimensionless spin
{
 in the \textbf{NS} spin   range
$]M,a_{\diamond}[$ and  on the orbit    $\bar{r}\in]r_e,r_{\epsilon}^+]$. Moreover, the frequency $\omega_{Z}(\bar{r}, a)$
decreases with the spin in the range of values  $a\in]M,a_{\diamond}[$ and on  $\bar{r}\in]0,r_{e}]$}.
This situation is distinctly different for \textbf{NS} with   $a>a_{\diamond}$, for which in the ergoregion an increase of the spin corresponds to a decrease of  $\omega_Z$.
This is an important distinction between different \textbf{NS} regimes.

{
We note that   $r_e(a_s)=r_+(a_s)$ for the spin $a_s=\sqrt{2 \left(\sqrt{2}-1\right)} M$ (see Fig.\il\ref{wplotn}).
Moreover, in  $\mathbf{NSII}$ naked singularity  spacetimes   with spin  $a_{\diamond}=\sqrt{2} M:\;r_e=r_{\epsilon}^+$,
we obtain that $\omega_{Z}^{\epsilon}=\omega_e$--Fig.\il\ref{rewe}. Remarkably,  the spin $a_{\diamond}$ is the maximum point of the frequency
$\omega_{Z}^{\epsilon}(a_{\diamond})=\omega_e(a_{\diamond})\equiv\omega_{Z}^{\epsilon-Max}=0.176777$ and also the   maximum point of the frequency  $\omega_{+}^{\epsilon}$ (see Fig.\il\ref{Table:asterisco2}). In other words, in naked singularity spacetimes with
$a=a_{\diamond}$, where   $r_e=r_{\epsilon}^+$,  the ZAMOs frequency  at the ergosurface $\omega_{Z}^{\epsilon}$ reaches a maximum value which is equal to $\omega_e$, defined through the radius in  Eq.\il(\ref{Eq:cii-Scli}); moreover, the frequency $\omega_{+}^{\epsilon}$ reaches its maximum value at the ergosurface.}
 \begin{figure}[h!]
\centering
\begin{tabular}{cc}
\includegraphics[scale=.3]{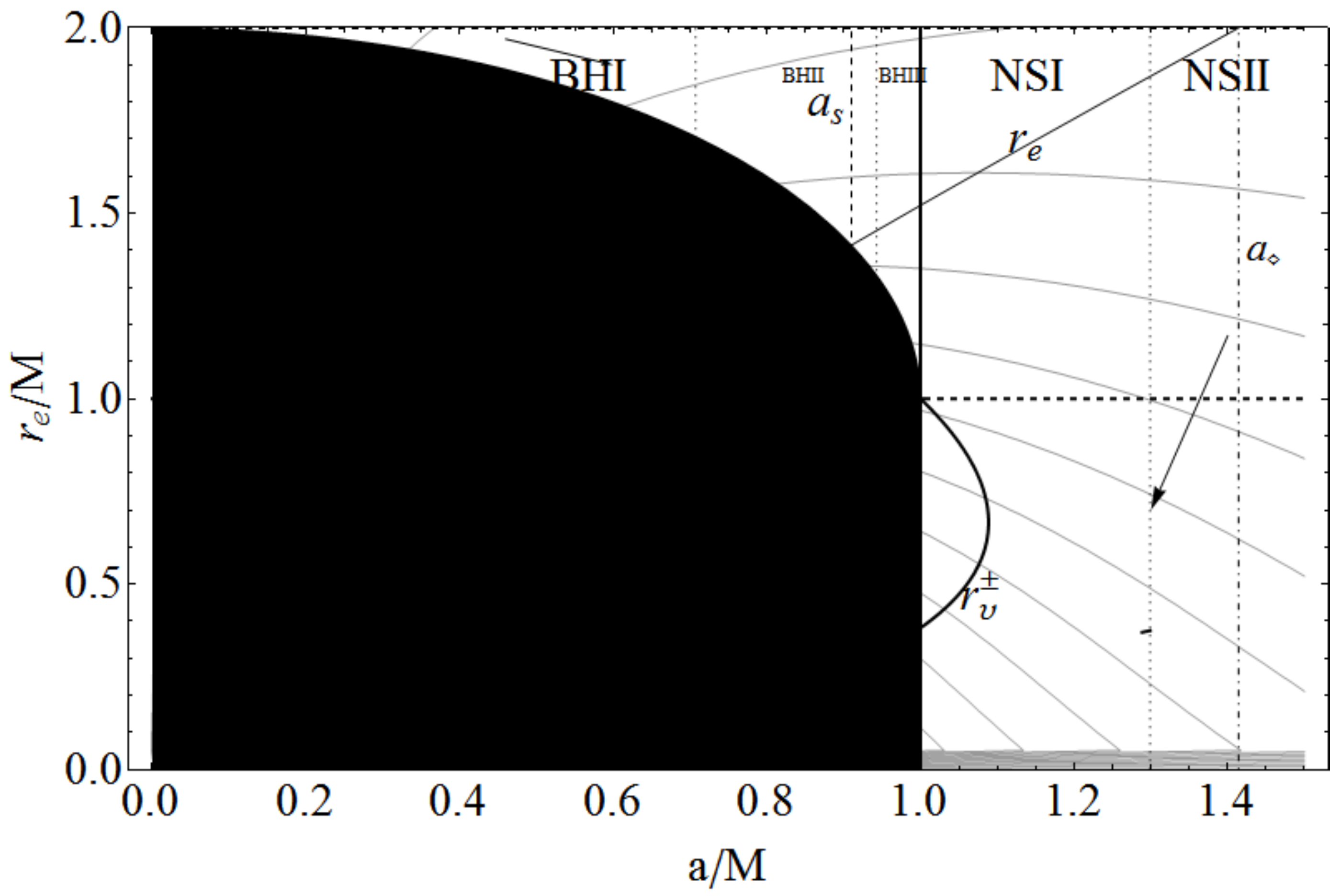}
\end{tabular}
\caption[font={footnotesize,it}]{The plot shows   the orbits (gray curves) of constant ZAMOs velocity $\omega_{Z}=$constant in the \textbf{BH} and \textbf{NS} regions. The radius $r_e$ and the spin $a_s:\; r_e=r_+$ are marked by dashed lines. The arrows show the increasing of the  angular velocity.}
\label{wplotn}
\end{figure}
 \begin{figure}[h!]
\centering
\begin{tabular}{ccc}
\includegraphics[scale=.3]{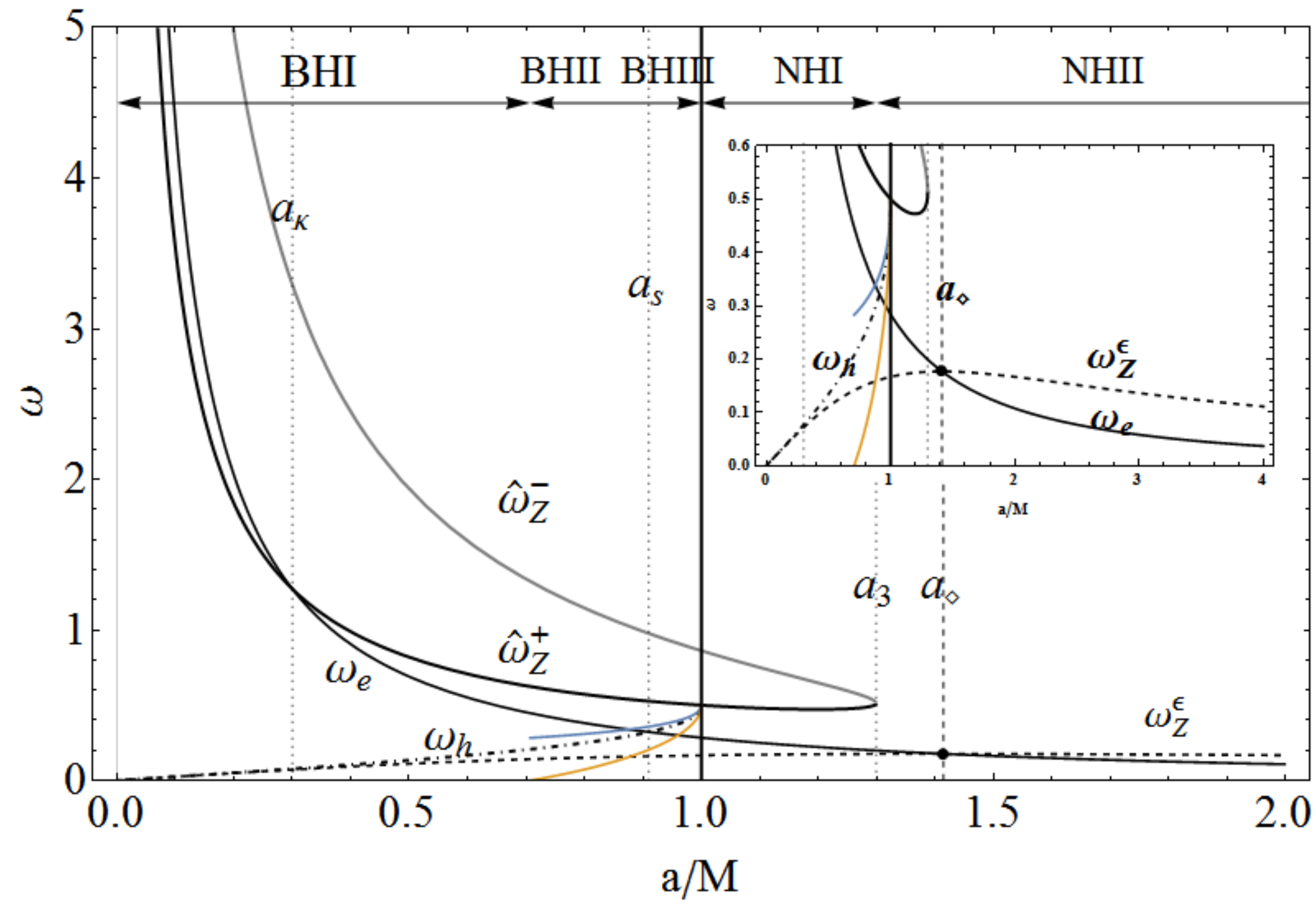}
\\\includegraphics[scale=.17]{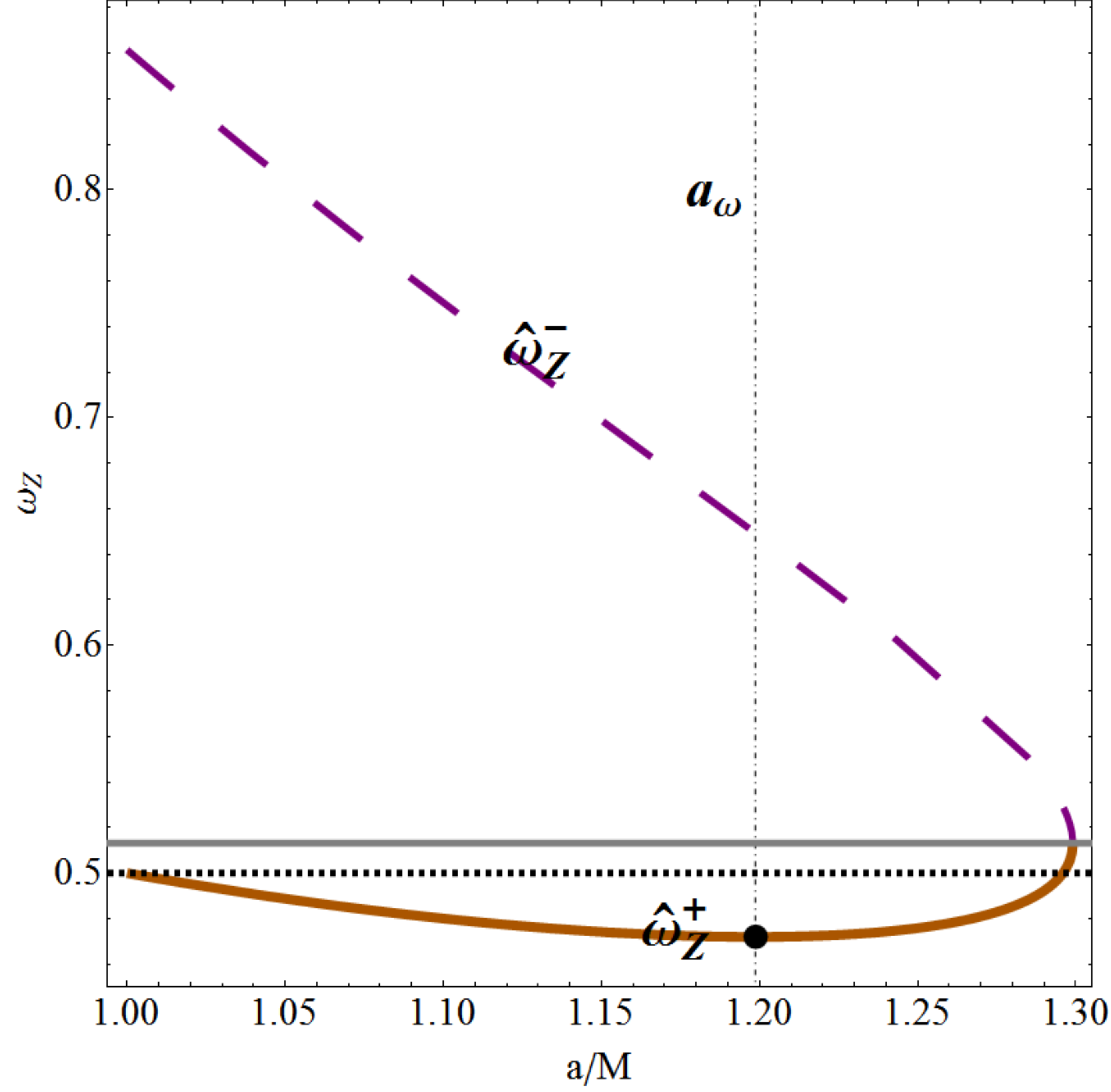}\\
\includegraphics[scale=.2]{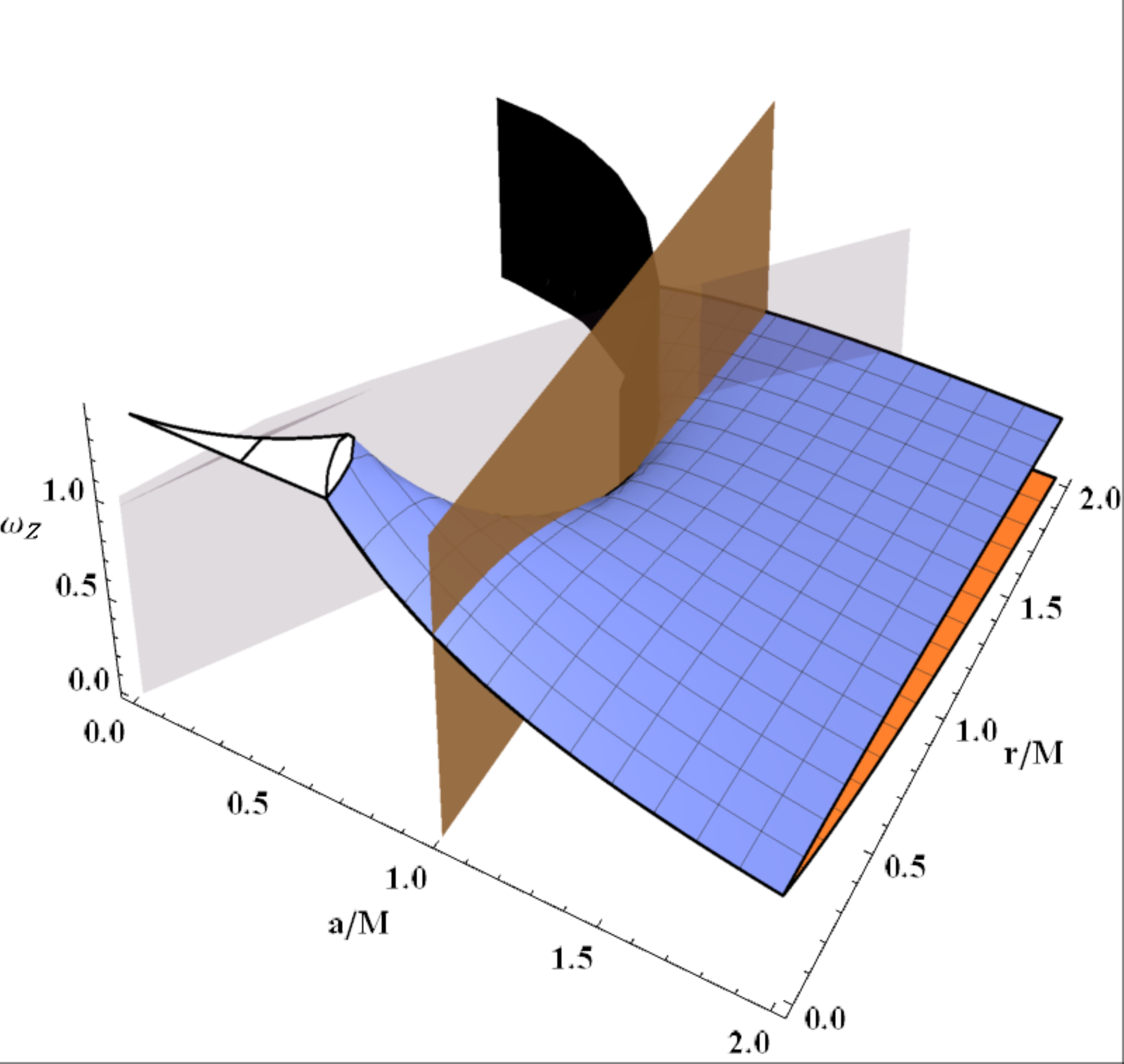}
\end{tabular}
\caption[font={footnotesize,it}]{\footnotesize{Upper panel: The angular velocity $\omega_e\equiv\omega_{Z}(r_e)$ as a  function of $a/M$.
The angular velocities $\omega_{Z}^{\epsilon}\equiv\omega_{Z}(r_{\epsilon}^+)$ (dashed curve), $\omega_{h}\equiv \omega_{\pm}
(r_+)=\omega_{Z}(r_+)$ (dot-dashed curve), $\hat{\omega}_{Z}^{\pm}\equiv\omega_{Z}(\hat{r}_{\pm})$ as functions of the spacetime rotation
$a/M$ for different \textbf{BH} and \textbf{NS} classes.
Dotted lines are $a_{\kappa}\approx0.3002831060M:\; \omega_e=\hat{\omega}^+_{Z}$, $a_s\approx0.91017M: \omega_e=\omega_{h}$, ${a}_3: \hat{\omega}^+_{Z}=\hat{\omega}^-_{Z}={8}/{9 \sqrt{3}}$, and finally the spin $a_{\diamond}=\sqrt{2}M: \omega_{Z}^{\epsilon}=\omega_2$ (dashed line) which is a maximum for $ \omega_{Z}^{\epsilon}$ (the maximum point is marked with a point).
The inset plot is a zoom. The radius $r_e/M$ is a maximum for $\omega_e$. The angular velocities $\omega_{\pm}$ on the \textbf{BH} photon orbit $r_{\gamma}\in \Sigma_{\epsilon}^+$ are also plotted (colored lines). Center panel:  $\hat{\omega}_{Z}^{\pm}\equiv\omega_{Z}(\hat{r}_{\pm})$ as functions of $a/M$ for different \textbf{NS} classes. The minimum point of the ZAMOs frequency $\hat{\omega_Z}^+$ is marked with a point at spin $a_{\omega}=1.19866M$. Bottom panel: The ZAMOs angular velocity $\omega_{Z}$ is plotted as a function of the spin $a/M$ and the radius $r/M$. The plane $a=M$ and  the horizon surface $r=r_+$ are black surfaces. The gray surface denotes the orbit $r_e$.  For both \textbf{NS}  and \textbf{\textbf{BH}} spacetimes, the ZAMOs have a maximum frequency which is a function of $a/M$.  The  black thick curve corresponds to  $\mathcal{E}=0$. The black region denotes the region inside the outer horizon $r<r_+$.} }
\label{rewe}
\end{figure}

\textbf{ZAMOs energy}

The circular motion of  test particles can be described easily by using the effective potential approach \cite{MTW}.
The exact  form  of such an effective  potential  in the Kerr spacetime is well known in the literature (see, for
example, \cite{Pu:Kerr,Pu:KN}).
The effective potential function $\mathcal{V}_{eff}^{+}$
represents the value of $\mathcal{E}/\mu$ that makes $r$ into a turning
point $(\mathcal{V}_{eff}=\mathcal{E}/\mu)$, $\mu$ being the particle mass; in other words, it
 is the value of $\mathcal{E}/\mu$
(in the case of photons,  $\mu$ shall depend on an affine parameter and the impact parameter
$\ell\equiv \mathcal{L}/\mathcal{E}$ is relevant for the analysis of trajectories)
at which the (radial) kinetic energy of the particle vanishes. This can easily be obtained from the geodesic equations with the appropriate constraints or through the normalization conditions of the four-velocities, taking into account the constraints
and the constants of motion \cite{MTW}.
Here we consider specifically an effective potential associated  to the ZAMOs.

The orbits $\hat{r}_{\pm}$ are critical points of the effective potential, i. e., $\hat{r}_{\pm}:\, \partial_r\left.V_{eff}\right|_Z^2=0$.
Here we consider for the ZAMO
$\left.V_{eff}\right|_Z^2=\tilde{\kappa } g_{\phi\phi}[\omega_{*}^2-\omega_Z^2]$  where $\tilde{\kappa}$ is a factor related to the normalization condition of the {ZAMO} four-velocity   ($\tilde{\kappa }=-1$ for timelike ZAMOs, where $u^{\phi}=-\omega_Z u^t$ and $u^t=-\epsilon \mathcal{E}/ g_{\phi\phi}[\omega_*^2-\omega_Z^2]$,   $\epsilon=1$ according to Eq.\il(\ref{Eq:after}); in the ergoregion
$\left.V_{eff}\right|_Z^2>0$, but $\left.V_{eff}\right|_Z^2=0$ for $r=0$ and $r=r_+$).
The energy $E$ of  the ZAMOs  is always positive for both \textbf{BH} and \textbf{NS} spacetimes, and  it grows with the source spin;
 in fact,  solutions for $\left.V_{eff}\right|_{Z}=0$ are not possible because  this would correspond to  the  case of a
null angular momentum with  null energy.
The energy  on  the orbits $\hat{r}_{\pm}$ where $\mathcal{L}=0$ is always positive.
In  \textbf{BH} geometries, the potential $\mathcal{V}_{eff}$, at $\mathcal{L}=0$,  increases  with the distance from the source
and has no critical points as a function of $r/M$.
%
The most interesting case is  then for the slow  naked singularity spacetimes of the first class,  \textbf{NSI} with $a\in]M, {a}_1]$, where there is  a closed  and  connected orbital region of circular orbits with  $r\in]\hat{r}_-,\hat{r}_+[$. The radii $\hat{r}_{\pm}$ are ZAMOs orbits, and in this region the potential  decreases with the orbital radius.
However, in the outer region $r\in]r_+,\hat{r}_-[\cup]\hat{r}_+,2M[$,
the potential  increases with the radius.
This implies that the  radii $\hat{r}_{\pm}$  are possible circular ZAMOs orbits.
 In fact,  $\hat{r}_-$ is an  \emph{unstable} orbit  and  $\hat{r}_+$ is a \emph{stable} orbit.
Thus, in any geometry of   this set, there is a stable orbit for the ZAMOs   with angular velocity
$\hat{\omega}_{Z}^{\pm} \equiv{\omega}_{Z}(\hat{r}_{\pm})$ different from zero,  where
$\hat{\omega}_{Z}^{-}<\hat{\omega}_{Z}^{+}$ (see Fig.\il\ref{rewe}).

 In \cite{ergon},  we investigated the orbital nature of the static limit. Here,
in  Fig.\il\ref{Fig:L0V0Zamos1}, the velocity $\omega_{Z}$    and  the ratio $\mathcal{R}^{\epsilon}\equiv \mathcal{E}^{\epsilon}_{-}/\mathcal{L}^{\epsilon}_-$ (that is, the inverse of the specific angular momentum defined as  $u_{\phi}/u_t$) are considered  as functions
 of the source spin at the static limit.
We explore  the relation between the  ZAMOs and  the stationary observes, where $\omega_{Z}=(\omega_++\omega_-)/2$, for
 \textbf{NSI} sources at the static limit.
A maximum value, $\mathcal{R}^{\epsilon}=0.853553M$, is reached at $a=2M\in \mathbf{NSII}$. Also,
a maximum value $\omega_{Z}^{\epsilon-Max}=0.176777$ exists for the ZAMOs angular velocity  at $a=a_{\diamond}\in \mathbf{NSII}$.
This  ratio is always greater than the angular momentum of the ZAMOs at the static limit.

In  \textbf{BH} spacetimes, the  angular velocity for stationary observers is limited by the value $\omega_{h}$ which occurs
for the radius  $r_+$. We can evaluate the deviation of this velocity in a neighborhood of the radius $r_+$,
since the four-velocity of the observers rotating with $\omega$ (where $u^a\equiv\xi_t+\omega \xi_{\phi}$) must be  timelike outside the horizon  and therefore it has to be
 \(
 \mathcal{R}={\mathcal{E}}/{\mathcal{L}}>\omega_{h}
 \)
in that   range ({the event horizon of a  Kerr black hole rotates with angular velocity $\omega_{h}$ \cite{Wald}}).
This limit cannot be extended to the case of naked singularities.
However,  one can set similarly the   threshold   $ \mathcal{E}>\omega_a \mathcal{L}$ in the case of circular orbits,
  where the frequency limit is restricted to the values  $\omega_a\in[1,{a}_{\mu}^{-1}M[$ as
$\omega_a\in[\omega_0(a=M),\omega_0(a_{\mu})[$.
\begin{figure}[h!]
\centering
\begin{tabular}{cc}
\includegraphics[scale=.3]{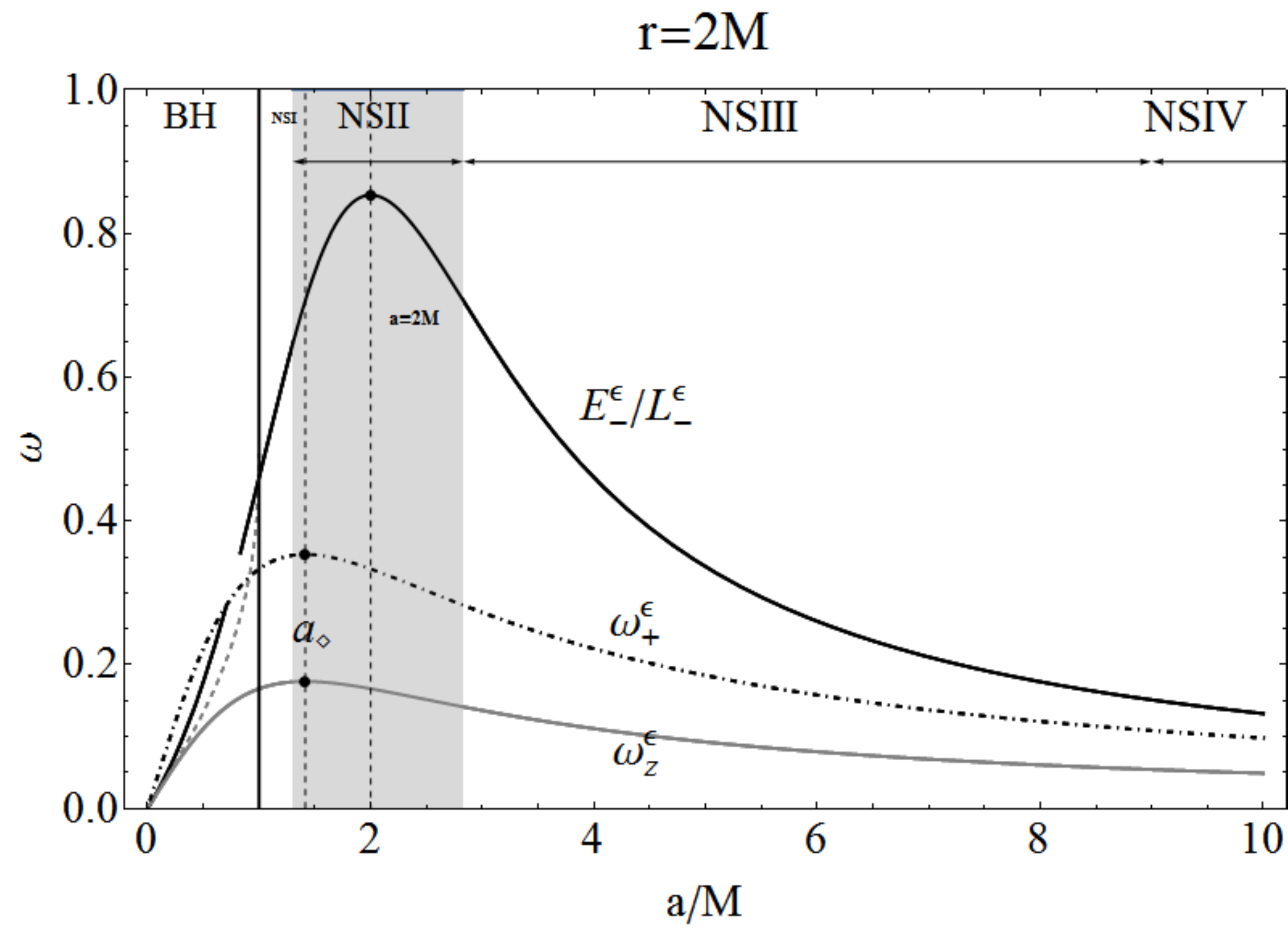}
\\
\includegraphics[scale=.3]{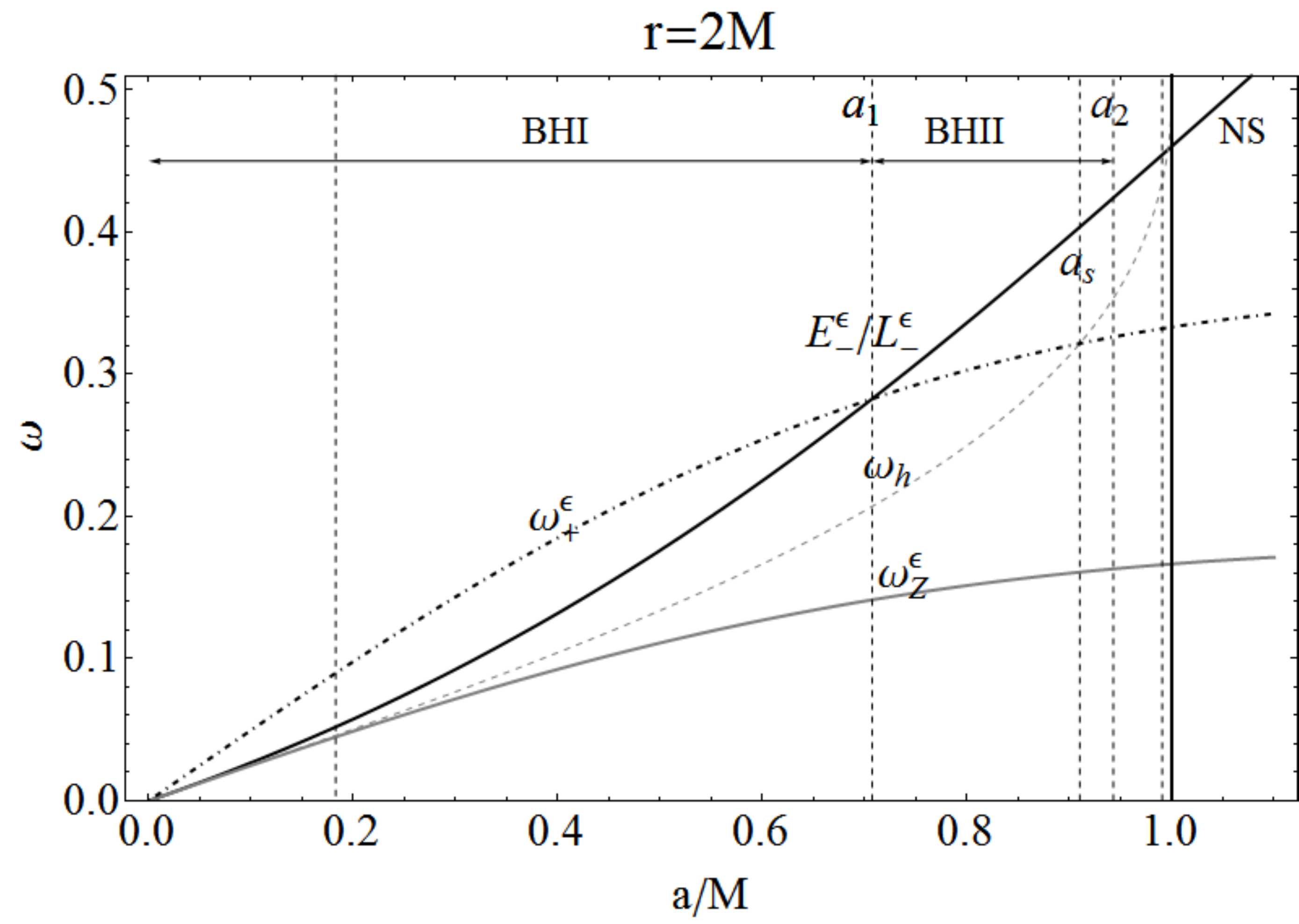}
\end{tabular}
\caption[font={footnotesize,it}]{Upper panel: The ratio $\mathcal{E}^{\epsilon}_{-}/\mathcal{L}^{\epsilon}_-$ and  the angular momentum of the ZAMOs $\omega_{Z}^{\epsilon}$ as a function of $a/M$ in the static limit $r=r_{\epsilon}^+$.
The angular momentum $\omega_+^{\epsilon}\equiv\omega_+(r_{\epsilon}^+)$ which is a boundary frequency for the stationary
observer (outer light surface) is  plotted (gray curve). The radius $r_{\epsilon}^+$  is defined by the condition
$\omega_-(r_{\epsilon}^+)=0$, $\omega_{h}$ is the ZAMOs angular velocity  on $r=r_+$, i.e.  $\omega_{\pm}(r_{\pm})=\omega_{h}$. The maxima are denoted by points. The \textbf{NSII} region is in light-gray. A zoom of this plot in the \textbf{BH} region is in the bottom panel. }
\label{Fig:L0V0Zamos1}
\end{figure}

\section{Summary and Conclusions}
\label{sec:con}
In this work, we carried out a detailed analysis of the physical properties of stationary observers moving in the ergoregion along equatorial circular orbits
in the gravitational field of a spinning source, described by the stationary and axisymmetric Kerr metric.
We derived the explicit value of the angular velocity of stationary observers and analyzed all possible regions where circular motion is allowed, depending on the radius and the rotational Kerr parameter. We found that in general the region of allowed values for the frequencies is larger for naked singularities than for black holes. In fact, for certain values of the radius $r$, stationary observers can exist only in the field of naked singularities. We interpret this result as a clear indication of the observational differences between black holes and naked singularities. Given the frequency and the orbit radius of a stationary observer, it is always possible to determine the value of the rotational parameter of the gravitational source. Our results show that in fact the probability of existence of a stationary observer is greater in the case of naked singularities that in the case of black holes. Moreover, it is possible to introduce
a classification of rotating sources by using their rotational parameter which, in turn, determines the properties of stationary observers.
Black holes and naked singularities turn out to be split each into three different classes in which stationary observers with different properties can exist. In particular, we point out the existence of weak ({\bf NSI}) and strong ({\bf NSIII}) naked singularities,
corresponding to spin values close   to or distant from the limiting case of an extreme black hole, respectively.

Light surfaces are also a common feature of rotating gravitational configurations. We derived the explicit value of the radius for light surfaces on the equatorial plane of the Kerr spacetime. In the case of black holes, light surfaces are confined within a restricted radial   and frequency range. On the contrary, in the naked singularity case, the orbits and  the frequency ranges are larger than for black holes.
Again, we conclude that light surfaces can be found more often in naked singularities. The observation and measurement of the physical parameters of a particular light surface is sufficient to determine the main rotational properties of the spinning gravitational source.
{We believe that the study of light surfaces (defining the ``throat'' discussed in Sec.\il\ref{Sec:1-st})   has important applications regarding the possibility of directly observing a black hole in the immediate vicinity of an event horizon
(within the region defined by  the  static limit), as this seems to be  possible in the immediate future through, for example,
the already active   Event Horizon Telescope  (\textbf{EHT}) projects\footnote{
 http://www.eventhorizontelescope.org/}}.

We also analyzed the conditions under which a ZAMO can exist in a Kerr spacetime. In particular, we computed the orbital regions and the energy of ZAMOs. The frequency of the ZAMOs is always positive, i.e., they rotate in the same direction of the spinning source as a consequence of the dragging of inertial frames. The energy is also always positive. The most interesting case is that of slowly rotating
naked singularities  ({\bf NSI}) where there exists a closed and disconnected orbital region. This particular property could, in principle, be used to detect naked singularities of this class. We derived the particular radius at which the frequency of the ZAMOs is maximal, showing that the measurement of this radius could be used to determine whether the spinning source is a black hole or a naked singularity and its
class, according to the classification scheme formulated here.
To be more specific, from Table\il\ref{Table:asterisco2} we infer that the existence of stationary observers in black hole spacetimes  is limited from above by the frequency
$\omega_+^{\epsilon}$,  which is the highest frequency
on the static limit, implying the frequency lower bound $\omega=0$ -- see also
Fig.\il\ref{Fig:Trav-ling}. In this figure, we also show the maximum frequency,
$\omega_{\epsilon}^+$, at the static limit for a naked singularity with
$a=a_{\diamond}=\sqrt{2}M \in \mathbf{NSII}$. This spin  plays an important role for the variation of the ZAMOs frequency  in \textbf{NSs} in terms of the singularity dimensionless spin -- see Fig.\il\ref{rewe} and Fig.\il\ref{Fig:L0V0Zamos1}.
On the other hand, for strong \textbf{BHs}, with $a>a_1$, the frequency is bounded from
 below by $\omega=\omega_{\epsilon}^+$ and from above by $\omega_{n}$,  as the radial upper bound is $r_s^+$.
A  similar situation occurs  for  \textbf{NSs}, provided that  $\omega_n$ is replaced with the limiting frequency
$\omega_0$.
The special role   of the \textbf{BH} spin $a_1$ is related to the presence of the  photon  circular orbit in the \textbf{BH} ergoregion, which is  absent in \textbf{NS} geometries;
consequently,   as seen in Table\il\ref{Table:asterisco2}, there is no distinction between the
naked singularities classes.
However, the analysis of the frequencies   in  Fig.\il\ref{Fig:Trav-ling}
shows differently that there are indeed distinguishing features in the corresponding ergoregions.
In the case of naked singularities, the frequency range  of stationary observers has as a boundary the outer light-surface, $r=r_s^+$, then it narrows as the spin increases, and finally vanishes near  the static limit.

The frequency of the orbits on the static limit, in fact, converges
to the limit $\omega_0=M/a$, which is an important  frequency threshold for the \textbf{NS} regime.
The presence of a maximum for the special \textbf{NS} geometry with  $a=a_{\diamond}$
 on the static limit is  symptomatic for the nature of this source -- see Figs.\il\ref{Fig:Trav-ling}, \ref{Fig:Trav-inr-B} and
\il\ref{Fig:L0V0Zamos1}.
The  study of the surfaces   $r_s^{\pm}$ on the plane $(r,\omega)$, for different values of the spin-mass ratio, shows a clear difference between the allowed regions in naked singularities and black holes (gray region in Fig.\il\ref{Fig:Trav-inr-B}).
 There is an  open ``throat''  between the spin values  $a\lesssim M$ (strong \textbf{BHs}) and  $a\gtrapprox M$ (very weak \textbf{NSs}), with an opening of the cusp (at $r=0$ in these special coordinates) for the  frequency $\omega=0.5$.
We note a change in the situation for spins in   $a/M\in]1, 1.0001]$;  this region is in fact extremely sensitive to a change of the source spin;  the throat of $r_s^{\pm}$ has, in this special spin range,  a saddle point around $(r=M, \omega=1/2)$ between $[a_{\mu},a_3]$, which is not present  in stronger singularities.
 The spins in this range are related to   the negative state energy and the radii $r_{\upsilon}^{\pm}$,
where the orbital energy is $\mathcal{E}=0$ -- Fig.\il\ref{Fig:L0V0Zamos}.
Particularly, we point out the spin   $a=a_{\sigma}=1.064306M$, where
$r_{\upsilon}^-=r_{\Delta}^-=0.5107M$, for which at  $r_{\Delta}^{\pm}$ there is a critical point of the frequency amplitude $\Delta\omega^{\pm}$.
In
  \textbf{BH} geometries,
the frequencies increase with the spin and with the decrease of the radius towards the horizon. The curves $r_s^{\pm}$
continue to increase with the presence of a transition throat at $r=M$  that increases, stretching and widening.
This throat  represents a ``transition region'' between \textbf{BH} and super-spinning sources from the viewpoint of stationary observes.
The regions outlined here play a distinct role in the collapse processes with possible spin oscillations and different behaviors for  weak, very weak, and strong naked singularities.
  As the spin increases, the frequencies of \textbf{NSs} observes move to lower values,
  widening the throat.   This  trend,  however, changes with the spin,  enlightening some  special thresholds.

This analysis shows firstly  the importance of the limiting frequency $\omega_0=M/a$, determining  the main properties of both frequencies  $\omega_{\pm}$ and  the radii $r_{s}^{\pm}$; it is also   relevant in relation to ZAMOs dynamics in \textbf{NS} geometries. In this way, we may see $\omega_0$ as an  extension of the frequency  $\omega_h$ at the horizon for
 \textbf{BH} solutions--Fig.\il\ref{Fig:Trav-ling}.
In the \textbf{NS}  regime, all the curves $r_s^{\pm}$ converge to the same ``focal point''
$r=0$, regardless of the type of naked singularity,  but as  $\omega_h$ is  the limiting frequency at the \textbf{BH} horizon, each source is characterized by only one
$\omega_0\neq0$ frequency.
 The greater is the spin, the lower is the frequency $\omega_{\pm}$ at fixed radius, and  particularly in the neighborhood of the singularity ring, according to the limiting value
$\omega_{0}$.   The frequency range   at fixed $r/M$  narrows   for higher dimensionless
\textbf{NS}  spin $a/M$.
  This  feature distinguishes between strong, weak and very weak naked singularities.
From  Figs.\il\ref{Fig:Trav-inr-B} it is  clear also  that the throat  of the light-surfaces
$r_{s}^{\pm}$, in the plane $r-\omega$, for different spins $a/M$ closes for  $a\approx M$, which is a spin transition region that includes the  extreme Kerr solution.
This region has been   enlarged in
Fig.\il\ref{Fig:Trav-inr-B}-bottom. Figures\il\ref{Fig:QPlot}  and \ref{Fig:QPlot1} show from  a different perspective    the transition between the \textbf{BH} region, gray region in Figs.\il\ref{Fig:Trav-inr-B}, and the \textbf{NS} region  for different spins.
Any spin oscillation in that region generates a tunnel in the light-surface\footnote{Since any simulation of stellar collapse  returns to the \textbf{BH} regime,
  there must be some (retroactive) mechanism that closes the observer tunnel,  as  even light does not run away in the forbidden region at $r<M$. Moreover,  hypothetical super-luminary matter would violate the bonding of the tunnel wall.}. The transition region is around
$ \omega_{\pm}\approx1/2 $, which is a special value related to the spin $a=2M$  of strong naked singularities--see
Figs.\il\ref{Fig:Pilosrs2} and \ref{Fig:Conf3Dna}. In this region, as  in  the neighborhood of the ring singularity  $(r=0)$,
 the orbital range reaches  relatively small values\footnote{It is worth to mention that  predicted quantum effects   close to the singularities could play a major  role in this region.
 However, we recall that the extreme limit  $a=M$ in this model  is never faced, as we continue to see the spacetime  for all  \textbf{NSs} using a Boyer-Lindquist frame.
It is well known that approaching the horizon at $a=M$,   the radial coordinate velocity
appears as never penetrating  the black
hole, spiraling as $t$ goes to infinity. This is the consequence of a coordinate singularity which can be avoided by using Kerr coordinates or Eddington-Finkelstein coordinates.}.
This shows the existence of limitations for a spin transition in the parameter region of very weak naked singularities,  pointed out also in \cite{Pu:Charged,Pu:class,Pu:Neutral,Pu:KN,Pu:Kerr}.

On the other hand,  in  the strong \textbf{NSs} regimes, a spin threshold  emerges at
$a=2M$ and $a=M$ (see Figs.\il\ref{Fig:Pilosrs2}, \il\ref{Fig:Trav-ling} and \il\ref{Fig:Trav-inr-B}).
In Fig.\il\ref{Fig:L0V0Zamos1}, we analyze the properties at the static limit
$r_{\epsilon}^+$.   The  maximum value  of  $\mathcal{E}_-/\mathcal{L}_-$ is then reached in the ergoregion of the  \textbf{NSII} class\footnote{
 The throat depth  in the region would lead to an immediate change of the observers properties and it  is reasonable to ask if this may  imply an  activation instead of a ``positive feedback'' phenomenon.
  We recall that in this scenario, we are not considering a change of symmetries which would have an essential role. Then it is important to emphasize that in these hypothetical  spin transitions, the external boundary of the ergoregion   remains unchanged,
  but not the frequency at the static limit.}.
Around $a=a_3$  the throat  width  becomes more or less constant. The situation is different for $a>a_3$  and  $a>2M$ and then for  $a_4$,  where the frequencies range  narrows, and near  $r=r_{\epsilon}^+$ becomes restricted to a small range of a few mass units in the limit of large spin  $a/M$.
In strong and very strong  \textbf{NSs}, the  wide  region is inaccessible for stationary observers, whereas it is accessible in the \textbf{BH} case. This  significantly separates  strong and weak \textbf{NSs}, and distinguish them from the \textbf{BH} case.
Interestingly,
the saddle point
  around $r=M$, which narrows the throat of frequencies even in the case of \textbf{NS} geometries for
$a\in ]M,a_{\sigma}[$, could perhaps be viewed as a trace of the presence of $r_+$, which is
absent for   $a >M$.  For $a=a_{\sigma}$, where the  saddle point disappears,
the  shape of the $r_{s}^{\pm}$ tube is different.
This, on the other hand, would suggest  that the existence of the flex in the case of very weak
\textbf{NSs} would prevent a further increasing of the spin. This does not hold for a transition to stronger  \textbf{NSs}, $a\geq a_{\sigma}$,
where no saddle point is present--Fig.\il\ref{Fig:L0V0Zamos}-bottom.  Obviously, the consequences of the hypothetical transition processes should also take into account the transient phase times.
{Very weak naked singularities show a ``rippled-structure'' in the frequency profiles of $\omega$ with respect to $r/M$ and $a/M$,  as appears in Figs.\il\ref{Fig:Conf3Dna}, \ref{Fig:Trav-inr-B}, \ref{Fig:QPlot}, and \ref{Fig:QPlot1}. The  significance of this structure
is  still to be fully investigated, but it may be seen perhaps as a fingerprint-remnant of the \textbf{BH} horizon. This may open an interesting perspective  for the study of \textbf{NS} geometries.}

{An interesting application of our results would be related to the characterization of the  optical phenomena in the Kerr naked singularity and black hole  geometries, such as the \textbf{BH} raytracing and the determination of the \textbf{BH} silhouette (shadow).
The light escape cones are a  key element for such phenomena.
Light escape cones of local observers
(intended as sources)  determine the  portion of radiation emitted by a source that  could escape
to infinity and the one which is  trapped.
This is related to the study of the radial motion of photons because the boundary of the escape cones is given by
directional angles associated to unstable spherical photon orbits.
Light escape cones can be identified in locally non-rotating frames, in frames associated  to circular geodesic  motion
and in radially free-falling observers \cite{RAG,Schee:2008kz,Stuchlik:2010zz,taka,Schee:2008fc}. We want to point out, however,
that light escape cones do not define the properties of the light-cone causal structure, and are not directly related to stationary observers; they rather depend on the photon orbits. A thoroughout analysis of the photon circular motion  in the region of the
ergoregion can be found in
\cite{ergon}.  In Figs.\il\ref{Fig:Pilosrs1}, \ref{Fig:Pilosrs2}, \ref{Fig:Trav-ling} and \ref{rewe}, we show the photon orbit $r_{\gamma}$
and the limiting frequencies crossing this radius; this enlightens the relation with the frequency $\omega_n$.
We consider there in more detail the relation  between the quantities $\omega_Z$ $\omega_*$, the constants of motion $\mathcal{L}$ and
$\mathcal{E}$ and the effective potential, briefly addressed also in Sec.\il \ref{Sec:saz}.}

In general, we see that it is possible to detect black holes and naked singularities  by analyzing the physical properties (orbital radius and frequency) of stationary observers and ZAMOs.
 Moreover, the main physical properties (mass and angular momentum) of the spinning gravitational source can  be determined by measuring the parameters of stationary observers. This is certainly important for astrophysical purposes since the detection and analysis of compact astrophysical objects is one of the most important issues of modern relativistic astrophysics.
In addition, the results presented in this work are relevant  especially for investigating non-isolated singularities, the energy extraction processes, according to Penrose  mechanism, and the gravitational collapse processes which lead to the formation of black holes.

\begin{acknowledgements}
D.P. acknowledges support from the Junior GACR grant of the Czech Science Foundation No:16-03564Y.
This work was partially supported
by UNAM-DGAPA-PAPIIT, Grant No. 111617.
D.P. thanks Dr. Jan Schee for interesting discussion on the light escape cone and stationary observers.
\end{acknowledgements}


\end{document}